\documentclass[12pt]{article}
\pdfoutput=1
\usepackage{multirow}
\usepackage{calc}
\usepackage{amsmath,amssymb,amsthm,amscd}
\numberwithin{equation}{section}
\usepackage[bf]{caption}
\usepackage{longtable}
\usepackage{array}
\usepackage{enumerate}
\usepackage{float}
\usepackage{subfig}
\usepackage{multicol}
\usepackage{multirow}
\usepackage{epsfig}
\usepackage{graphicx}
\usepackage{pict2e}
\usepackage{color}
\usepackage{authblk}
\usepackage{cite}
\usepackage[all]{xy}
\usepackage[width=1.09\textwidth]{caption}
\usepackage{bbm}
\usepackage{hyperref}

\usepackage[utf8]{inputenc}
\usepackage[T1]{fontenc}
\usepackage[english]{babel}
\usepackage{xspace}
\usepackage{pifont}
\usepackage[OT2,T1]{fontenc}

\DeclareSymbolFont{cyrletters}{OT2}{wncyr}{m}{n}
\DeclareMathSymbol{\Sha}{\mathalpha}{cyrletters}{"58}

\def\cO{{\mathcal O}}

\newcommand{\beq}{\begin{equation}}
\newcommand{\eeq}{\end{equation}}
\newcommand{\bea}{\begin{eqnarray}}
\newcommand{\eea}{\end{eqnarray}}

\newcommand{\nn}{\nonumber}

\begin{document}

\baselineskip=15pt
\begin{titlepage}

\begin{center}
\vspace*{ 2.0cm}
{\Large {\bf F-theory on Quotient Threefolds with (2,0) Discrete Superconformal Matter }}\\[12pt]
\vspace{-0.1cm}
\bigskip
\bigskip 
{
{{Lara B.~Anderson}$^{\,\text{a}}$}, {{Antonella~Grassi}$^{\,\text{b}}$}, {{James~Gray}$^{\,\text{a}}$} and {{Paul-Konstantin~Oehlmann}$^{\,\text{a}}$}
\bigskip }\\[3pt]
\vspace{0.cm}
{\it 
 ${}^{\text{a}}$ Physics Department,~Robeson Hall,~Virginia Tech,~Blacksburg,~VA 24061,~USA\\
${}^{\text{b}}$ Department of Mathematics,~University of Pennsylvania,~209 S 33rd St., Philadelphia,\\
~PA 19104,~USA
}
\\[2.0cm]
\end{center}

\begin{abstract}
\noindent
We explore $6$-dimensional compactifications of F-theory exhibiting $(2,0)$ superconformal theories coupled to gravity that include discretely charged superconformal matter. Beginning with F-theory geometries with Abelian gauge fields and superconformal sectors, we provide examples of Higgsing  transitions which break the $U(1)$ gauge symmetry to a discrete remnant in which the matter fields are also non-trivially coupled to a $(2,0)$ SCFT. In the compactification background this corresponds to a geometric transition linking two fibered Calabi-Yau geometries defined over a singular base complex surface. An elliptically fibered Calabi-Yau threefold with non-zero Mordell-Weil rank can be connected to a \emph{smooth} non-simply connected genus one fibered geometry constructed as a Calabi-Yau quotient. These hyperconifold transitions exhibit multiple fibers in co-dimension 2 over the base.
\end{abstract}

\end{titlepage}
\clearpage
\setcounter{footnote}{0}
\setcounter{tocdepth}{2}
\tableofcontents
\clearpage

\section{Introduction}
\label{sec:introduction}
Superconformal theories (SCFTs) in various dimensions have been the subject of intense and recent interest. In particular, the study of $(1,0)$ and $(2,0)$ SCFTs as they arise in F-theory in $6$-dimensions has proven to be a rich arena in which to characterize, explore and in some cases classify \cite{Heckman:2015bfa, DelZotto:2014hpa,Heckman:2013pva}, possible theories. Furthermore, these $6$-dimensional theories provide higher-dimensional insight into many lower dimensional theories via compactification.

While recent investigations have explored in detail the structure of so-called non-Abelian ``superconformal matter" \cite{DelZotto:2014hpa} in F-theory, comparatively less work has been dedicated to investigating possible Abelian sectors and associated \emph{discrete symmetries}. It is a goal of this work to take some initial steps in this direction by considering collections of F-theory vacua exhibiting superconformal structure, Abelian gauge symmetries and on certain branches in the vacuum space, discrete symmetries. Indeed, in the spirit of \cite{DelZotto:2014fia} we will explore $6$-dimensional superconformal theories coupled to gravity, in this case leading to non-trivial, discretely charged superconformal matter. In particular, we will provide examples of F-theory compactifications in which the structure of the global, compact Calabi-Yau (CY) threefold \emph{enforces the existence of discretely charged superconformal matter.}

\vspace{7pt}

To accomplish this goal, we must explore discrete symmetries as they arise in the background geometry of F-theory compactifications themselves. It is well known that in string compactifications, such geometric symmetries can also frequently lead to discrete symmetries in the associated physical effective theories. As we will outline in subsequent sections, there are two primary origins for such discrete symmetries in F-theory geometry. The first origin is through genus one fibered Calabi-Yau geometries which admit multi-sections only and hence are linked to a discrete symmetry manifested geometrically via a non-trivial Tate-Shafarevich (TS) group \cite{Morrison:2014era,Klevers:2014bqa,Anderson:2014yva,Mayrhofer:2014haa,Cvetic:2015moa}  (or more generally a discrete symmetry linking a set of Calabi-Yau torsors \cite{Bhardwaj:2015oru}). The second origin is through discrete automorphisms of the full Calabi-Yau compactification geometry itself (see \cite{DelZotto:2014fia} for examples of singular geometries of this type in F-theory, in the following Sections we will also explore \emph{smooth} CY geometries admitting such automorphisms\footnote{It should be noted that there exist many genus one fibered CY geometries admitting such discrete automorphisms. A first step towards a database of such geometries will be appearing soon \cite{Anderson:2016ler,Anderson:2016cdu,Anderson:2017aux,website, us_to_appear}.}). The second of these possibilities will prove to lead to many examples of novel and previously unexplored effective theories and it will be the primary focus of this work to explore the F-theory effective physics of such compactifications. Furthermore, as we will argue below, the two geometric origins for discrete symmetries are intrinsically linked. In particular, in all known cases of smooth fibered CY geometries quotiented by a free discrete automorphism, it has been observed that they are in fact genus one fibered, with multi-sections of order $n>1$ \cite{Donagi:1999ez} only.

It will prove useful in investigating these discrete symmetries to begin with their unbroken, Abelian origin. As was done in the case of CY geometries with multi-sections \cite{Braun:2014oya,Morrison:2014era,Anderson:2014yva}, the simplest window into the effective physics arises from considering Higgsing transitions in which Abelian gauge symmetries are broken to discrete subgroups via giving vevs to certain charged matter fields. This Higgsing transition is related in the compactification background to a geometric (i.e. conifold-type) transition linking an elliptically fibered Calabi-Yau manifold with non-trivial Mordell-Weil (MW) group to a genus one fibered manifold with a multi-section only. In this work, the same Higgsing transitions will arise, but here we will consider the new scenario in which certain $U(1)$ charged matter fields lie on curves in the base that are shrinkable to orbifold singularities. Those singular points correspond to superconformal matter charged under the Abelian symmetry which can be Higgsed to a discrete remnant.

Such transitions will be illustrated in detail in subsequent sections, but before we begin, it is worth observing a few facts about the global CY geometries that will lead to discretely charged superconformal matter. Although Calabi-Yau manifolds admit no continuous isometries, it is well known that they do admit freely acting, discrete automorphisms and we will explore numerous examples of such symmetries in the following sections. Unlike previous work \cite{Apruzzi:2017iqe}, we will consider not just isometries of the base to the genus one fibered F-theory compactification geometry, but rather symmetries which extend non-trivially to the full Calabi-Yau threefold. These manifolds might be expected to intrinsically manifest a discrete symmetry in their associated effective theories via the fact that they are non-simply connected with a discrete first fundamental group. 

A standard approach in the literature to build non-simply connected CY geometries is to quotient a simply connected CY manifold by a freely acting discrete symmetry.  Let $\Gamma$ be a discrete, freely acting automorphism of a smooth, simply connected Calabi-Yau threefold $X$. Then the quotient $\tilde{X}=X/\Gamma$ is also a smooth CY threefold with a nontrivial first fundamental group, $\pi_1(X)=\Gamma$, if the symmetry in question is manifest for a sufficiently general complex structure. The fact that the quotient manifold $\tilde X$ remains\footnote{Note that this property does not hold for quotients of CY manifolds of even (complex) dimension which in general lead to $Ind(\cO_{\tilde{X}}) \neq 0$ (for example, the Enriques quotient of $K3$).} Calabi-Yau \cite{beauville} follows from the fact that the canonical divisor is invariant under the action of the group and is preserved with $h^{j,0}(\tilde{X})=0, \ 0<j<3$. Since $X$ is simply connected, it follows that $\pi_1({\tilde X})=\Gamma$. Studies of such quotient CY manifolds are numerous in the literature see for example \cite{Bini:2012vya}, with careful classifications occurring in \cite{Braun:2010vc} (quotients of complete intersection manifolds in products of simple projective spaces) and in \cite{Batyrev:2005jc} (which characterized the $16$ CY 3-folds constructed as toric hypersurfaces \cite{Kreuzer:2000xy} which exhibit non-trivial fundamental group). Such geometries have played an important role in heterotic model building (see \cite{Donagi:1999ez,Donagi:2000zf,Andreas:1999ty} as well as \cite{Anderson:2013xka,Anderson:2012yf,Anderson:2011ns} for recent examples), but have not yet been fully explored in F-theory.

To employ CY quotient geometries in F-theory compactifications, it is necessary that $\tilde{X}$ admits a genus one fibration. One way to guarantee such a fibration $\tilde{\pi}: \tilde{X} \to \tilde{B}$ is to require that $X$ itself is either elliptically or genus one fibered over a base $B$ ($\pi: X \to B$) and that the discrete automorphism $\Gamma$ preserves the fibration structure. Examples of ``upstairs" fibered geometries, $X$, exhibiting appropriate discrete automorphisms leading to fibered ``downstairs" geometries $\tilde{X}$, will be studied in detail in the following sections.

For now, we will begin by noting that CY quotient geometries appear to exhibit an interesting and interlinking set of geometric features. Our goal will be to characterize these features and try to understand their impact of the associated $6$-dimensional effective physics of F-theory over such backgrounds. In this context of quotient CY threefolds, novel geometrical properties that can arise include:

\begin{itemize}
\item The smooth manifold $\tilde{X}$ is non-simply connected (i.e. has a non-trivial fundamental group, $\pi_1(\tilde{X}) \neq 0$).
\item The quotient exhibits torsion in homology. In general for a CY 3-fold, $X$, non-trivial torsion can appear in the form of finite Abelian groups $A(X)=\text{Tors}(H^2(X, \mathbb{Z}))$ and $B(X)=\text{Tors}(H^3(X, \mathbb{Z}))$. The latter group, $B(X)$ (the cohomological Brauer group) is known to play a role in generating discrete gauge groups, $\mathbb{Z}_k$, in $5$-dimensional compactifications of M-theory \cite{Mayrhofer:2014haa}. In general, for CY quotient geometries $A(\tilde{X}) \neq 0$ while $B(\tilde{X})$ may or may not be trivial.
\item Discrete automorphisms $\Gamma$ which preserve the fibration structure of $X$ under quotienting (leading to a genus one fibered threefold $\tilde{X}$) induce actions on the fiber and base of $X$ respectively. In general, although $\Gamma$ is fixed point free, the induced action $\Gamma_B$ \emph{will exhibit fixed points}, leading to fibrations over generically \emph{singular} base geometries $\tilde{\pi}: \tilde{X} \to \tilde{B}$. 
\item Such orbifold-type singularities in the base geometry have been demonstrated to lead to superconformal theories (SCFTs) coupled to gravity in the associated $6$-dimensional theories. Singular base surfaces and their associated ``superconformal matter"  have been studied in \cite{DelZotto:2014fia,DelZotto:2015isa}.
\item Since $\tilde{X}$ is a smooth fibration over a singular base manifold, the fibers of $\tilde{X}$ over the fixed points in $\tilde{B}$ must necessarily differ dramatically from those of a Weierstrass model. As we will see below, in many cases the action will give   \emph{multiple fibers}  over co-dimension $2$ points in $\tilde{B}$; the multiple fibers  were also classified by Kodaira \cite{KodairaMultiple}. A singular fiber $E_p=\sum n_i E_i$ is multiple if the greater common divisor of the $\{n_i\}$ is non trivial, that is $E_s= m E'_s$, where $E'_s$ is an effective (reduced) curve; $E_s$ is  singular (non-reduced). In this work $E'_s$ is a smooth genus one curve. Multiple fibers have already appeared in the F-theory context \cite{deBoer:2001wca,Bhardwaj:2015oru}, where similar quotients of elliptically fibered geometries have been considered.
\item Finally, note that the existence of  multiple fibers prohibits the existence of a \emph{section} to the fibration $\tilde{\pi}: \tilde{X} \to \tilde{B}$. As a result, any \emph{smooth} CY quotient over a singular base surface, must admit at best a multi-section of order $n$. Such geometries are well-known to lead to discrete symmetries in the associated $6$-dimensional compactifications of F-theory. This raises the interesting question, how are the discrete symmetries associated to the multi-section and CY torsors (i.e. the symmetries linking the set of CY fibrations that share the same Jacobian, $J(X)$) related to those associated to $\pi_1(X)$ and the torsion described above?
\end{itemize}
In the following sections we will explore the links between these geometric features and their 
associated F-theory physics. The main approach in that exploration is to tune in a section on the quotient geometry $\widehat{X}$ resulting in $\Gamma$ fixed points to collide with the CY hypersurface. Those singular points correspond to Lens spaces upon resolution to a fibration that is smooth and simply connected. Physically, those phases correspond to the tensor branches of the superconformal theories including Abelian and possible non-Abelian gauge enhancements which we provide in a number of explicit examples.
\vspace{7pt}

The structure of this work is as follows. In Section \ref{section2} we begin by outlining the effective, $6$-dimensional physics of F-theory with discretely charged superconformal matter. In Section \ref{section3} we provide an overview of the main components of this work, including the explicit Calabi-Yau manifolds underlying these constructions. It should be noted that in order to make this work relatively self-contained, Sections \ref{sec:Quotientreview} to \ref{sec:scft_points} provide a brief review of the main ingredients of our discussion -- CY quotient geometries, hyperconifold transitions, the F-theory physics of multi-section geometries, and superconformal points. The reader familiar with these topics can skip directly to Section \ref{sec:6dAnomalies}. Section \ref{section4} provides explicit examples/constructions, while Section \ref{sec:Summary} is a summary of our results. Assorted technical details are provided in the appendices.

\section{Coupling Discrete Symmetries to Superconformal Theories}
\label{section2}
To begin, it is useful to consider the physical ingredients of interest -- namely a $6$-dimensional SUGRA theory with a discrete symmetry coupled to a (2,0) SCFT subsector -- in the simplest possible set up. To realize this in an F-theory compactification, the most straightforward possibility takes the form of a generic, singular Weierstrass model. We will begin with such a geometry before describing the rich network of linked, compact, smooth (i.e. fully resolved) threefolds giving rise to such physics in Sections~\ref{section3} and~\ref{section4}.

In light of the recent classification of $6$-dimensional SCFTs \cite{Heckman:2013pva,Heckman:2015bfa} via F-theory, it is natural to try to recouple those theories back to gravity (see e.g. \cite{DelZotto:2014fia}). In doing so, the superconformal theory itself is of course lost (by the introduction of the $6$-dimensional Planck scale), however the SCFT can appear as a strongly coupled subsector with locally enhanced supersymmetry.

In terms of the F-theory geometry such a subsector can be understood as M5 branes probing isolated $\mathbb{C}^2/\Gamma$  singularities where $\Gamma$ is a finite subgroup of $U(2)$. Furthermore those models can readily be coupled to additional ADE gauge groups by engineering a divisor in the base that admits an ADE singularity in the F-theory fiber. Especially interesting are then the cases when those divisors hit the singular point and therefore modify the SCFT.

The categorization of SCFTs within F-theory arises from a simple geometric interpretation of the tensor branch of the theory via the resolution of singular points (by a chain of $\mathbb{P}^1$'s in the base of an elliptically fibered CY threefold\footnote{In the following we will always assume that the (2,0) theory admits a tensor branch. Theories with terminal singularities on the other hand have recently been considered in \cite{ArrasGrassiWeigand,Garcia-Etxebarria:2015wns}.}). The power of F-theory lies in the automatic identification of the ADE singularities in the elliptic fiber over the resolution $\mathbb{P}^1$'s that dictates gauge groups and matter representations of the former (2,0) theory. After this transition, a field theory description is available where all anomalies are canceled (via the Green-Schwarz (GS) mechanisms). In this way one can relate the anomaly polynomial of the (2,0) SCFT with that of the tensor branch \cite{Ohmori:2014kda,DelZotto:2014fia }. 

Due to the central role of anomalies in the classification of these theories, it is clear that they are even more constrained when recoupled to gravity on a compact base where gravitational anomalies must also be satisfied. This has been investigated in \cite{DelZotto:2014fia} by considering a singular base complex surface ($\mathbb{P}^2/\mathbb{Z}_3$) coupled to $SU(N)$ theories (realized by fiber singularities). It should further be noted that parallel to the classification of SCFTs, significant progress on Abelian (discrete) gauge symmetries has been made in global F-theory compactifications\cite{Braun:2014oya,Morrison:2014era,Mayrhofer:2014laa,Mayrhofer:2014haa,Anderson:2014yva,Klevers:2014bqa,Cvetic:2015moa,Oehlmann:2016wsb}.
Hence it is natural to ask if and how Abelian (discrete) symmetries can be linked to strongly coupled (2,0) subsectors and we turn to this question now.

\subsection{Discrete symmetries in Weierstrass models over a $\mathbb{P}^2/\mathbb{Z}_3$ base}
To illustrate these ideas concretely, we will begin by considering the generic Weierstrass model over a simple $\mathbb{P}^2/\mathbb{Z}_3$ base as in \cite{DelZotto:2014fia}. The base is given by the coordinates
\begin{align}\label{z3action}
(y_0, y_1 , y_2 ) \in \mathbb{P}^2/\mathbb{Z}_3 \, \text{ with }\, (y_0,y_1,y_2) \sim ( \lambda y_0, \Gamma_3 \lambda y_1, \Gamma_3^2  \lambda  y_2)  \, ,
\end{align}
where $\lambda \in \mathbb{C}^*$ and $\Gamma_3$ is a third root of unity. Clearly the discrete $\Gamma_3$ action leads to three codimension two singular fixed points located at 
\begin{align}
(y_0,y_1,y_2) = (\underline{1,0,0}) \, , 
 \end{align} 
 where the underline denotes permutations and the $\mathbb{C}^*$ scaling can be used to set the residual coordinate to one.
The most generic Weierstrass model on such a base has to have the form \cite{DelZotto:2014fia}
\begin{equation}
Y^2=X^3+f_{12}(y)X +g_{18}(y) \, ,
\end{equation}
with
\begin{align}
\label{eq:WSF}
f_{12} = \sum_{l+m+n=12} = y_0^l y_1^m y_2^n f_{l,m,n} \, , \qquad g_{18} = \sum_{l+m+n} g_{l,m,n} y_0^l y_1^m y_2^n \, ,
\end{align}
such that $f$ and $g$ are $\mathbb{Z}_3$ invariant sections.
The complex structure coefficients in \eqref{eq:WSF} can be readily verified to give 95 parameters. Subtracting the three $\mathbb{C}^*$ scalings results in $92$ free complex degrees of freedom. 

This generic theory admits no gauge symmetry but admits three orbifold fixed points in the base. The (2,0) theories hosted at the orbifold fixed points in this case are referred to as $A_2$ theories (so-called as this is the type of geometry seen after blowing-up of the singular orbifold base to a smooth dP$_6$ surface). 

Upon resolution of the singularities in the base, the resulting Weierstrass model over the blown-up base stays smooth and there is no additional gauge symmetry. Physically we can think of the contributions from the singular points as that coming from stacks of three coincident M5 branes minus a free (2,0) tensor \cite{DelZotto:2014fia}. Thus this actually contributes to the anomaly as two free (2,0) tensors.
As a result, the only remaining anomalies that must be checked are the gravitational ones, given as
\begin{align}
\label{eq:gravanomalies}
&\text{grav}^4:\quad &H-V +29T + 30 n_s -273 = 0 \, , \\
& (\text{grav}^2)^2:\quad & 9- T - n_s - (K_b^{-1})^{2}=0 \, ,  
\end{align}
where as usual $H,V$ and $T$ refer to the number of hyper, vector and tensor multiplets and $n_s$ denotes the multiplicity of (2,0) tensors (each of which can be thought of as a $(1,0)$ tensor and neutral hyper and in our example there are  $n_s = 3 \times 2$ of these). $K_b^{-1}$ refers to the anticanonical class of the base complex surface. Note that we do not have any $(1,0)$ tensor multiplets in this geometry since $T= h^{1,1}(B) -1 $. Nor do we have any vector multiplets. After the inclusion of the universal hypermultiplet, it follows that $H=93$ so that the first anomaly is solved. Furthermore, the reducible gravitational anomaly is canceled by noting that 
\begin{align}
K_b^{-1} = 3 H_b \, \text{  and  } \, (K_b^{-1})^2 = 3 \, ,
\end{align}
on the quotient geometry.

As described above, moving to the tensor branch of the $(2,0)$ theory amounts to resolving the three fixed points of the singular base variety. Each singularity requires the addition of two $\mathbb{P}^1$s and hence yields a smooth (non generic)  dP$_6$ base as depicted in Figure~\ref{fig:dP6ResBase1}. For this new phase of the theory (equivalently geometry) we have $n_s = 0$ and $T=6$ as well as $99$ neutral hypermultiplets to cancel all anomalies. 

 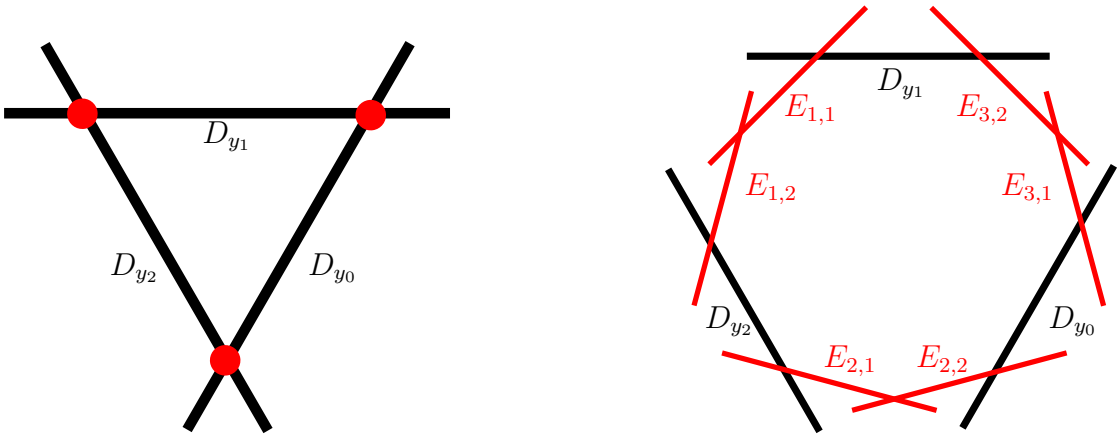
\begin{figure}
 \hspace{0.6cm}
\begin{picture}(50,140)
 \put(75,130){$D_{y_1}$}
 \put(40,80){$D_{y_2}$}
 \put(115,80){$D_{y_0}$}
\put(0,20){\includegraphics[scale=0.6]{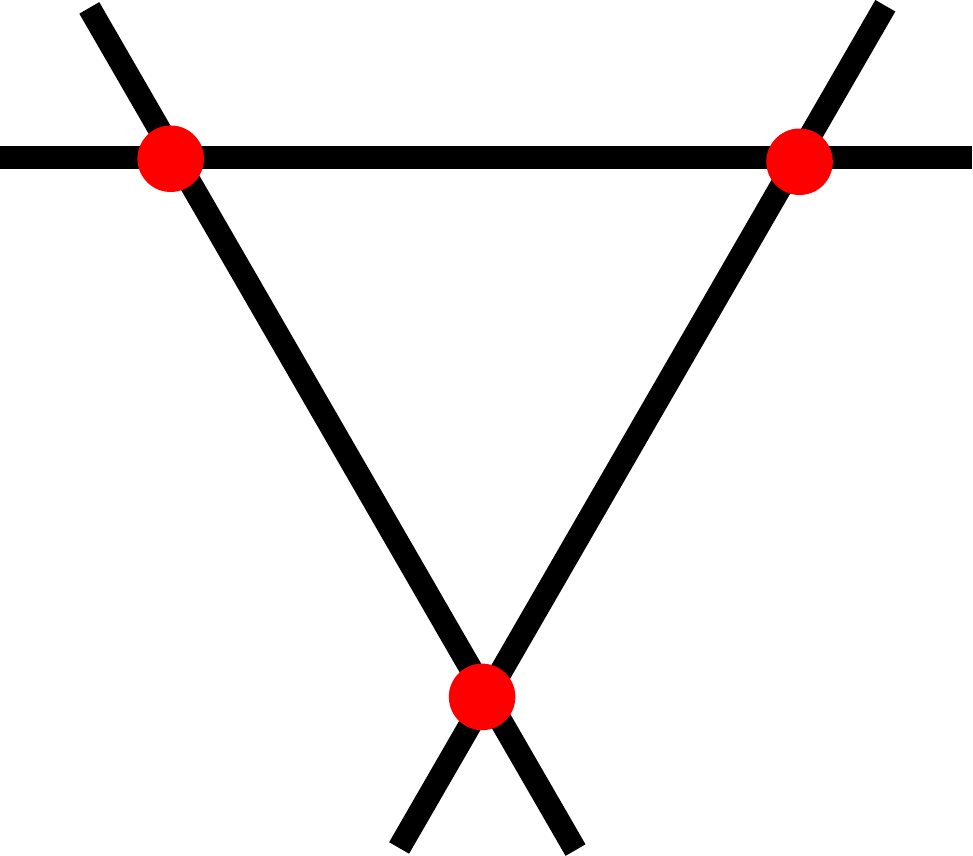}}

\put(250,20){\includegraphics[scale=0.6]{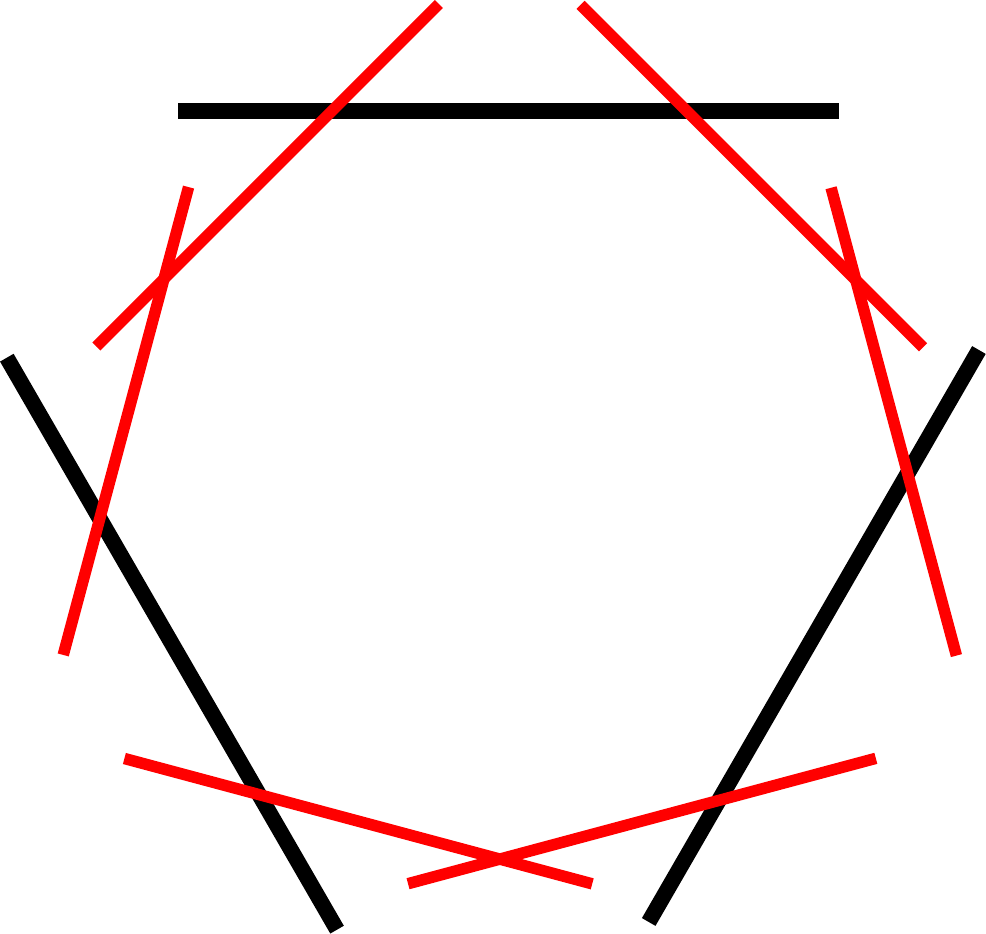}}
\put(330,150){$D_{y_1}$}
\put(265,60){$D_{y_2}$}
 \put(395,60){$D_{y_0}$}

\put(310,45){$\textcolor{red}{E_{2,1}}$}
\put(345,45){$\textcolor{red}{E_{2,2}}$}

 \put(280,110){$\textcolor{red}{E_{1,2}}$}
 \put(295,140){$\textcolor{red}{E_{1,1}}$}

 \put(377,110){$\textcolor{red}{E_{3,1}}$}
 \put(360,140){$\textcolor{red}{E_{3,2}}$}

\end{picture}

\caption{\label{fig:dP6ResBase1}{\it Intersections of toric base divisors for a  $\mathbb{P}^2/\mathbb{Z}_3$ base, before and after blow-up. Singular points and their blow-up divisors are highlighted in red.}}
\end{figure} 
 
\subsection{Tuning discrete symmetries: multi-section geometries}
\label{sec:Sec2Discrete}
With this generic Weierstrass model in hand, it is now possible to consider the addition of a discrete symmetry and to ask how it can be coupled to the $(2,0)$ theory. Engineering discrete symmetries is a priori possible by considering genus one fibrations with multi-sections and we review the basic geometry briefly here.

Discrete symmetries in F-theory can be associated to sets of genus one fibrations that share a common Jacobian. Thus, there exist collections of linked CY geometries -- more precisely, $n-1$ equivalent genus one fibrations that have no section but only n-sections. Those geometries can be collected together to elements of the Tate-Shafarevich (or more generally the Weil-Ch$\hat{\text{a}}$telet group of CY torsors\cite{Bhardwaj:2015oru}). Each element of the group is a genus one fibration with the same axio-dilaton profile $\tau$ as in the Jacobian threefold and therefore describes equivalent F-theory physics. However in the dual $5$-dimensional M-theory compactifications over the same CY geometries, each background can be distinguished by $n$ different discrete choices of three-form flux, $C_3$ (where the Jacobian CY threefold with fiber $Jac(\mathcal{C})$ denotes the trivial choice). 

The key to understanding the physical relationship between the collection of CY geometries lies in the charged matter visible in the geometry with a section. It is this background where the physical theory is most easily determined.  The Weierstrass model of the corresponding Jacobian admits codimension two non-crepant resolvable singular I$_2$ fibers where matter charged under a discrete remnant of a broken $U(1)$ is localized. In the physical theory, a geometric transition between the geometry with enhanced Mordell-Weil group and those with multi-sections arises via a Higgsing transition in which a non-minimally charged hypermultiplet, $\mathbf{1}_{n}$, acquires a vev and breaks a $U(1)$ symmetry to a discrete subgroup.

The Jacobian Weierstrass models which can connect the theories with Abelian gauge symmetry and those with discrete gauge groups come in a highly specialized form, as one can see from the Weierstrass coefficients $f$ and $g$ given for three relevant cases in Appendix~\ref{app:cubicinWSF}. From the point of view of a generic Weierstrass model over a given base, these Jacobians take the form of a tuned point in complex structure moduli space. For concreteness, we will illustrate these ideas here with the tuning of a discrete $\mathbb{Z}_3$ symmetry corresponding to the Jacobian of a geometry in which the fiber is a cubic in $\mathbb{P}^2$, giving rise to a multi-section of order 3. Such a fiber is given by the vanishing polynomial
\begin{align}
\label{eq:cubic_v1}
&p=s_1 x_0^3 + s_2 x_0^2x_1 + s_3x_0x_1^2 + s_4x_1^3 + s_5x_0^2 x_2 + s_6 x_0 x_1 x_2+
 s_7 x_1^2 x_2 + s_8 x_0 x_2^2 + s_9 x_1 x_2^2 + s_{10} x_2^3 \, ,
\end{align}
where $s_i$ are functions of the base coordinates.

For the case of such a cubic fiber, the Weierstrass coefficients $f$ and $g$ of the Jacobian can be expressed in terms of the ten sections $s_i$ of the base, given in Appendix~\ref{app:cubicinWSF}. Constructing this model over the $\mathbb{P}^2/\mathbb{Z}_3$ base it is important to note, that \emph{only $f$ and $g$ have to be $\mathbb{Z}_3$ invariant sections but the individual $s_i$ do not need to be}. We will return to this point in a moment, but for now  we begin by choosing the invariant combination for all sections that transform as $s_i \in K_b^{-1}$. From the action given in \eqref{z3action} the $s_i$ have to be of the form
\begin{align}
\hat{s}_i = a_{i,1} y_0^3 + a_{i,2} y_1^3 + a_{i,3} y_2^3 + b_{i} y_0 y_1 y_2 \, .
\end{align} 
 Hence there are $40$ non-vanishing coefficients $a_{i,j}$ and $b_i$. Subtracting the $\mathbb{C}^*$ scaling of the base for all ten $\hat{s}_i$, minus the one of the Weierstrass function yields $29$ complex structure moduli. Using the Weierstrass coefficients $f$ and $g$ the discriminant
 \begin{align}
 \label{eq:discriminant}
 \Delta = 4 f^3 + 27 g^2 \, ,
 \end{align}
 is a long non-factorized degree twelve polynomial in the $\hat{s}_i$ and therefore no gauge enhancement is present. By anomaly considerations alone, such a theory must contain $63$ additional discrete charged singlets counted by codimension two I$_2$ singular fibers. Indeed, for this type of fibration, the amount of discrete charged hypermultiplets has been computed \cite{Klevers:2014bqa} for a general base and its multiplicity is given as
\begin{align}
\label{eq:z3charged}
 \begin{array}{ccc}
 H_{\mathbf{1}_{1}}:  &      3 ( 6 (K_b^{-1})^2 - \mathcal{S}_7^2 + \mathcal{S}_7 \mathcal{S}_9    - \mathcal{S}_9^2 + K^{-1}_b (\mathcal{S}_7 + \mathcal{S}_9))  &\overset{\mathrm{\mathcal{S}_9 =\mathcal{S}_7 = K_b^{-1}}}{=}63  \, .
\end{array}
\end{align}
The $\mathcal{S}_7$ and $\mathcal{S}_9$ are the bases classes of their respective sections $s_7$ and $s_9$ in the Weierstrass form and equivalent to the canonical class $K_b^{-1}$ as stated before reproducing the correct number of discretely charged fields.\\

Up to this point we have tuned a $\mathbb{Z}_3$ gauge symmetry by choosing a special form of the elliptic fiber over the $\mathbb{P}^2/\mathbb{Z}_3$ base, but we have not considered its impact on the three orbifold (2,0) points.
If the discrete gauge symmetry were associated to a particular divisor in the geometry, we could simply check its intersection with the (2,0) points which would hint at a possible modification of the $A_2$ theory. However, such an understanding for a $\mathbb{Z}_3$ divisor is still lacking\footnote{Some progress was made by taking the IIB limit of Mordell-Weil U(1) symmetries and discrete gauge symmetries in \cite{MayorgaPena:2017eda}.}. Instead, we can try to explicitly check for any modification of the $(2,0)$ theory by going onto the tensor branch and  looking for additional gauge and matter degrees of freedom over the resolution divisors $E_{i,j}$ for $i=1..3$ and $j=1,2$ defined by $e_{i,j} =0$. 

The resolution of the orbifold fixed points in $\mathbb{P}^2/\mathbb{Z}_3$ yields a dP$_6$ base surface and  in such a case the Weierstrass sections $s_i$ given above are replaced by the generic four monomials in the anticanonical class of dP$_6$:
   \begin{align}
   \acute{s}_i = e_{2,1} e_{2,2}^2 e_{3,1}^2 e_{3,2}  y_0^3 a_{i,1} +e_{1,1}^2 e_{1,2}  e_{3,1} e_{3,2} ^2 y_1^3 a_{i,2} +   
    e_{1,1} e_{1,2}^2 e_{2,1}^2 e_{2,2}  y_2^3 + e_{1,1} e_{1,2} e_{2,1} e_{2,2} e_{3,1} e_{3,2} y_0 y_1 y_2 b_{i} 
          \end{align} 
From the counting, we find the same $40$ non-vanishing coefficients reduced by five $\mathbb{C}^*$ scalings which results in $35$ complex structure coefficients, a counting we
will reconfirm in the equivalent genus one geometry in Appendix~\ref{app:P2Dp6Direct}.
 
 Analyzing the discriminant $\Delta$ over the blow-up loci $e_{i,j}=0$ reveals no codimension one nor two singularities as the sections $\acute{s}_i$ are non-vanishing over any of the resolution divisors above. On the other hand, there remains the unchanged relations of the base sections and their intersections as
 \begin{align}
 \mathcal{S}_7 = \mathcal{S}_9 = K_b^{-1} \,  \text{ with }\, (K_b^{-1})^2=3 \, .
 \end{align}
Thus, we find once again that all gravitational anomalies in  Eq.~\eqref{eq:gravanomalies} are canceled. Moreover it is clear that there are no additional gauge symmetries nor matter multiplets appearing over the (2,0) tensor branch.\\ 

To summarize, beginning with a generic Weierstrass model over $\mathbb{P}^2/\mathbb{Z}_3$ it is possible to tune the complex structure to make manifest a connection to a multi-section geometry with a $\mathbb{Z}_3$ symmetry. In doing so, we find that generically the three superconformal points in the base geometry carry through this tuning largely unaffected. That is, we have thus far considered a supergravity with three (2,0) $A_2$ points which we coupled to a discrete symmetry without coupling/charging the $A_2$ points to the discrete symmetry. It remains to ask then, what happens when such a coupling does occur? We turn to this possibility next.
   
\subsection{Coupling discrete multiplets to the tensor branch} 
\label{sec:A2DiscreteCoupled}
In the following our goal will be to minimally tune the complex structure moduli of the Weierstrass model over dP$_6$ given above such that we find discrete charged singlets residing exactly over the resolution divisors. Once this tuning has been achieved in the tensor branch (i.e. resolved base geometry) of the theory, we can then take the singular limit to go back to the strongly coupled theory. The global features of the associated $\mathbb{Z}_3$ multi-section geometry, the singular Jacobian and its resolution will be described in detail in Section~\ref{sec:example1}, however here we will begin with a brief overview of the physics associated to a simple, tuned Weierstrass model. It should be noted that such a tuning is not in the smooth moduli space of deformations of the generic Weierstrass model over dP$_6$ and instead will correspond (under resolution of singularities) to a topologically distinct Calabi-Yau threefold.\\ 

In order to tune discrete charged singlets over the resolution divisors $E_{i,j}$ we have to tune the sections $s_i$ such, that we obtain an  I$_2$ fiber at $e_{i,j}=0$ plus another constraint. A strategy to search for such a model is to tune the $\acute{s}_i$ to factor as
\begin{align}
\bar{s}_k = \sum_{i,j} e^{n_{i,j,k}}_{i,j} d_k \, ,
\end{align}
with powers $n_{i,j,k}$ and the $d_k$ some residual polynomials, such that the discriminant becomes of the form
\begin{align}
\label{eq:tunedWSF}
\Delta = \left(\sum_{i,j} e_{i,j} \right) (P(d_k) + Q(d_k) \mathcal{O}(\sum_{i,j} e_{i,j}) ) \, ,
\end{align}
and transforms as a section of $K_b^{-12}$ of the dP$_6$ base.
In this way we obtain an I$_1$ fiber over $E_{i,j}$ enhanced to the desired I$_2$ loci over $E_{m,n}$ as well as $P=0$. An exhaustive scan for such solutions is beyond the scope of this work but one solution is given as
\begin{align}
\begin{array}{lll}
\bar{s}_1= e_{1,1} e_{2,1} e_{3,1} d_1\, , &  \bar{s}_2= e_{1,1} e_{1,2} e_{2,1} e_{2,2} e_{3,1} e_{3,2} d_2\, , &  \bar{s}_3=
 e_{1,1} e_{1,2} e_{2,1} e_{2,2} e_{3,1} e_{3,2} d_3 \, ,   \\  \bar{s}_4 = e_{1,2} e_{2,2} e_{3,2} d_4 \, , & \bar{s}_5= e_{1,1} e_{2,1} e_{3,1} d_5 \, , &  \bar{s}_6= d_6 \, , \\ \bar{s}_7 = e_{1,2} e_{2,2} e_{3,2} d_7  \, , & \bar{s}_8 =
 d_8 \, , & \bar{s}_9 = d_9\, ,  \, \qquad  \bar{s}_{10}= d_{10} \, ,
 \end{array}
\end{align}
where the residual polynomials $d_i$ are given explicitly in Appendix~\ref{app:sections}.
In this case we find $31$ generic coefficients that get reduced by the $\mathbb{C}^*$ scalings of the dP$_6$ base to $26$ complex structure degrees of freedom.  It should be noted that under this tuning, the equivalences between various sections $\bar{s}_i$, no longer hold. That is, written as bundle relations $\mathcal{S}_7 \neq \mathcal{S}_9 \neq K_b^{-1}$ which can be explicitly read off from the expressions given in Appendix~\ref{app:sections}. In order to check for the multiplicities we can use the toric intersections of the dP$_6$ base as given in Figure~\ref{fig:dP6ResBase1} or equivalently by its Stanley-Reisner ideal:
\begin{align}
\begin{split}
SRI: \{  &
y_0 y_1, y_0 y_2, y_0 e_{1,1}, y_0 e_{1,2}, y_0 e_{2,1}, y_0 e_{3,2}, y_1 e_{2,2}, y_2 e_{2,2}, e_{1,1} e_{2,2}, e_{1,2} e_{2,2}, e_{2,2} e_{3,1},\\& e_{2,2} e_{3,2}, y_1 e_{3,1}, y_2 e_{3,1}, e_{1,1} e_{3,1}, e_{1,2} e_{3,1}, e_{2,1} e_{3,1}, y_1 y_2, y_1 e_{1,2}, y_1 e_{2,1}, y_2 e_{1,1}, \\ &e_{1,1} e_{2,1}, e_{1,1} e_{3,2}, y_2 e_{3,2}, e_{1,2} e_{3,2}, e_{2,1} e_{3,2}, e_{1,2} e_{2,1} 
\} \, ,
\end{split}
\end{align}
  and using the linear equivalences of dP$_6$
\begin{align}
\begin{split}
[y_0]& \sim[y_2 + 1/3 e_{1,1} + 2/3 e_{1,2} + 1/3 e_{2,1} - 1/3 e_{2,2} - 2/3 e_{3,1} - 1/3e_{3,2}] \, ,\\
[y_1]& \sim[y_2 - 1/3 e_{1,1} + 1/3 e_{1,2} + 2/3 e_{2,1} + 1/3e_{2,2} - 1/3e_{3,1} - 2/3e_{3,2}]\, ,
\end{split}
\end{align}
to deduce the relations
\begin{align}
\begin{split}
\label{eq:BlowUpBaseclasses}
\mathcal{S}_9 \sim &[3y_2 + 1/3 e_{1,1}+ 5/3e_{1,2}+ 4/3 e_{2,1} + 2/3e_{2,2}- 2/3e_{3,1}- 1/3e_{3,2}]  \, ,\\
\mathcal{S}_7 \sim& [3y_2 + 2/3e_{1,1} + 7/3e_{1,2} + 5/3e_{2,1}+ 4/3e_{2,2} - 1/3e_{3,1} + 1/3e_{3,2}] \, ,
\end{split} 
 \end{align}
 which admit the following linear equivalences
  \begin{align}
  \begin{split}
 2 K_b^{-1} - \mathcal{S}_7 - \mathcal{S}_9 \sim &   [e_{1,1}+e_{2,1}+e_{3,1}]\, , \\
 2 \mathcal{S}_7 - \mathcal{S}_9-K_b^{-1} \sim & [e_{1,2}+e_{2,2}+e_{3,2}]\, , \\
 2 \mathcal{S}_9 - \mathcal{S}_7-K_b^{-1} \sim & [ e_{1,1}+e_{2,1}+e_{3,1}e_{1,2}+e_{2,2}+e_{3,2}]\, ,
 \end{split}
 \end{align}
 and intersections
 \begin{align}
  \mathcal{S}_7 K_b^{-1} =  \mathcal{S}_9 K_b^{-1} = (K_b^{-1})^2=3 \, , \quad  
 \mathcal{S}_7 \mathcal{S}_7  =  \mathcal{S}_9 \mathcal{S}_9 = 1\, , \quad  
 \mathcal{S}_7 \mathcal{S}_9  = 2 \,  .
 \end{align}
 This information is enough to deduce the discrete charged hypermultiplets via \eqref{eq:z3charged} which yields the following spectrum
\begin{align}
 \begin{tabular}{|l|l|} \hline
 Representation & Multiplicity \\ \hline
      $\mathbf{1}_1$ & $72$  \\ \hline
      $\mathbf{1}_0$   & $27$ \\ \hline
      $\mathbf{T}$ & $6 $ \\ \hline
   \end{tabular} \, .
\end{align}
By direct comparison with the matter content given in the previous Subsection we find that nine neutral hypermultiplets have been exchanged for nine discrete charged hypermultiplets and hence all anomalies are canceled again. Moreover these additional hypers are \emph{located exactly over the six resolution divisors} as the discriminant is precisely of the form \eqref{eq:tunedWSF} with the polynomial
\begin{align}
P= (-d_{10} d_6^3 -   d_6 d_7 d_8^2 +  d_4 d_8^3 + d_6^2 d_8 d_9) \, ,
\end{align}
which is a section in the homology class of the base
\begin{align}
[P_1] \sim [y_0+y_1+y_2 + 3 K_b^{-1}]\, .
\end{align}
The multiplicity of the matter can be evaluated by using the intersections
\begin{align}
E_{i,1}E_{i,2}=1\, ,\qquad E_{i,j}K_b^{-1} = 0 \, , \qquad  E_{i,j}[y_0 + y_1 + y_2] = 1 \, .
\end{align}
In Section~\ref{section4} we consider the fully resolved geometry and confirm the factorization of the genus one fiber over the above loci explicitly. 

In summary we have presented here a tuned fibration with exactly nine additional discrete charged hypermultiplets located over the three loci of the former $\mathbb{Z}_3$ fixed point as depicted in Figure~\ref{fig:dP6ResBase1}. Those additional hypermultiplets come at the cost of nine complex structure degrees of freedom as dictated by anomaly cancellation.
 \begin{figure}
\begin{picture}(50,140)
 \put(75,130){$D_{y_1}$}
 \put(40,80){$D_{y_2}$}
 \put(115,80){$D_{y_0}$}
\put(0,20){\includegraphics[scale=0.6]{z3orbifoldbase}}

\put(250,20){\includegraphics[scale=0.6]{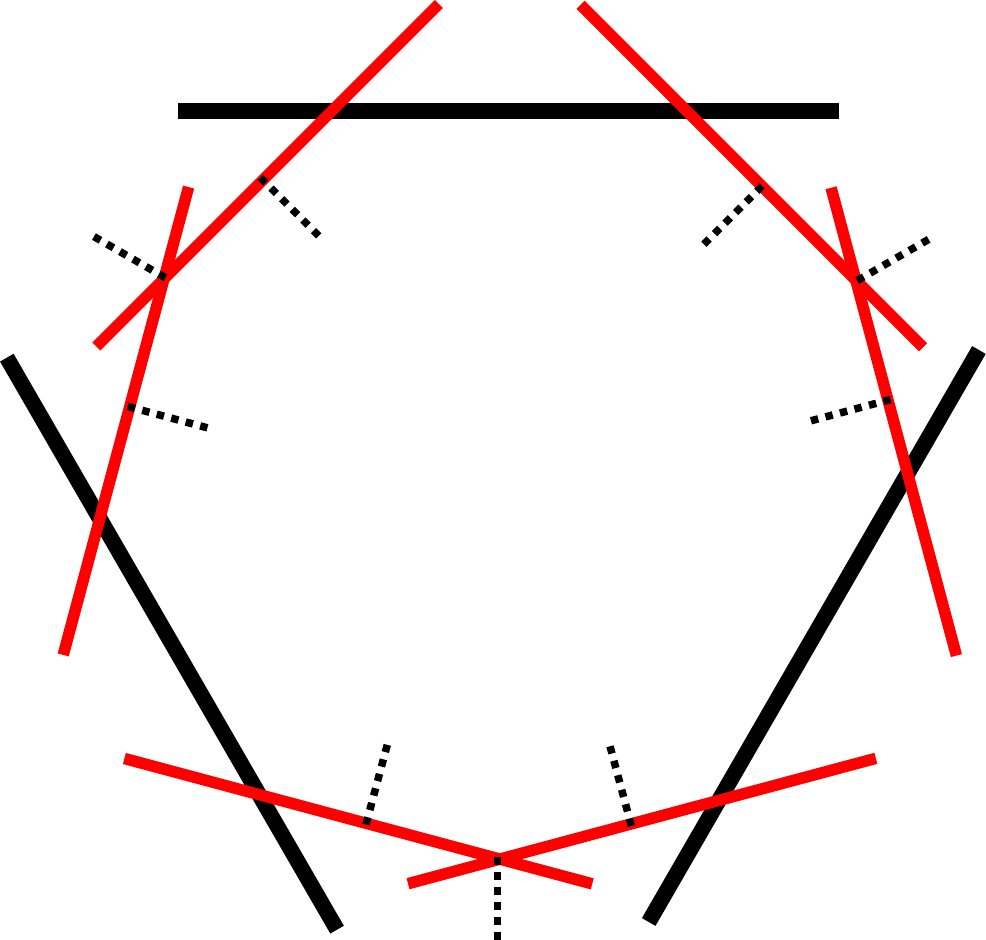}}

\put(248,139){\framebox{$\mathbf{1}_1$}}
\put(305,139){\framebox{$\mathbf{1}_1$}}
\put(286,105){\framebox{$\mathbf{1}_1$}}

\put(371,107){\framebox{$\mathbf{1}_1$}}
\put(352,137){\framebox{$\mathbf{1}_1$}}
\put(411,139){\framebox{$\mathbf{1}_1$}}

\put(327,8){\framebox{$\mathbf{1}_1$}}
\put(308,60){\framebox{$\mathbf{1}_1$}}
\put(347,60){\framebox{$\mathbf{1}_1$}}

\end{picture}

\caption{\label{fig:dP6ResBase1}{\it Matter locations of the second $\mathbb{Z}_3$ model and its tensor branch where additional discrete charged hypermultiplets reside. Note that the blow-ups are at non-generic points leading to a toric dP$_6$.}}
\end{figure}
\subsubsection{Going back to the strong coupling geometry}
Our goal remains to understand how the charged matter described above interacts with the superconformal sectors of the theory in the limit of a singular base. To accomplish this, we must consider the geometry above as we go back to strong coupling by blowing down the exceptional divisors $E_{i,j}$ within the dP$_6$ base. In doing so we note that the sections $\bar{s}_i$ are now all degree three polynomials in the $\mathbb{P}^3/\mathbb{Z}_3$ coordinates $y_i$ and therefore transform in their anticanonical class $K_b^{-1}$ just as in our first model presented in Subsection \ref{sec:Sec2Discrete}. However, unlike in that geometry, here the sections are generically {\it not} invariant under the $\mathbb{Z}_3$ action anymore but transform homogeneously under the $\mathbb{Z}_3$ action
\begin{align}
(\bar{s}_i) \rightarrow  \left( 
\begin{array}{ccccc}
s^{(0)}_1 \, ,  & \Gamma_3^2 s^{(2)}_2  & \Gamma_3 s^{(1)}_3&   s^{(0)}_4  &  \Gamma_3s^{(1)}_5 \\ 
 s^{(0)}_6  &\Gamma_3^2  s^{(2)}_7  & \Gamma_3^2s^{(2)}_8  &\Gamma_3  s^{(1)}_9  &  s^{(0)}_{10}   
 \end{array} \right)
\end{align}
where the superscript denotes the power of $\Gamma_3$ it transforms under. However from the Weierstrass coefficients we see that $f(\bar{s}_i), g(\bar{s}_i)$ and $\Delta(\bar{s}_i)$ are still $\mathbb{Z}_3$  invariant combinations and therefore our Weierstrass fibration is well defined. Note that we can add three more polynomial deformations $\delta a_i y_i^3$ to the polynomial $s_{10}^{(0)} = b y_0 y_1 y_2$ as
\begin{align}
s_{10}^{(0)} \rightarrow s_{10}^{(0)} + \delta a_0 y_0^3 + \delta a_1 y_1^3 + \delta y_2^3 \, ,
\end{align}
that respects the $\mathbb{C}^*$ scaling and $\mathbb{Z}_3$ transformations of the base and must be added as complex structure degrees of freedom. Summing up we obtain $26+3=29$ independent complex structure deformations. Turning back to the charged spectrum we have to read off the classes of sections $s_7$ and $s_9$ again that are of the form
\begin{align}
\begin{split}
s^{(2)}_7 =&a_{14} y_0 y_1^2 + a_{23} y_0^2 y_2 + a_{30} y_1 y_2^2\, , \\
s^{(1)}_9 =& a_{16} y_0^2 y_1 + a_{25} y_1^2 y_2 + a_{32} y_0 y_2^2\, , 
\end{split}
\end{align}
which vanish over the orbifold singularities and hence their associated classes $\mathcal{S}_7$ and $\mathcal{S}_9$ are not Cartier divisors. As both sections are degree three polynomials  we denote their classes as $\mathcal{S}_7 = \mathcal{S}_9 = K_b^{-1}$ by abuse of notation. We must however keep in mind that they are not Cartier divisors, consistent with their change in intersection numbers upon blow-up, a fact we discuss in more generality in Section~\ref{sec:6dAnomalies}. 
Using then the self-intersection $(K_b^{-1})^2 = 3$, we compute the spectrum by Equation~\eqref{eq:z3charged} 
of Subsection~\ref{sec:Sec2Discrete} to 63 discrete charged hypermultiplets, exactly the same amount as in the first model.
Therefore we find the count of all massless degrees of freedom of both models to match exactly,
including the contribution of the three $A_2$ theories consistent with all anomalies. \\

This spectrum provides a puzzle that we will investigate in the following Sections: Both theories are 6-dimensional SUGRA theories coupled to a $\mathbb{Z}_3$ discrete symmetry and three strongly coupled $A_2$ (2,0) SCFTs with the same amount of massless degrees of freedom. One is thus tempted to say that they are identical theories with the only difference, in terms of the Weierstrass model, that the sections $\bar{s}_i$ are $\mathbb{Z}_3$ invariant in the first model and covariant in the second. Intriguingly however, when we go to the tensor branches of both theories we find additional discrete charged states in the second theory which is not the case in the first and hence the second $A_{2}$ is charged under the discrete symmetry. \\

To alleviate this puzzle, in the remainder of this work, we will consider the fully resolved genus one fiber of this (and other models) that can be described as the following hypersurfaces:
\begin{enumerate}
\item First Model: Genus one fibration as anticanonical hypersurface $P \subset (\mathbb{P}^2 \times (\mathbb{P}^2/\mathbb{Z}_3)) $. The discussion of the smooth tensor branch geometry can be found in Appendix~\ref{app:P2Dp6Direct}.
\item Second Model: Genus one fibration as anticanonical hypersurface $P \subset (\mathbb{P}^2 \times \mathbb{P}^2)/\mathbb{Z}_3 $.
\end{enumerate}
In fact the quotient in the second model extends to a full free action on the genus one fibered Calabi-Yau with a multiple fibers over the fixed points. In addition we show that the tensor branch transition in the second case is obtained by a resolution of a Lens space. The above and other examples are presented in Section~\ref{section4}. As we will see, the structure of the global Calabi-Yau geometry encodes important differences in the superconformal sectors of the theories.

 \section{Quotient Manifolds and Hyperconifold Transitions}\label{section3}
In this section we consider the general construction of smooth genus one fibered geometries that have the generic properties presented in the previous section. As will be described below, the geometries that encapsulate the special structure of discretely charged superconformal matter described in Section~\ref{sec:A2DiscreteCoupled} have a number of remarkable features, most importantly they can be described as a smooth quotient of a Calabi-Yau threefold by a freely acting discrete symmetry. In addition, they are non-simply connected and the genus one fibrations exhibit multiple fibers in co-dimension 2 as described in Section~\ref{sec:introduction}. In addition, unlike in cases previously considered, transitions between a multi-section geometry displaying a discrete symmetry and the "un-Higgsed" $U(1)$ geometries are not realized as ordinary conifold transitions, but rather as so-called ``hyperconifold transitions" in the sense of \cite{Davies:2009ub}.

To clearly define these compact quotient geometries, we first review the general constraints for the covering space threefold $X$, and recall the properties that Calabi-Yau quotient geometries obey. In addition, in the context of $U(1)$ Higgsing transitions in 6-dimensional F-theory it is useful to provide a brief review of the physics associated to genus one fibrations and we do this in Section~\ref{sec:F-theoryReview}. There we will highlight the appearance of $\mathbb{Z}_n$ discrete gauge symmetries and $(2,0)$ superconformal sectors, realized as $\mathbb{Z}_n$ singularities in the base. 

With these results in hand we are in a position at last to study in detail the tensor branch of those (2,0) theories which differs by those of standard $A_{n-1}$ theories by a coupling to the discrete gauge symmetry, which is why we denote them as $\mathcal{A}_{n-1}$. The tensor branch of the $\mathcal{A}_{n-1}$ theories is obtained by hyperconifold resolutions that replaces a Lens space in the threefold with a chain of $\mathbb{P}^1$'s in the base with discrete charged hypers over them. In Section~\ref{sec:anomalies} we show that the full 6-dimensional anomaly cancellation is satisfied on the quotient geometry including the contribution of $\mathcal{A}_{n-1}$ (2,0) subsectors.

\subsection{Construction of  genus one fibered quotient threefolds}
 \label{sec:Quotientreview}
In this section we briefly review the construction of non-simply connected, smooth torus-fibered Calabi-Yau threefolds, $\widehat{X}$, and their properties. Manifolds such as these have played an important role in smooth heterotic model building where symmetry breaking is achieved via discrete Wilson lines (see e.g. \cite{Donagi:1999ez,Andreas:1999ty,Anderson:2009mh}), but have not yet been systematically employed in F-theory. At present, no systematic characterization of such CY threefold geometries exist, but classifications have been completed for several important datasets, including toric hypersurfaces \cite{Batyrev:2005jc} and also complete intersections in products of projective spaces \cite{Braun:2010vc,Constantin:2016xlj}. 

In these known constructions $\widehat{X}$ is obtained by using a freely acting discrete automorphism $\Gamma_n$ to quotient a covering Calabi-Yau manifold $X$ as
\begin{align}
\widehat{X} = X/\Gamma_n \, .
\end{align}
The topology of $\widehat{X}$ is fully specified by that of $X$, with $\text{Ind}(\widehat{X})=\text{Ind}(X)/|\Gamma|$ and $H^i(\widehat{X},T\widehat{X})=H^{i}_{inv}(X,TX)$ and the Chern classes and intersection numbers likewise descending (see \cite{Anderson:2009mh} for a brief review).

In the present context, since we hope to employ such geometries in F-theory, we also require in addition that both $X$ and $\widehat{X}$ exhibit a fibration structure 
\begin{align}
\begin{array}{lll}
T^2& \rightarrow& X \\
&&\downarrow \pi \\
&&B_{\text{cov}} \, .
\end{array}
\end{align}
Those conditions put some constraints on the action of $\Gamma_n$ and its form which we will review \cite{Donagi:1999ez,Andreas:1999ty} in the following. Concretely, it is necessary for $\Gamma_n$ to preserve the holomorphic volume form, to preserve the fibration $\pi$ and to act freely on $X$ such that $\widehat{X}$ is also smooth and non-simply connected. In particular, to  accomplish the second requirement we will assume that  $\Gamma_n$ is a  composition of a fiber and base action
\begin{align}
\Gamma_n = \Gamma_{n,f} \circ \Gamma_{n,b} \, ,
\end{align}
that are compatible with the fibration, as
\begin{align}
\begin{array}{lll}
T^2/\Gamma_{n,f} & \rightarrow& X/\Gamma_n \\
&&\downarrow \pi \\
&&B_{\text{cov}}/\Gamma_{n,b} \, .
\end{array}
\end{align}
In general, the action on the base $\Gamma_{n,b}$ will be not free and admit fixed points\footnote{Note, an exception involves cases in which the base of the fibration is an Enriques surface.}, leading to \emph{singular base manifolds for the genus one fibration}.

The smoothness of the total CY geometry can be preserved despite the above singularities in the base by novel structures in the fiber. In the examples considered here, the fibers above the orbifold fixed points become \emph{multiple fibers} -- that is the fiber is a non-reduced curve of the form $n{\cal E}$ where $n>1$ and ${\cal E}$ is a smooth genus one curve (equivalently, the fiber above the orbifold fixed points in the base is \emph{everywhere singular}). We will explore this in more detail in the examples of Section \ref{section4} and in Appendix \ref{app:GIODescription}.

For this work we restrict ourselves to cyclic group actions for $\Gamma$ and, following the characterization in \cite{Donagi:1999ez}, we  consider separately the case of an elliptic fibration with (1) an elliptic fibration with a rational section and (2) a genus one fibration with multi-section:
\begin{enumerate}
\item $X$ is an elliptic fibration with a zero section $\sigma(s_0)$.  In this case, since the action of the symmetry must preserve the ``horizontal'' and ``vertical'' decomposition of divisors within $X$, we expect that the discrete symmetry should map sections to sections. That is, $\Gamma_{n,f}$ should act as a translation acting on the fiber \cite{BarthPetersVandeVen,Bhardwaj:2015oru}. The fiber over the fixed points is smooth, and  this translation is possible provided that the fibration has additional n-torsion sections $\sigma(s_m)$;  the $\Gamma_{n,f}$ acts as
\begin{align}
\Gamma_{n,f} (\sigma(s_m)) = \sigma(s_{m+1})\quad \text{for } \, m \text{ mod } n \, . 
\end{align}
In other words, we require that $\Gamma_{n,f}$ is an homomorphism of the torsion part of  Mordell-Weil group of the elliptic fiber; for example 
\begin{align}
s_{0}  \to s_0 +  \underbrace{ s_{1}\ldots +s_{1}}_{\text{m times}} \, ,
\end{align}
with $'+'$ denoting addition under the Mordell-Weil group law and $s_1$ as the torsion generator.
 We note that the torsion part of the Mordell Weil group leads to the presence of some ADE gauge algebra $\mathcal {G}$ with some $\mathbb{Z}_n$ sub-center \cite{MirandaTorsion, Aspinwall:1998xj,Mayrhofer:2014opa}.  The global gauge group $G$ of the fibration $X$ is
 modded by the sub-center and becomes non-simply connected with first fundamental group of  $\pi_1 (G) = \mathbb{Z}_n$.\\\\
As  this translation does not preserve the section \cite{Donagi:1999ez}
 the resulting quotient geometry $\widehat{X}$ only admits a multi-section $s^{(n)}$ of order $n$ resulting in a genus one fiber, $\mathcal{C}$, as
\begin{align}
s^{(n)}: s^s \sim n \cdot s_1 \, , \text{ with } s^{(n)} \cdot \mathcal{C} = n \,.
\end{align}
 Moreover the cyclic group action identifies all resolution divisors $E_i$ of the gauge algebra $\mathcal{G}$, that are intersected by some torsional section \cite{Donagi:1999ez} reducing the total rank of the gauge group of $X$. Note that this quotient\footnote{Similar observations have been made in the context of Little String theories after a fiber-base duality \cite{Bhardwaj:2015oru} and used in \cite{Apruzzi:2017iqe}.}  reduces the total gauge group. On the other hand, the presence of multi-sections suggests the presence of a discrete symmetry on $\widehat{X}$.

\item $X$ is a genus one fibration with a multi-section of order $n$ and no section.  Once again, the discrete symmetry must factor into an action on the fiber and base in such a way that the fibers should not acquire fixed points. The natural candidate is a discrete $\mathbb{Z}_n$ rotation that acts cyclically on all n-sections, as the action should be free, as illustrated in Figure~\ref{fig:multisecrotation} and the fiber becomes a {\it multiple} fiber of order n in all known examples \cite{us_to_appear}.
 \begin{figure}
\begin{picture}(0,100)
\put(100,10){\includegraphics[scale=1]{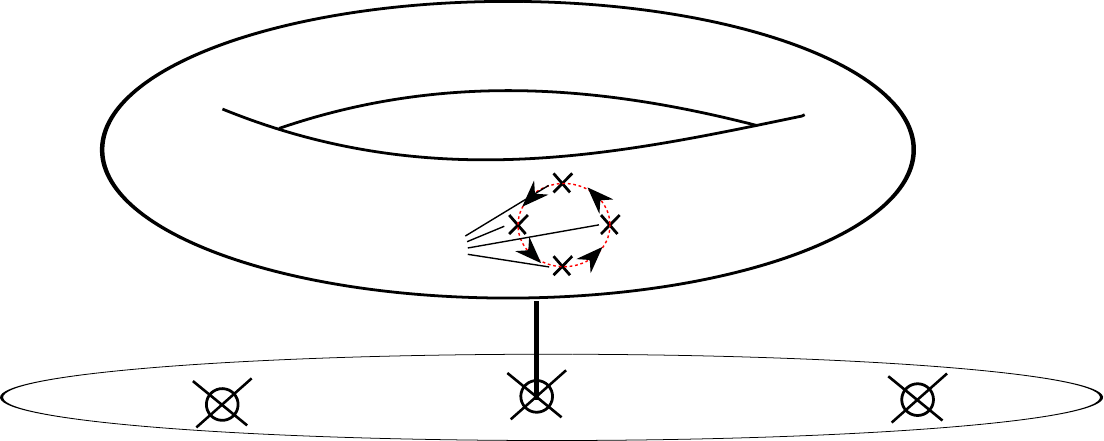}}
 \put(215,60){$s^{(n)}$}
    
\end{picture}
 \caption{\label{fig:multisecrotation}{\it Depiction of the fiber rotation over a fixed point in the base. $\gamma_{n,f}$ acts as a discrete fiber rotation that rotates the $n$ solutions of the n-section $s^{(n)}$ on the covering space to avoid fixed points on the fibers (note that locally this is equivalent to a translation on the fiber).}}
 \end{figure}
\end{enumerate}
Thus to summarize, taking a free $\mathbb{Z}_n$ quotient on $X$ that preserves the fibration yields a geometry which is
\begin{itemize}
\item genus one fibered,
\item fibered over a singular base manifold $B_{\text{orb}}=B/\Gamma_{n,b}$,
\item non-simply connected with $\pi_1(X/\mathbb{Z}_n) = \mathbb{Z}_n$.
\end{itemize}
For simplicity, in this work we will focus primarily on quotients of the second kind, although many of the following results and relations can be extended to quotients of the first type as well.

\subsection{Hyperconifolds and Lens spaces}

\label{sec:LensReview}
The results of the previous section make clear that in order to study F-theory on quotient Calabi-Yau geometries it is necessary to consider multi-section geometries. In order to describe the physics of such backgrounds, we must also be prepared to discuss transitions linking elliptic fibrations with section to genus one multi-section fibrations -- physically realized as a Higgsing process that breaks a $U(1)$ theory to a discrete remnant. In this Subsection, we investigate such geometric transitions within quotient geometries.

For the quotient geometries of the previous section, by construction $\widehat{X}$ is smooth and the fibers over the singular fixed points in the base are fixed point free and so-called ``multiple fibers'' (non-reduced, everywhere singular curves). This smooth multi-section geometry can be connected to a fibration with section via a geometric transition.

This geometric transition must include a singular geometry from which both the genus one and elliptically fibered geometries are ``visible'' -- as a deformation or resolution of the singularity, respectively. Beginning with the multi-section fibration, this singular point can be reached via a complex structure deformation that allows a $\mathbb{Z}_n$ fixed point in the ambient space to hit the CY hypersurface. This tuning and the subsequent resolution of the singularities is known as a {\it hyperconifold transition} \cite{Davies:2009ub} which we review here for completeness, following \cite{Davies:2013pna}. 

A standard conifold transition \cite{Candelas:1989ug} can be represented in local coordinates $y_i \in \mathbb{C}^4$ as
\begin{align}
\label{eq:conifold}
p= y_1 y_4 - y_2 y_3 = 0 \, .
\end{align}
This nodal defining relation represents the cone over $S^3 \times S^2$ which can either be deformed to an $S^3$ for $p\rightarrow p+s$ or to an $S^2$ via a small resolution. \\ 

In the case at hand, it is possible to consider such a conifold transition under the action of a discrete symmetry, $\mathbb{Z}_n$ and a quotienting of both sides. The inclusion of a $\mathbb{Z}_n$ action on the coordinates and the subsequent quotient makes this a hyperconifold transition \cite{Davies:2011is,Davies:2013pna} by the additional action
\begin{align}
\label{eq:hyperconifold}
(y_1,y_2,y_3,y_4) \sim (\Gamma y_1,\Gamma^k y_2,\Gamma^{-k} y_3,\Gamma^{-1} y_4) \, ,
 \end{align} 
 with $\Gamma = e^{2\pi i /n}$ and $n$ an $k$ being co-prime. Hence the above action does not result in a standard three-sphere, when we go to the deformed phase, but a free-quotient of it, namely a Lens space $L(n,k)$. 
 
 This difference can be seen by a matrix parametrization of \eqref{eq:conifold} as
 \begin{align}
 p= \text{det}(W)\, \quad \text{ with  }\quad  W=\left( \begin{array}{cc}y_1 & y_2 \\ y_3 & y_4  \end{array}  \right) \, ,
 \end{align}
 and then rewriting  $W$ in terms of a triple $(r,X,v)$ \cite{Evslin:2007ux} with radial coordinate $r$, the matrix $X\in SU(2) \sim S^3$ and $v \in \mathbb{C}^2$ with $|v|=1$ representing a point on $\mathbb{P}^1$. In this parametrization we can write
 \begin{align}
 W = r X v v^\dagger \, ,
 \end{align}
 with the transformation law in Eq.~\eqref{eq:hyperconifold} acting as
 \begin{align}
 X\rightarrow \left(\begin{array}{cc} \Gamma & 0 \\ 0 & \Gamma^{-k}    \end{array}   \right) X \left(\begin{array}{cc} 1 & 0 \\ 0 & \Gamma^{k-1}    \end{array}   \right)\, ,\qquad v\rightarrow \left(\begin{array}{cc} 1 & 0 \\ 0 & \Gamma^{1-k}    \end{array}   \right)v \, .
 \end{align}
 Put differently, the action on the $S^3$ in terms of complex coordinates $z_0, z_1 \in \mathbb{C}^2$ with $|z_0|^2+|z_1|^2 = 1$ can be written as
 \begin{align}
 (z_0, z_1) \rightarrow (\Gamma z_0, \Gamma^{-k} z_1) \, .
 \end{align}
which is a free action\footnote{The action on the $\mathbb{P}^1$ is a simple rotation.} and defines the aforementioned Lens space $L(n,k)$. 

\begin{figure}
\begin{center}
\includegraphics[scale=0.4]{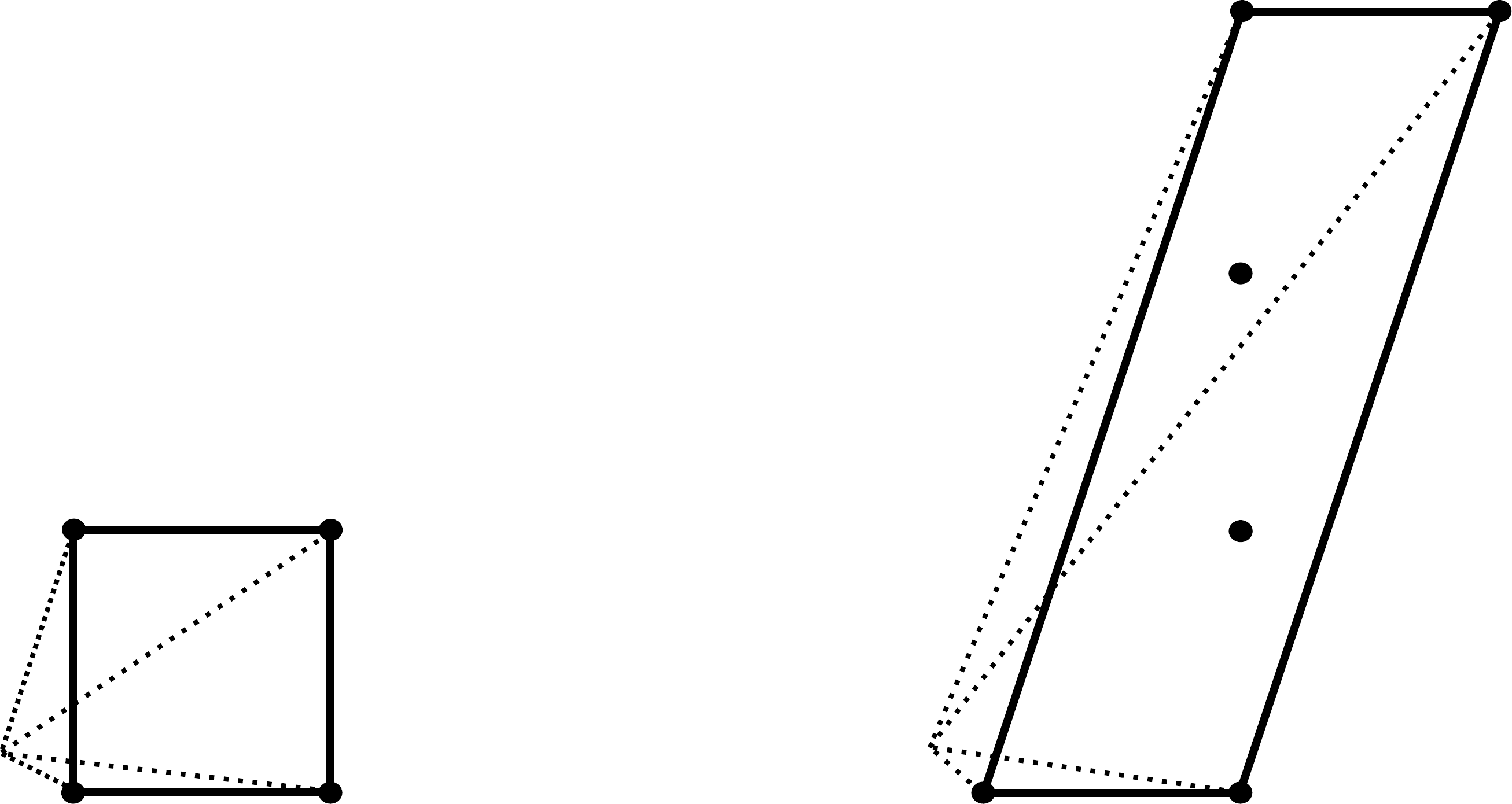}
\caption{\label{fig:toricconifold}{\it The toric diagram of the conifold on the left and its $\mathbb{Z}_3$ quotient on the right. The deformation phase corresponds to the twisted $S^3$ of the Lens space $L(3,1)$. Its resolution requires two exceptional divisors, represented by points in the interior of the parallelogram.}}
\end{center}
\end{figure}

The resolution side of the hyperconifold can also be considered which leads to the addition of $n$ resolution divisors. This can be seen by first considering the toric diagram of the hyperconifold when going to homogenous coordinates of \eqref{eq:conifold} with
\begin{align}
y_1 = z_1 z_3 \, , y_2 = z_1 z_4\, , y_3 = z_2 z_4\, , y_4 = z_2 z_4 \, , 
\end{align}
which admits the $\mathbb{C}^*$ scaling
\begin{align}
  (z_1, z_2, z_3, z_4) \sim   (\lambda z_1, \lambda^{-1} z_2,\lambda z_3, \lambda^{-1} z_4)\, ,
\end{align}
encoded in the toric diagram depicted in Figure~\ref{fig:toricconifold} where each coordinate represents a $1$-dimensional cone $v_i \in N=\mathbb{Z}^3$. The toric diagram is spanned by the fan of $1$-dimensional cones
\begin{align}
\Sigma_1: \{ v_1 = (1,0,0)\, , v_2 = (1,1,0) \, , v_3 = (1,0,1) \, , v_4= (1,1,1) \, \} \, .
\end{align}
The quotient that acts on the conifold in Equation~\eqref{eq:hyperconifold} can be understood \cite{Davies:2013pna} as a refinement of the base lattice $N$ which we represent by the new basis 
\begin{align}
N': \{ (1,0,0), (0,1,0), (0,-k/n, 1/n) \} \, ,
\end{align}
such that in this new basis $\Sigma_1$ has coordinates 
\begin{align}
\Sigma'_1 = \{(1,0,0), (1,1,0),(1,k,n),(1,k+1,n)\} \, ,
\end{align}
depicted schematically in Figure~\ref{fig:toricconifold}.
As the volume of the parallelogram in the $(y,z)$ plane has volume $n$, it is easy to see that we need $n-1$ additional exceptional divisors $E_i$ for a full resolution of the space. Performing the full regular star triangulation introduces $2n$ additional 3-dimensional cones to the original diagram.

This toric description provides everything we need to obtain the change in the Hodge and Euler numbers in a hyperconifold transition $\widehat{X}\rightarrow \widetilde{X}$. As the Euler number is the number of top dimensional cones, we find 
\begin{align}
\begin{array}{rl}
 \chi(\widehat{X})-\chi(\widetilde{X}) =&2n \, , \\
 h^{1,1}(\widehat{X}) - h^{1,1}(\widetilde{X})=&1-n \, , \\
  h^{2,1}(\widehat{X}) - h^{2,1}(\widetilde{X}) =&1 \, , \\
  \pi_1(\widetilde{X})  =& \pi_1(\widehat{X}) / \pi_1(L(m,k))=\mathbf{1} \, .
\end{array}
\end{align}
The change in complex structures is derived from $\chi=2(h^{1,1}-h^{2,1})$.
Moreover, in going to the resolution phase $Y$ we have lowered\footnote{In general it can happen that only an $\mathbb{Z}_m$ singularity hits $\widehat{X}$ and therefore reduces only a subgroup of the fundamental group on $\widehat{X}$: $\pi_1 (\widetilde{X}) = \mathbb{Z}_{n/m}$.} or eliminated the fundamental group of $\widehat{X}$ by deleting the Lens space $L(n,k)$ and adding in resolution divisors \cite{Davies:2013pna}. That is, $\widetilde{X}$ is in general simply connected. 

It is also possible to relate some triple intersection numbers across a hyperconifold transition
\begin{align}
\widehat{d}_{K,L,M} = \widehat{D}_K \widehat{D}_L \widehat{D}_M\, , \quad \text{ and }\quad
   d_{K,L,M} = D_K D_L D_M , \,
\end{align}from those on $\widehat{X}$ to those on $\widetilde{X}$.
  For this we distinguish the following sets of divisors on $\widehat{X}$ and $\widetilde{X}$ respectively
\begin{align}
\widehat{D}_M : \{\widehat{D}_\alpha, \widehat{D}_m \} \, , \quad \text{ and }\quad D_M: \{ D_\alpha, D_m, E_i \} \, .
\end{align}
The Cartier divisors $\widehat{D}_\alpha$  on $\widehat{X}$  do not intersect the conifold point upon tuning and stay in the same homology class $[D_\alpha]$ on $\widetilde{X}$ and therefore do not change intersection numbers. As the $\widehat{D}_{\alpha}$ miss the conifold, they also miss the resolution divisors $E_i$ on $\widetilde{X}$ and therefore we have intersections
 \begin{align}
 d_{i,\alpha,\alpha'} = d_{i,j,\alpha}=0 \,, \qquad \widehat{d}_{\alpha,\alpha',\alpha''}= d_{\alpha,\alpha',\alpha''}\, .
 \end{align} 
 The divisors $\widehat{D}_m$ on the other hand are not Cartier and have altered intersection numbers upon blow-up. In particular the zero-section $D_0$ or its multi-section analog \cite{Mayrhofer:2014haa,Klevers:2014bqa} is a divisor of this type which we need in order to deduce the intersection pairing of the base 
 \begin{align}
 d_{0,\alpha,\beta} = \Omega_{\alpha,\beta} \, ,
 \end{align}
relevant for 6-dimensional anomaly cancellation in Section~\ref{sec:6dAnomalies}. From that point of view it is clear that 
the intersection matrix on $Y$ obtains a block diagonal form as
 \begin{align}
 d_{0,\alpha,i} = \Omega_{i,\alpha}=0\, , \quad \text{ and } \quad d_{0,i,i'}= \Omega_{i,i'} \, .
 \end{align} 
  
Having summarized the properties of generic free quotients of CY  manifolds and hyperconifolds, we now have to specialize to the case of a genus one fibered quotient geometry. Suppose there exists a projection $\pi: \widehat{X} \to B_{\text{orb}}$ down to a possibly singular base $B_{\text{orb}}$. In this case the second base homology $H_2(B_\text{orb},\mathbb{Z})$ is generated by the divisors $D^{b}_m$ in the image of the projection $\pi$. Some of these can be non-Cartier divisors that, however, can be Cartier on $B_{\text{orb}}$ i.e. they can avoid orbifold singularities on $B_{\text{orb}}$. Note however that, by construction the full fibration is smooth but with multiple fibers over the fixed points. 

As we are considering elliptically or genus one fibered threefolds  we distinguish  how the local fan $\Sigma^{(n)}_{(1)}$ of the hyperconifold restricts to the base under the projection, $\pi$:
\begin{enumerate}
\item $\Sigma^{(n)}_{(1)}$ restricts to a local fan  
\begin{align}
\Sigma^{(n),b}_{(1)}: \{v_1 =(1,0), v_n=(1,n)        \} \, ,
\end{align}
which is an $A_{n-1}$ singularity in the base. In this case all resolution divisors $E_i$ are horizontal and restrict to base resolution divisors increasing $h^{1,1}(B_{\text{orb}})$ resulting in $n-1$ additional tensor multiplets.
\item $\Sigma^{(n)}_{(1)}$ restricts to a single vertex in $B_\text{orb}$. In such a case the $A_{n-1}$ singularity is purely in the fiber and we have added an $SU(n)$ gauge symmetry with $E_i$ being vertical resolution divisors.
\item $\Sigma^{(n)}_{(1)}$ restricts to $m$ divisors in the base only, with $n-m$ resolution divisors of gauge algebras over them. This case is a combination of the last two cases.
\end{enumerate}
In this work we mainly consider transitions of the first type and comment on those of the second type in some examples.

\subsection{The F-theory physics of genus one fibrations}
\label{sec:F-theoryReview}
In the previous section we reviewed the geometry of hyperconifold transitions in quotient Calabi-Yau geometry. It is our goal to employ this geometry to model $U(1)$ Higgsing transitions that involve discretely charged superconformal matter. Before beginning this analysis though, it is useful to review briefly the physics of ``ordinary'' $U(1)$ Higgsing transitions in F-theory, realized via conifold-type transitions.

Cyclic symmetries $\mathbb{Z}_n$ are known to be generated in F-theory via compactification on a geometry with an n-section of the fiber $\mathcal{C}$ \cite{Morrison:2014era}. In all known examples, it can be observed that the n-section geometry can be linked by a chain of conifold transitions to an elliptic fibration with $n$ additional linearly independent rational sections (these sections will give rise to a rank $n$ sublattice of the Mordell-Weil group).  The transitions\footnote{Note that one U(1) un-Higgsing can involve multiple conifold points.} un-Higgs the $\mathbb{Z}_n$ to a $U(1)^n$ gauge symmetry.

\begin{figure}
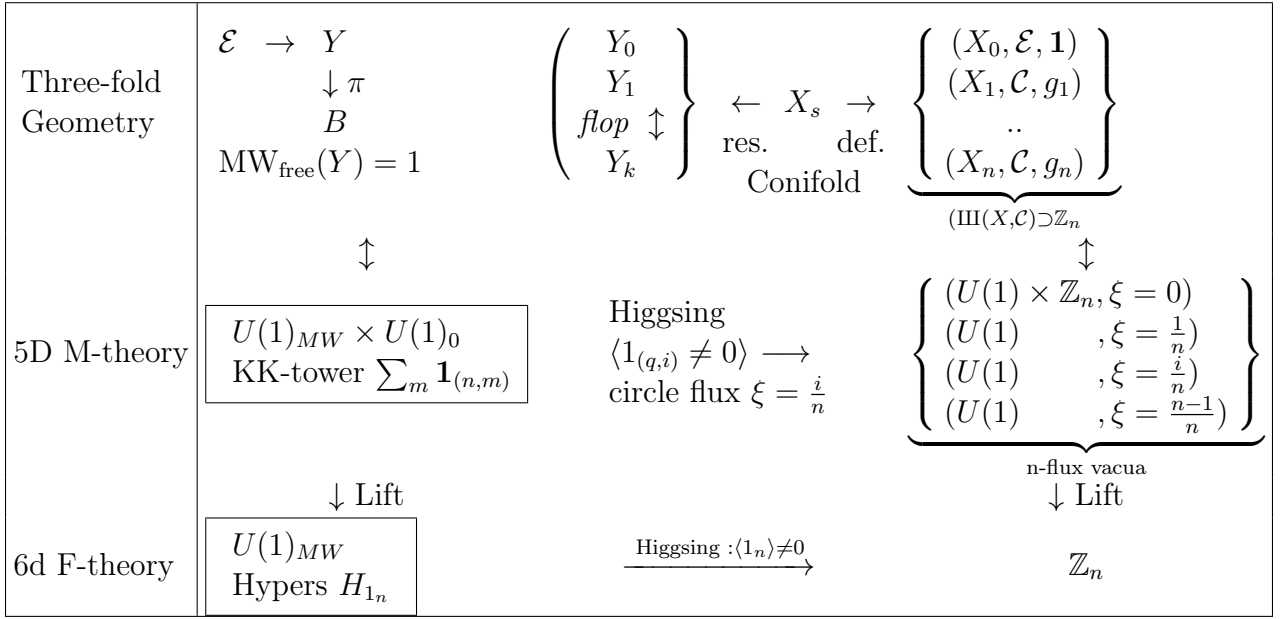

\begin{center}
 \renewcommand{\tabcolsep}{0.1cm} 
 \begin{tabular}{|l|llll|} \hline
 \begin{tabular}{l} 
Three-fold\\
 Geometry
  \end{tabular}
   & 
  $  \begin{array}{lll}
  \mathcal{E} &\rightarrow& Y  \\
   &&\downarrow \pi  \\
 && B 
 \\
 \multicolumn{3}{c}{\text{MW}_{\text{free}}(Y)=1}
 \end{array} $
 
 & 
 $ \left(  \begin{tabular}{c}
  $ Y_0 $ \\
 $ Y_1 $ \\
 $ \text{\it flop } \updownarrow $ \\
 $  Y_k $
 \end{tabular} 
 \right\} $
 & 
 \begin{tabular}{ccc}
 & &   \\
 & &   \\
   $\leftarrow$  &  $X_s$ & $\rightarrow$ \\
 
  res. & & def. \\
  \multicolumn{3}{c}{Conifold}
 \end{tabular}
& 
$ \underbrace{   \left \{  \begin{tabular}{c}
  $ (X_0, \mathcal{E}, \mathbf{1}) $ \\
 $ (X_1, \mathcal{C}, g_1) $ \\
 $ .. $ \\
 $  (X_n, \mathcal{C}, g_n) $
 \end{tabular} 
 \right\}}_{ (\Sha(X ,\mathcal{C}) \supset \mathbb{Z}_n} $
  \\
  
  & 
  
 \multicolumn{1}{c}{ $\updownarrow$}
  
  &
  
  &
  &
 \multicolumn{1}{ c|}{ $\updownarrow$}
  \\
  5D M-theory &    \framebox{
  \begin{tabular}{l}
  $U(1)_{MW} \times U(1)_0 $\\
 KK-tower $\sum_m \mathbf{1}_{(n,m)} $\end{tabular}}
  & 
  
\multicolumn{2}{c}{
  \begin{tabular}{l}
 Higgsing  \\
  $ \langle 1_{(q,i)}\neq 0    \rangle $ $\longrightarrow$\\
 circle flux $\xi = \frac{i}{n} $
  \end{tabular} }
  & 
$ \underbrace{\left\{ \begin{tabular}{l}
 $(U(1) \times \mathbb{Z}_n, \xi=0) $\\
  $(U(1)\phantom{\times \mathbb{Z}_n} \, \, \, , \xi=\frac{1}{n})   $\\
$(U(1)\phantom{\times \mathbb{Z}_n} \, \, \, , \xi= \frac{i}{n} )$ \\
$(U(1)\phantom{\times \mathbb{Z}_n} \, \, \, , \xi = \frac{n-1}{n} )$
  \end{tabular} \right\}}_{\text{n-flux vacua}} $

 \\
  & 
  
 \multicolumn{1}{c}{ $\downarrow$ Lift}
  
  &
  
  &
  &
 \multicolumn{1}{ c|}{ $\downarrow$ Lift}
  \\

 6d F-theory
  &
 \framebox{
 \begin{tabular}{l}
  $U(1)_{MW}$ \\ 
 Hypers $H_{1_n}$ 
 \end{tabular} }
 &
 \multicolumn{2}{c}{$\xrightarrow{\text{ Higgsing :} \langle 1_n \rangle \neq 0}  $ }
   &
   \multicolumn{1}{c|}{  $\mathbb{Z}_n$ }
   \\ \hline
 \end{tabular}
 \caption{\label{fig:genusonefibrations}{\it Graphical summary of the transition of genus one fibered threefolds $X$ towards an elliptically fibered geometry $Y$ with enhanced Mordell-Weil group described in \ref{sec:F-theoryReview}. Depicted are the geometric transition above and the $5$-dimensional M-theory and $6$-dimensional F-theory lifts below.}}
 \end{center}

 \end{figure}

In this picture, the last Higgsing $U(1)\rightarrow \mathbb{Z}_n$ is of particular interest, triggered by the vev of $q=n$ U(1) charged hypermultiplets in $6$-dimensions. There is an intricate and beautiful interplay between the threefold geometry, the physics of the 5-dimensional M-theory and its 6-dimensional F-theory uplift which we have depicted in Figure~\ref{fig:genusonefibrations}. The central geometric object is the singular geometry $X_s$ with a conifold singularity which admits both a small resolution and a deformation, leading to two topologically distinct, smooth threefolds. The resolution side represents a collection of several elliptic fibrations $Y_i$, related by flop transitions and a free Mordell-Weil group of rank one. In the 5-dimensional M-theory, where we have the additional circle $U(1)_0$, those vacua represent different realizations of holomorphic curves with the same $U(1)_{MW}$ charge but different KK $U(1)_0$ charges \cite{Mayrhofer:2014laa, Cvetic:2015moa}. Indeed, shrinking the differently realized holomorphic curves and then deforming realizes $n$ different sets of genus one fibered geometries $X_i$ with n-sections only, that all share the same Jacobian $\text{J}(X_i)=X_0$. 
The set of all these geometries can be collected to a group, together with an action on the geometries $g_i$ that forms the group of \emph{Calabi-Yau torsors}, known as the Weil-Ch$\hat{\text{a}}$telet (WC) group \cite{silverman}. Commonly in the F-theory literature, the set of CY torsors reduces to a subgroup of the WC group, known as the Tate-Shafarevich (TS) group $\Sha(X,\mathcal{C})$ which admits a $\mathbb{Z}_n$ subgroup. The difference between the Weil-Ch$\hat{\text{a}}$telet and Tate-Shafarevich groups is frequently negligible, but in the case of CY fibrations admitting multiple fibers in co-dimension 2, the difference can become significant. 

In 5-dimensional compactifications of M-theory on these sets of geometries, there exists a beautiful match between the CY torsors and the collection of holomorphic curves $C_i$ with charges $(n,i)$ under $U(1)_{\text{MW}} \times U(1)_0$ that can become massless and induce a non-trivial flux $
\xi = \frac{i}{n}$ along the circle in the resulting geometry. Thus we see that only one geometry, the Jacobian, admits a full $U(1)_0 \times \mathbb{Z}_3$ symmetry after Higgsing, triggered by the veved field which geometrically does not intersect the zero-section. The other geometries without sections correspond to $U(1)$ theories with non-trivial flux $\xi=\frac{i}{n}$ labeling the various M-theory vacua. 

It is important to note that only one geometry in this collection, the Jacobian, admits non-trivial torsional three-cycles \cite{Mayrhofer:2014laa} whereas the genus one geometries do not. The torsion appearing in the Jacobian plays a clear role in discrete flux backgrounds in M-theory \cite{Camara:2011jg,Mayrhofer:2014haa} and mathematically is an element of the cohomological Brauer group, $B(X)$. It is important to note that this finite group is one of only two types of cohomological torsion available in CY threefolds. The universal coefficient theorem guarantees that $\text{Tors}(H_i(X, \mathbb{Z})) \simeq [\text{Tors}(H^{i+1}(X,\mathbb{Z})]$ \cite{Batyrev:2005jc}. Moreover, there is no torsion in $H^0(\check{X}, \mathbb{Z})=H^6(\check{X}, \mathbb{Z})$ for CY manifolds. The non-trivial structure occurs then as $B(X) \subset H^3(\check{X}, \mathbb{Z})$ (which gives also rise to a finite group $B^*$ in $H^4(\check{X}, \mathbb{Z})$) and torsion $A(X)$ in $H^2(X, \mathbb{Z})$ where $A(X)$ is a finite Abelian group (with $A^*$ appearing as discrete torsion in $H^5(X, \mathbb{Z})$)\footnote{If  $(X,\check{X})$ are mirror pairs in toric hypersurfaces in \cite{Batyrev:2005jc}  $A(X)=B(\check{X})$ and $B(X)=A(\check{X})$; however  for a possible  general  counterexamples  see \cite{Braun:2017oak}}. In the case of Calabi-Yau quotient geometries only one of these torsion groups \emph{must be non-zero}, namely $A(X)=Hom(\pi_1(X), \mathbb{Q}/{\mathbb{Z}})$. It is important to make that distinction as the non-simply connected threefolds we will consider in the following sections are all genus one fibered but already exhibit torsional\footnote{In the context of Type IIA strings and M-theory, the presence of $A(X)$ torsion
can be shown to be in correspondence to discrete gauge symmetries \cite{Camara:2011jg}.} cycles ($A(X)$), unlike in the covering spaces in which only the Jacobian contains torsion. Finally in 6-dimensional F-theory vacua, the theory is only sensitive to the $\tau$ function, which coincides for all sets of the TS-group and thus all elements of $\Sha(X,\mathcal{C})$ lift to the same 6-dimensional F-theory physics with a $\mathbb{Z}_n$ discrete symmetry. 

We are nearly ready to consider geometric transitions linking quotient geometries and discuss the physics of their associated $U(1)$ Higgsing transitions. This will take the schematic form of quotients of both genus one fibered CY manifolds and their associated (singular) Jacobians. The commutative relationship between these processes can be illustrated as follows:
 
  \begin{centering}
\begin{equation}\label{Jac_quotient}
  \begin{array}{lllll}
  &X &\stackrel{\phi}{\longrightarrow}&J(X)&\\
  \Gamma &\downarrow&&\downarrow \,  \Gamma& \\
  &\widehat{X}&\stackrel{ \phi }{\longrightarrow}&J(\widehat{X})&
 \end{array}  \;,
\end{equation} 
\end{centering}
with the Jacobian map $\phi$ from the genus one to the elliptic fibration. Before turning to this however, we must first address some of the physics of the superconformal sectors associated to the singular base geometries $B_{orb}$, and we turn to this now.

 \subsection{$\mathbf{A}_{n}$ (2,0) super conformal points}\label{sec:scft_points}
 There is a vast literature concerning (2,0) SCFTs and their properties. These theories have a highly non-trivial and rich structure when coupled to various flavor symmetries, whose full review is beyond the scope of this work. Instead we review the $\mathbb{Z}_n$ quotient singularities and their physics for the simplest cases. In such a case we can view the $A_{n-1}$ singularity as a stack of M5 branes that probe the singularity and support the $(2,0)$ theory in terms of $n-1$ free $(2,0)$ tensors $T_{(2,0)}$. Such a tensor consists of an anti-self-dual tensor, two negative chirality tensorini and five real bosons that can be understood as the transverse directions of the M5 brane. In terms of a (1,0) theory those tensors can be decomposed into
\begin{align}
T_{(2,0)} = H_{\mathbf{1}_0} \oplus T_{(1,0)} \, .
\end{align} 
Hence a $(2,0)$ tensor contributes equivalently as a   $(1,0)$ tensor and a neutral hypermultiplet in the anomaly polynomial. This is precisely the same as the contribution from the M5 brane induced R-symmetry anomaly in \cite{Harvey:1998bx} canceled by the 8 form contribution 
\begin{align}
I^{A_{n-1}}_8 = \frac{n-1}{192} \left( tr R^4 - \frac14 (tr R^2)^2             \right) \, .
\end{align}
Note that we generically can have multiple $\mathbb{Z}_n$ singularities at the same time, as well as $\mathbb{Z}_m$ ones, if m divides $n$. 
Thus, if we have $m_i$  $\mathbb{Z}_{n_i}$ singularities this will yield the total amount of (2,0) tensors 
\begin{align}
T_{(2,0)}= \sum_i m_i (n_i-1) \, ,
 \end{align} 
which will modify the gravitational anomalies as
 \begin{align}
 H-V+29T_{(1,0)}-30 T_{(2,0)} -273 = 0 \, , \qquad 9-T = a \cdot a + T_{(2,0)}\, .
 \end{align}
 Note that the number of perturbative $T_{(1,0)}$ can be computed from the number of 
 K\"ahler moduli of the base minus the overall volume.
 \begin{align}
 T_{(1,0)} = h^{1,1}(B)-1 \, .
 \end{align}

\subsection{Anomalies and $ \mathcal{A}_n $ (2,0) theories}
\label{sec:6dAnomalies}
In this section we turn to the physics of 6-dimensional F-theory compactifications on smooth quotient geometries, supporting $\mathcal{A}_{n-1}$ points and discuss their tensor branches. As it will turn out, those tensor branches differ
by those of regular $A_{n-1}$ (2,0) points by a coupling to the discrete symmetry.
  For this we review first the cohomology lattice of the orbifold base, that encodes the Green-Schwarz coefficients in the anomalies which will be crucial for our argument. For simplicity, we consider smooth quotients that do not change the K\"ahler deformations\footnote{Similar generalizations of quotient theories on the level of the anomaly lattice were performed in \cite{Apruzzi:2017iqe}.}  of the CY.
 In the following we will show that such quotients simply lower the global matter spectrum and introduce additional free (2,0) tensors consistent with anomaly cancellation. The tensor branch of these theories however can be computed from a hyperconifold transition using the features as reviewed above and reveals additional discrete charged hypermultiplets and therefore differs from the one of an $A_{n-1}$ theory. Hence we denote those superconformal subsectors as $\mathcal{A}_{n-1}$ theories.

\subsubsection{The 6-dimensional anomaly lattice}
An important ingredient in the description of 6-dimensional F-theory physics is the second homology lattice of the F-theory base $H_2(B, \mathbb{Z})$, whose intersections captures the Green-Schwarz anomaly coefficients of the SUGRA theory  \cite{Park:2011ji}.
The homology lattice is identified with the string charge lattice and satisfies tight constraints, being integral and unimodular \cite{Park:2011ji}. However it is well known that in the case of singularities, even orbifolds, $H_2$ is not necessarily integral and hints at fractional instanton charges of the strongly coupled sectors \cite{DelZotto:2014fia}. 
 For our purposes we distinguish three related base homologies 
\begin{align}
H_2(B_{\text{cov}}, \mathbb{Z}) \, , \qquad H_2(B_{\text{orb}},\mathbb{Z})\, , \qquad H_2 (B_{\text{res}},\mathbb{Z}) \, ,
 \end{align} 
 related by 
 \begin{align}
B_{\text{res}} \xrightarrow{\widehat{\pi}}  B_{\text{orb}} =  B_{\text{cov}}/\Gamma_{n,b} \, ,
 \end{align}
  via the blow-down map $\widehat{\pi}$. The homology lattice of the resolved geometry, we identify as the tensor branch of the strongly coupled orbifold geometry.
 For any of these bases we can expand a divisors $D$ in terms of a basis $e_M \in H_2 (B,\mathbb{Z})$ as
 \begin{align}
 D = \sum_M d^M e_M \, ,
 \end{align}
 for $M = 1, \cdots, T$. Thus the $SO(1,T)$ intersection matrix
 \begin{align}
  \Omega_{M,N} = e_M \cdot e_N \, ,
\end{align}
 can be used in order to raise and lower indices and take the intersection product of two divisors
\begin{align}
D \cdot D' = \Sigma_{M,N} d^M d^N \, .
\end{align}
Fixing some basis $ e_M$ on the covering base $B_{\text{cov}}$, we find that after quotienting by $\Gamma_{n,b}$ they become $\widehat{e}_M$ with intersections on $B_{\text{orb}}$ for our choice of the basis given as
\begin{align}
\widehat{e}_M \cdot \widehat{e}_N =  \frac1n e_M \cdot e_N \, ,
\end{align}
which is  not integral in that base but fractional and hence these divisors are non-Cartier on $B_{\text{orb}}$.

On the other hand, the orbifold base is linked to a smooth base $B_{\text{res}}$ by gluing in
resolution divisors $e_i$, whose second homology is again integral and unimodular as here we have a well defined SUGRA description, representing the tensor branch of the super conformal points. 
As reviewed in Section~\ref{sec:LensReview} the intersection matrix on $B_{\text{res}}$  becomes block diagonal
\begin{align}
\Omega_{M,N} = \Omega^s_{\alpha,\alpha'} \oplus \Omega^r_{i,j}\, ,
\end{align}
with respect to the basis of divisors $e_\alpha$ of the quotient geometry and $e_i$ the resolution divisors. For the arguments that follow the Cartier divisors on $B_\text{orb}$ are again of particular importance, as they have an unchanged homology class in $H_2(B_\text{res},\mathbb{Z})$ and unchanged intersection numbers if they contain components of the resolution divisors \cite{DelZotto:2014fia}.

\subsection{Anomaly cancellation on the quotient geometry}
\label{sec:anomalies}
We turn now to anomaly cancellation on the quotient geometry. The connection to the anomalies is made by the identification of the Green-Schwarz coefficients as intersections of vertical divisors in the base \cite{Park:2011ji}. The full consistency conditions are listed in Appendix~\ref{app:Anomalies} but the central objects are the base divisors
\begin{align}
a \sim K_b \,, \quad b \sim [\mathcal{S}_{ADE}] \, , \quad b_{mn} = \pi (\sigma(s_m) \cdot \sigma(s_n)) \, ,
\end{align}
with $\mathcal{S}_{ADE}$ the base divisor of some ADE fiber and its U(1) analog $b_{mn}$ which is the N\'eron-Tate height pairing \cite{Morrison:2012ei} of Shioda maps $\sigma(s_m)$ associated to an enhanced Mordell-Weil group. As we are considering compact geometries, we have to satisfy in particular the gravitational anomaly
\begin{align}
H-V+29T  -273= 0 \, ,
\end{align}
which gives a strong constraint on the global spectrum of the gauge group.

We start from a torus fibered CY $X$ which is fully resolved and where all anomalies are canceled. Applying the freely acting quotient, as stated in Section~\ref{sec:Quotientreview} we obtain a smooth threefold $\widehat{X}$ where over the fixed points in the base there are at most multiple but non-reducible fibers without a gauge enhancement. This amounts to the requirement that ADE divisors $b$ are Cartier and do not cross a singularity in $B_{\text{orb}}$ and in analogy
we also demand the same for the height pairings $b_{mn}$.

As the fundamental domain of $B_{\text{cov}}$ is reduced by $n$, the quotient reduces the amount of hypermultiplets by $n$. In the following we want to show full gauge anomaly cancellation before we consider the gravitational anomalies. For this we introduce the notation of the base divisors for the covering and orbifold theory
\begin{align}
\begin{array}{lll}
a, b,b_{mn} \in H_2(B_{\text{cov}},\mathbb{Z})\, & \text{ and } & \widehat{a}, \widehat{b}, \widehat{b}_{mn} \in H_2(B_{\text{orb}},\mathbb{Z})\,.
\end{array}
\end{align}
All anomalies are summarized in Appendix~\ref{app:Anomalies} and here we include only a selection that will be useful in the following arguments. We begin with the mixed gravitational Abelian anomaly
\begin{align}
    -\textstyle{\frac{1}{6}}\sum_{\underline{q}} x_{q_r, q_s} q_r q_s&=a\cdotp b_{rs} \, ,  
\end{align}
 where $x_\mathbf{R}$ denotes the multiplicity of the hypermultiplets in the representation $\mathbf{R}$. Anomaly cancellation in the quotient geometry is thus satisfied as we have
 \begin{align}
 \widehat{a} \cdot \widehat{b}_{rs} = \frac1n a \cdot b_{rs} \, ,
 \end{align}
which cancels the contribution of the $\widehat{x}_{q_r,q_s} =   x_{q_r,q_s}/n$ reduced amount of hypers. Similarly we can proceed for the non-Abelian gauge anomalies  as follows
\begin{align}
 \textstyle{ -\frac{1}{6}}\left( A_{adj_\kappa}-\sum_{\mathbf{R}} x_{\mathbf{R}} A_{\mathbf{ R}}\right)&=a \cdotp \left( \frac{b_{\kappa}}{\lambda_\kappa}\right) \, ,  \nn \\
  \textstyle{\frac{1}{3}} \left( \sum_{\mathbf{R}} x_{\mathbf{R}} C_{\mathbf{R}}-  C_{adj_\kappa}  \right) &= \left( \frac{b_\kappa}{\lambda_\kappa}\right)^2  \, .
\end{align}
Here we keep in mind that $b$ on $B_{\text{cov}}$ is a genus g curve with
\begin{align}
g = 1 + \frac12\left(b\cdot b + b\cdot a \right) \, ,
\end{align}
which supports $g$ adjoint hypermultiplets. Thus after taking the quotient the genus is changed to
\begin{align}
\label{eq:genusquotient}
\begin{split}
\widehat{g} =& 1 + \frac12 \left(\widehat{b} \cdot \widehat{b} +\widehat{b}\cdot \widehat{a} \right) \, ,\\  
&= 1 + \frac{1}{2n}\left(b \cdot b + b \cdot a                           \right) \, , \\
&= 1 + \frac1n (g - 1)\, ,
\end{split}
\end{align}
or equivalently that $\widehat{g}-1 = \frac1n (g-1) $. Then, pulling out the sum over the adjoint representation we find
\begin{align}
\begin{split}
   &\textstyle{ -\frac{1}{6}}\left( -\sum_{\mathbf{R}} \widehat{x}_{\mathbf{R}} A_{\mathbf{ R}}-A_{adj_\kappa} (\widehat{g}-1)     \right) -\widehat{a} \cdotp \left( \frac{\widehat{b}_{\kappa}}{\lambda_\kappa} \right) \, , \\
 =&   \textstyle{ -\frac{1}{6 n}}\left( -\sum_{\mathbf{R}} x_{\mathbf{R}} A_{\mathbf{ R}}-A_{adj_\kappa}  (g-1)     \right) -\frac1n a \cdotp \left( \frac{b_{\kappa}}{\lambda_\kappa} \right) =0 \, ,
  \end{split}
\end{align}
and hence the above the above gauge anomalies are canceled. Similar arguments hold for all other gauge anomalies as well.\\\\
 Finally we have to consider the gravitational anomaly which is where we will find an additional contribution. We start with the reducible anomaly
\begin{align}
h^{1,1}(B)-1 = T = 9-a \cdot a \, .
\end{align}
Note that the K\"ahler moduli of the base remain unchanged, while the self intersection of the canonical class on the other hand does change, and hence we obtain a mismatch of tensor multiplets
\begin{align}
\begin{split}
\Delta T &= a \cdot a - \widehat{a} \cdot \widehat{a} \, , \\
 &=a \cdot a \left(\frac{n-1}{n} \right) =T_{(2,0)} \, .
 \end{split}
\end{align}
Upon the blow-up this mismatch gets resolved by the introduction of the additional K\"ahler parameters which corresponds to the tensor branch of the superconformal matter points.
Secondly we have to consider the contribution of the irreducible gravitational anomaly that is
\begin{align}
\label{eq:quotientanomaly}
H_{\text{neut}} + H_{\text{charged}} + H_{\text{adjoint}} - V + 29 T -273 + \Delta_{\text{strong}}= 0 \, ,
\end{align}
where $\Delta_{\text{strong}}$ is the contribution of the strongly coupled sector and we have split up the contribution of the different types of hypermultiplets. Again, the amount of adjoint representations are counted by the genus $g_\kappa$ of the ADE curves whereas the multiplicity of them is reduced by the
neutral \emph{Cartan}-like states that we already count as complex structure deformations in $H_{\text{neut}}$. Thus the contribution of the adjoints comes with the multiplicity of {\it root}-like states  
\begin{align}
H_{\text{adjoint}} = \sum_\kappa \underbrace{(\text{dim(adj)}_\kappa-\text{rank}(G_\kappa))}_{\text{root}(G_\kappa)}g_\kappa\, .
\end{align}
In the covering geometry we do not have a strongly coupled sector and expect $\Delta_{\text{strong}} =0$. However,   $\Delta_{\text{strong}}$ should be non-zero in the quotient geometry. Since we can fix the rest of the spectrum in the quotient
in terms of the unquotiented theory we can compute the contribution $\Delta_{\text{strong}}$ exactly.\\
We focus again on the case where the gauge group and tensor multiplets stay unaltered by the quotient. Here we first compute the change in complex structure as follows.
\begin{align}
\begin{split}
\widehat{h}^{2,1} &= h^{2,1} +\widehat{h}^{1,1} - h^{1,1} -\frac12 \Delta \chi \,  \\
 &=  h^{2,1} - \frac12 \chi(X) \left( \frac{1-n}{n}\right) \, 
 \end{split}
\end{align}
Then Equation~\eqref{eq:quotientanomaly} in the quotient geometry becomes
 \begin{align}
 \frac{H_{\text{charged}}}{n} + H_{\text{neut}} + \frac12 \chi(X) \left(\frac{n-1}{n}\right) + \sum_\kappa \text{root}(G_\kappa)\left( \frac{g_\kappa-1}{n} +1 \right)- V +29 T+\Delta_{\text{strong}} -273 = 0
\end{align}
Subtracting the anomaly of the covering three-fold we eliminate the $H_{\text{neut}}$ and obtain
 \begin{align}
 \left( \frac{1-n}{n} \right)\left(H_{\text{charged}} - \frac12 \chi + \sum_\kappa \text{root}(G_\kappa)(g-1) \right) + \Delta_{\text{strong}}=0 \, .
 \end{align}
Inserting the contribution of the Euler number $ \chi = 2( \text{rank}(G) +  T_{(1,0)} - H_{neut}-3)$ 
we can therefore express the contribution of the strongly coupled sector as
\begin{align}
\begin{split}
\Delta_{\text{strong}} =&  
  \left( \frac{1-n}{n} \right) \left(V + T_{(1,0)}+3 -H_{\text{charged}} - H_{\text{adjoint}}- H_{\text{neut}} \right) \, , \\
  =& 
    \left( \frac{1-n}{n} \right) \left( 30 T_{(1,0)} -270         \right) \, , \\
      =& 
   30 T_{(2,0)}\, ,
   \end{split}
\end{align}  
where we have used the two gravitational anomalies again. Indeed we find exactly the contribution of $T_{(2,0)}$ (2,0) tensor multiplets stemming from the various fixed points which renders the theory consistent with all gauge and gravitational anomalies.

It should be stressed again that we have considered a very special kind of quotient in the above considerations that preserves smoothness and the dimension of all K\"ahler deformations $h^{1,1}(X)$. The above arguments suggest that we already seem to have captured all of the degrees of freedom appearing in the theory. However, there is a question as to whether the $(2,0)$ sector is charged under the discrete symmetry or not. We address this question in the next section.
\subsection{The $\mathcal{A}_n$ hyperconifold tensor branch}
\label{sec:Tensorbranch}
Let us reconsider what kind of theories we have constructed: these are theories that have to have discrete symmetries, originating from the genus one fibrations, and fixed points in the base carrying free (2,0) hypermultiplets. As we have shown before, the degrees of freedom of the free (2,0) hypermultiplets are enough to cancel all gravitational anomalies such as
\begin{align}
\label{eq:anomaly}
\widehat{H} - \widehat{V} +29 \widehat{T} + 30 T_{(2,0)}-273 = 0 \, .
\end{align}
  Hence there is no reason to assume that these are not regular $A_n$ theories, in particular as the fiber is non-singular and non-reducible over the fixed points, apart from being multiple. However in the following we will perform a hyperconifold transition, which corresponds physically to going to the tensor branch of the theory. This necessarily introduces additional $n$ discrete charged hypermultiplets consistent with all anomalies.
  Note again, that we require that all blow-up divisors of the hyperconifold restrict onto the base and all ADE gauge divisors and U(1) height pairings to be Cartier divisors not intersecting the singularity.
 This implies that, even after the resolution, they stay in the same homology class with unaltered intersections resulting in the same Green-Schwartz coefficients. This on the other hand implies 
 that the multiplicity of states charged under the continuous part of the gauge group does not change\footnote{Other possible transitions are those that are anomaly equivalent when a divisor develops ordinary double point singularities \cite{Klevers:2017aku}.} . 
As a result, the only change we can observe is in the irreducible gravitational anomaly \eqref{eq:anomaly} given as
\begin{align}
\widehat{H} + \Delta H -1 + \widehat{V} + 29(\widehat{T}+n-1) + 30(T_{(2,0)} - n -1) -273 =0\, ,
\end{align}
using that a hyperconifold reduces the complex structures by one and introduces additional $n-1$ Tensor multiplets. However the irreducible anomaly is not canceled anymore as we are missing $\Delta H = n  $ neutral hypers missing from the free $(2,0)$ tensors. This mismatch can not be compensated by any additional hypermultiplet charged under a continuous gauge symmetry as the associated gauge divisors are all Cartier and therefore their anomalies are not modified. However there is still a discrete gauge symmetry present and hence the only possible way to cancel the gravitational anomaly is by introducing $n$, $\mathbb{Z}_n$ charged hypermultiplets. 

Hence, we argue that the free quotient introduces $\mathbb{Z}_n$ discrete gauge symmetries and that the (2,0) free tensors, that live over the Lens spaces in the base are not regular $A_{n-1}$ theories, although they have the same degrees of freedom, but are coupled to the discrete gauge symmetry which is visible in their tensor branch. These are what we denote as $\mathcal{A}_{n-1}$ theories. \\

\begin{figure}
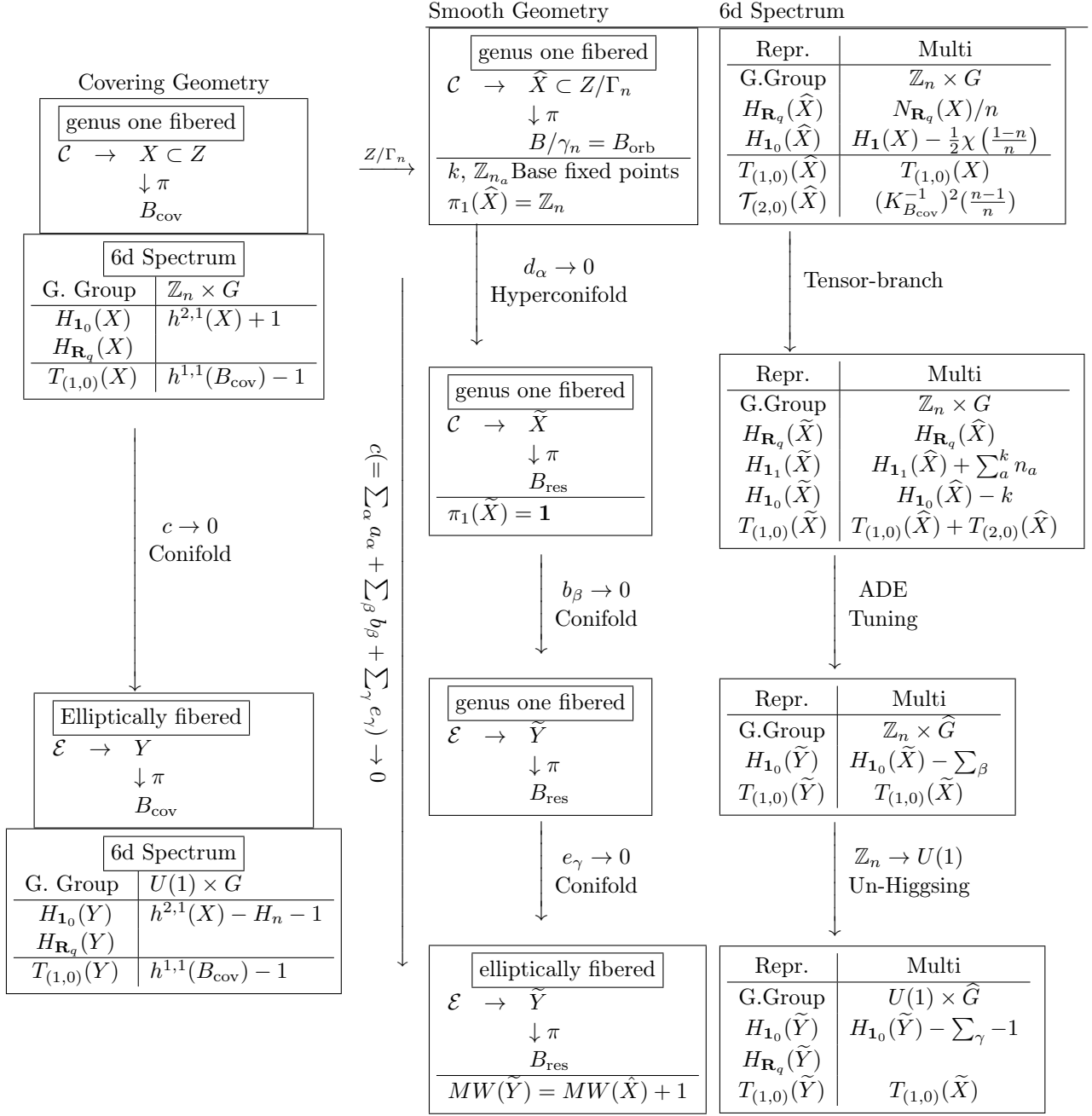

\begin{center}
{\footnotesize
\renewcommand{\tabcolsep}{0.05cm}
\begin{tabular}{lll}

\begin{tabular}{c}\vspace{-1.2cm} \\Covering Geometry  \\  
  \framebox{$
\begin{array}{ccl}
\multicolumn{3}{c}{\framebox{genus one fibered}} \qquad\\
\mathcal{C} & \rightarrow & X \subset Z   \\
&&  \downarrow \pi  \\
& & B_{\text{cov}} \\  
\end{array}
$} \\
  \framebox{$
\begin{array}{c|l}
\multicolumn{2}{c}{\framebox{6d Spectrum}} \\
$G. Group $&  \mathbb{Z}_n \times G  \\ \hline
H_{\mathbf{1}_0}(X) &  h^{2,1}(X)+1   \\
 H_{\mathbf{R}_q}(X) &    \\ \hline
 T_{(1,0)}(X) &  h^{1,1}(B_\text{cov})-1  \\
\end{array}
$} \\
\rotatebox{270}{$\hspace{-1.0cm}    \xrightarrow[\phantom{\text{\qquad \qquad \qquad Conifold \qquad \qquad}}]{ } $} \begin{tabular}{c}\\ \\ \\ \\   $c\rightarrow 0$ \\ Conifold        \end{tabular}   \\

 \begin{tabular}{c}
  \framebox{$
\begin{array}{ccl}
\multicolumn{3}{c}{\framebox{Elliptically fibered}} \qquad\\
\mathcal{E} & \rightarrow & Y     \\
&&  \downarrow \pi  \\
& & B_{\text{cov}} \\  
\end{array}
$} \\
  \framebox{$
\begin{array}{c|l}
\multicolumn{2}{c}{\framebox{6d Spectrum}} \\
$G. Group $&  U(1) \times G  \\ \hline
H_{\mathbf{1}_0}(Y) &  h^{2,1}(X)-H_n-1   \\
 H_{\mathbf{R}_q}(Y) &   \\ \hline
 T_{(1,0)}(Y) &  h^{1,1}(B_\text{cov})-1  \\
\end{array}
$} \\
 \end{tabular}

\end{tabular}
&
 
\begin{tabular}{c}
 $  \xrightarrow{Z/\Gamma_n}    $
  \\
\vspace{1cm}
   \\
         \rotatebox{270}{ $\hspace{-3cm} c( = \sum_\alpha a_\alpha + \sum_\beta b_\beta + \sum_\gamma e_\gamma)\rightarrow 0 $} $ \left\downarrow\rule{0cm}{5.5cm}\right.$  
  \end{tabular}   
     
           &
 
   \begin{tabular}{ l   c   l}
Smooth Geometry  & &  6d Spectrum \\ \hline
 
  \framebox{$
\begin{array}{ccl}
\multicolumn{3}{c}{\framebox{genus one fibered}} \\
\mathcal{C} & \rightarrow & \widehat{X} \subset  Z /\Gamma_n \\
&&  \downarrow \pi  \\
& &  B/\gamma_n=B_{\text{orb}} \\ \hline
\multicolumn{3}{c}{k , \,  \mathbb{Z}_{n_a}  $Base fixed points$ }\\
\multicolumn{3}{l}{ \pi_1(\widehat{X})= \mathbb{Z}_n }
\end{array}
$}

 & 
 &

 \framebox{$
\begin{array}{c|c }
\text{Repr.} & \text{Multi}   \\ \hline
$G.Group$ & \mathbb{Z}_n \times G \\
H_{\mathbf{R}_q}(\widehat{X}) & N_{\mathbf{R}_q}(X)/n    \\
H_{\mathbf{1}_0}(\widehat{X}) & H_{\mathbf{1}}(X)- \frac12 \chi \left( \frac{1-n}{n}\right) \\  \hline    
 T_{(1,0)}(\widehat{X}) & T_{(1,0)}(X) \\
 \mathcal{T}_{ (2,0)}(\widehat{X}) & (K_{B_{\text{cov}}}^{-1})^2(\frac{n-1}{n})  \\
\end{array}
 $ } \\

\quad\rotatebox{270}{$\hspace{-0.8cm}    \xrightarrow[\phantom{\text{Hyperconifold}}]{        } $} \begin{tabular}{c}\\   $d_\alpha \rightarrow 0$ \\ Hyperconifold      \end{tabular}     & & \qquad \rotatebox{270}{ $ \hspace{-0.8cm}    \xrightarrow[\phantom{\text{Hyperconifold}}]{ }$}  \begin{tabular}{c}\\ Tensor-branch    \end{tabular}  \\
  
   \framebox{$
\begin{array}{ccl}
\multicolumn{3}{c}{\framebox{genus one fibered}} \\
\mathcal{C} & \rightarrow & \widetilde{X}  \\
&&  \downarrow \pi  \\
& &  B_{\text{res}} \\ \hline
\multicolumn{3}{l}{ \pi_1(\widetilde{X}) = \mathbf{1} } \\
\end{array}
$}

 & & 

 \framebox{$
\begin{array}{c|c }
\text{Repr.} & \text{Multi}   \\ \hline
$G.Group$ & \mathbb{Z}_n \times G \\
H_{\mathbf{R}_q}(\widetilde{X} ) & H_{\mathbf{R}_q}(\widehat{X})    \\
H_{\mathbf{1}_1}(\widetilde{X} ) & H_{\mathbf{1}_1}(\widehat{X})  + \sum_a^k n_a   \\
H_{\mathbf{1}_0}(\widetilde{X} ) &H_{\mathbf{1}_0}(\widehat{X} ) - k \\      
 T_{(1,0)}(\widetilde{X} ) &  T_{(1,0)}(\widehat{X})+ T_{(2,0)}(\widehat{X})\\
\end{array}
 $ } \\

  \qquad \qquad \rotatebox{270}{$ \hspace{-0.7cm}     \xrightarrow[\phantom{\text{Conifold}\quad}]{     } $}  \begin{tabular}{c} \\  $b_\beta \rightarrow 0$ \\ Conifold \end{tabular}  & & \qquad \qquad \rotatebox{270}{ $  \hspace{-0.8cm}   \xrightarrow[\phantom{\text{Un-Higgsing }}]{ } $   } \begin{tabular}{c} \\    ADE \\ Tuning \end{tabular}  \\

   \framebox{$
\begin{array}{ccl}
\multicolumn{3}{c}{\framebox{genus one fibered}} \\
\mathcal{E} & \rightarrow & \widetilde{Y} \\
&&  \downarrow \pi  \\
& &   B_{\text{res}}\\  
\end{array}
$} 
 
 & 
 &

\framebox{$
\begin{array}{c|c }
\text{Repr.} & \text{Multi}   \\ \hline
$G.Group$ & \mathbb{Z}_n \times \widehat{G} \\
H_{\mathbf{1}_0}(\widetilde{Y}) & H_{\mathbf{1}_0}(\widetilde{X})-\sum_\beta   \\
  
 T_{(1,0)}(\widetilde{Y} ) & T_{(1,0)}(\widetilde{X} ) \\
\end{array}
 $ } \\

  \qquad \qquad \rotatebox{270}{$ \hspace{-0.7cm}     \xrightarrow[\phantom{\text{Conifold}\quad}]{     } $}  \begin{tabular}{c} \\  $e_\gamma \rightarrow 0$ \\ Conifold \end{tabular}  & & \qquad \qquad \rotatebox{270}{ $  \hspace{-0.8cm}   \xrightarrow[\phantom{\text{Un-Higgsing }}]{ } $   } \begin{tabular}{c} \\    $\mathbb{Z}_n \rightarrow U(1)$ \\ Un-Higgsing \end{tabular}  \\

   \framebox{$
\begin{array}{ccl}
\multicolumn{3}{c}{\framebox{elliptically fibered}} \\
\mathcal{E} & \rightarrow & \widetilde{Y} \\
&&  \downarrow \pi  \\
& &   B_{\text{res}}\\  \hline
\multicolumn{3}{c}{MW(\widetilde{Y})=MW(\hat{X}) +1}
  \\
\end{array}
$} 
 
 & 
 &

\framebox{$
\begin{array}{c|c }
\text{Repr.} & \text{Multi}   \\ \hline
$G.Group$ & U(1) \times \widehat{G} \\
H_{\mathbf{1}_0}(\widetilde{Y}) & H_{\mathbf{1}_0}(\widetilde{Y})-\sum_\gamma-1    \\
 H_{\mathbf{R}_q}(\widetilde{Y}) &   \\

 T_{(1,0)}(\widetilde{Y} ) & T_{(1,0)}(\widetilde{X} ) \\
\end{array}
 $ } \\

\end{tabular} 
\end{tabular}
}
\end{center}
\caption{\label{fig:QuotientDiagramBig}{\it Graphical summary of the geometry and physics of the covering and quotient geometries explained around sections \ref{sec:anomalies}-\ref{sec:Tensorbranch} when tuning in a section. On the left we perform a conifold which corresponds to an un-Higgsing , while on the quotient side right we perform the same transition where we have to go through several subtransitions: We first resolve the (2,0) point, tune in possible ADE groups, before we can un-Higgs the U(1)  on the quotient side. }}
\end{figure}

From here on we can perform several more conifold transitions to un-Higgs the $\mathbb{Z}_n$ symmetry to a U(1). This conifold transition which we inherit from the covering space $X$ as $c \rightarrow 0$ gets replaced by a transition $\widehat{c}\rightarrow 0$ on $\widehat{X}$ which leads to the un-Higgsed U(1). However, this tuning is often not as straight forward as on $X$, as the transition is now decomposed into several sub-transitions given as
\begin{align}
\widehat{c} = \sum_\alpha a_\alpha  + \sum b_\beta + \sum_\gamma e_\gamma \rightarrow 0 \, ,    
\end{align}
that all have to be tuned to zero, in order to get the desired un-Higgsing. We characterize those
transitions as 
\begin{itemize}
\item {\bf Hyperconifolds} $a_\alpha \rightarrow 0$ resolving the fixed points in the base.
\item {\bf ADE tunings} $b_\beta \rightarrow 0$ introducing ADE algebras over fixed points.
\item {\bf Residual tunings} $e_\gamma \rightarrow$ necessary to obtain the U(1) conifold transition.
\end{itemize} 
Upon the full transition $\widehat{c}\rightarrow 0$ we expect $n$ discrete charged singlets to become charged under the U(1) symmetry and hence, in the resolved geometry associated to the tensor branch I$_2$ fibers appear. Therefore we find that also the U(1) is automatically coupled to the $\mathcal{A}_{n-1}$ tensor branch, when the theory becomes un-Higgsed. In such a case, the corresponding U(1) height pairing $b_{11}$ is not Cartier anymore and is a fractional divisor 
when the base is taken singular which hints at the presence of U(1) charged superconformal matter. However as pointed out, by performing the tuning, it can become necessary to also tune $b_\beta \rightarrow 0$ which introduces additional non-Abelian gauge groups over the resolution divisors. In such a case we have an intricate coupling of the Abelian and non-Abelian gauge group in the tensor branch of the (2,0) theory.
  In Section~\ref{section4} we give several concrete examples of those theories and
the tunings to un-Higgs them. Finally we summarize the various operations and flows
of geometries in Figure~\ref{fig:QuotientDiagramBig}.

\section{Examples of Genus one Fibered Quotients}
\label{section4}
In this section we want to present concrete examples of the type of threefolds and transitions that we discussed in generality in the section before. While there are classifications of free quotients of CY manifolds available \cite{Batyrev:2005jc,Braun:2010vc}, we focus, for ease of exposition on the simplest examples that are toric hypersurfaces \cite{Batyrev:2005jc} in a $4$-dimensional ambient space. We give particular emphasis on the toric construction of the Calabi-Yau and the quotient action on the ambient space. To explore the physics on the quotient geometry we perform several hyperconifold transitions to smoothen out the base completely and confirm the additional discrete charged hypermultiplets over the resolution divisors explicitly. We contrast those geometries with canonical fibrations that dont have those hypermultiplets.
 
\subsection{Example 1: Threefold in $(\mathbb{P}^2 \times \mathbb{P}^2)/\mathbb{Z}_3$}
\label{sec:example1}
The first example is the bi-cubic hypersurface and its quotient manifold. This example connects directly to the Weierstrass model we presented in Section~\ref{section2} and represents a fully smooth genus one fibration. We identify the singularities in the base space, the behavior of the multi-sections, and give a discussion of the explicit hyperconifold transition, that corresponds to the tensor branch of the $(2,0)$ points in the base, and the additional discrete charged hypermultiplets. 
On both geometries, we perform conifold transitions to obtain a section in the covering and quotient geometry.

\subsubsection{The covering Calabi-Yau threefold}
The bi-cubic Calabi-Yau threefold is a generic hypersurface inside a $\mathbb{P}^2_F \times \mathbb{P}^2_B$ ambient space of degree $(3,3)$ in the fiber $\mathbb{P}^2_F$ and base $\mathbb{P}^2_B$ (see \cite{Gray:2014fla,Gray:2013mja,Gray:2014kda,Anderson:2015iia} for related recent constructions). The toric realization of the ambient space $Z$ of the CY threefold is encoded in the convex hull of the reflexive polytope $\Delta \in \mathbb{Z}^4$ given as
\begin{align}
\label{eq:DeltaCover}
\begin{tabular}{ccc|ccc}
$x_0$ & $x_1$ & $x_2$ & $y_0$ & $y_1$ & $y_2$ \\ \hline
1 & 0 & -1 & 0 & 0 & 0 \\
0 & 1 & -1 & 0 & 0 & 0 \\
0 & 0 & 0 & 1  & 0 & -1 \\
0 & 0 & 0 & 0 &1 & -1 \\
\end{tabular} \, ,
\end{align}
which yields the Stanley-Reisner ideal:
\begin{align}
\label{eq:SRIquotient}
SRI: \{x_0 x_1 x_2, y_0 y_1 y_2  \} \, .
\end{align}
We write the CY hypersurface in terms of the fiber coordinates $x_i$ as:
\begin{align}
\label{eq:cubic}
P = &s_1 x_0^3 + s_2 x_0^2x_1 + s_3x_0x_1^2 + s_4x_1^3 + s_5x_0^2 x_2 + s_6 x_0 x_1 x_2+
 s_7 x_1^2 x_2 + s_8 x_0 x_2^2 + s_9 x_1 x_2^2 + s_{10} x_2^3 \, ,
\end{align}
which is a generic cubic and therefore a genus one curve. The $s_i$ are generic sections of the canonical class of the base $s_i \in \mathcal{O}(-K_B = 3 H_B)$. Hence these sections are generic cubic polynomial with 10 monomials in the $y_i$ base coordinates just as the fiber.
The fiber $\mathcal{C}$ is a genus one curve, which admits no sections but only three-sections as: 
\begin{align}
[x_i] \cdot \mathcal{C} = 3 \, \quad \forall \,  i \, . 
\end{align}
Thus we have a smooth genus one fibered CY threefold 
\begin{align}
\begin{array}{ccl}
\mathcal{C} &\rightarrow& X \\
&&\downarrow \pi \\
&& \mathbb{P}^2   
\end{array} \, .
\end{align}
The Hodge and Euler numbers of $X$ can be computed as
\begin{align}
(h^{(1,1)}, h^{(2,1)})_\chi = (2,83)_{-162} \, .
\end{align}
 
\subsubsection{The F-theory physics of the covering space}
 
The F-theory physics of these kinds of threefolds has been considered already in \cite{Klevers:2014bqa,Cvetic:2015moa}. The three-sections have been identified as the generators of a discrete $\mathbb{Z}_3$ symmetry of the $6$-dimensional theory. 
We can consider the associated singular Jacobian fibration $\hat{Y}$
\begin{align}
\begin{array}{ccl}
\mathcal{E} &\rightarrow& \hat{Y} \\
&&\downarrow \pi \\
&& \mathbb{P}^2   
\end{array} \, ,
\end{align}
which admits the same $\tau$ function as $X$ and the elliptic fiber $\mathcal{E}$ admits a zero-section. The coefficients $f$ and $g$ of the Weierstrass model in term of the $s_i$ can be found in Appendix~\ref{app:cubicinWSF}. 
As opposed to the genus one fibration $X$, the Weierstrass fibration $\hat{Y}$ is singular and admits $A_1$ singular fibers over certain codimension two points in the base. On $X$ on the other hand those singularities are absent but the fiber degenerates into two $\mathbb{P}^1$'s.\\
Thus in the F-theory physics those points are interpreted as loci of discrete charged hypers.
Accounting for those discrete charged states, we  summarize the full $6$-dimensional matter spectrum in Table~\eqref{eq:cubicspectrum}. For a generic base \cite{Klevers:2014bqa}, the spectrum is fully fixed by the three classes base classes $\mathcal{S}_7$, $\mathcal{S}_9$ and $K_b^{-1}$ that are the classes of the sections $s_7$, $s_9$ in the fiber equation \eqref{eq:cubic} and the anticannonical class of the base.
   \begin{align}
\label{eq:cubicspectrum}
 \begin{array}{c|c|c}
$6d Rep.$& $Base Intersection$ & $Multiplicity$ \\ \hline 
 \mathbf{1}_{1}  &    \begin{array}{c}  3 ( 6 (K_b^{-1})^2 - \mathcal{S}_7^2 + \mathcal{S}_7 \mathcal{S}_9    \\   - \mathcal{S}_9^2 + K^{-1}_b (\mathcal{S}_7 + \mathcal{S}_9))   \end{array}   & 189 \\ \hline
 \mathbf{1}_0  &  h^{2,1}(X) +1  & 84 \\  \hline
 \mathbf{V}  &  h^{1,1}(X)-h^{1,1}(B)-1 & 0\\
 \mathbf{T}  &   h^{1,1}(B)-1  & 0 \\ \hline
\end{array} 
\end{align}
   Here we made use of the aforementioned identification $\mathcal{S}_7 = \mathcal{S}_9 = K_b^{-1}$.
The given spectrum clearly satisfies the gravitational anomaly
\begin{align}
\label{eq:gravanomaly}
H-V+29T-273 = & 0 \, , \\
9-T = & (K_b^{-1})^2 \, .
\end{align}
\subsubsection{Un-Higgsing  to an elliptic fibration}
\begin{figure}
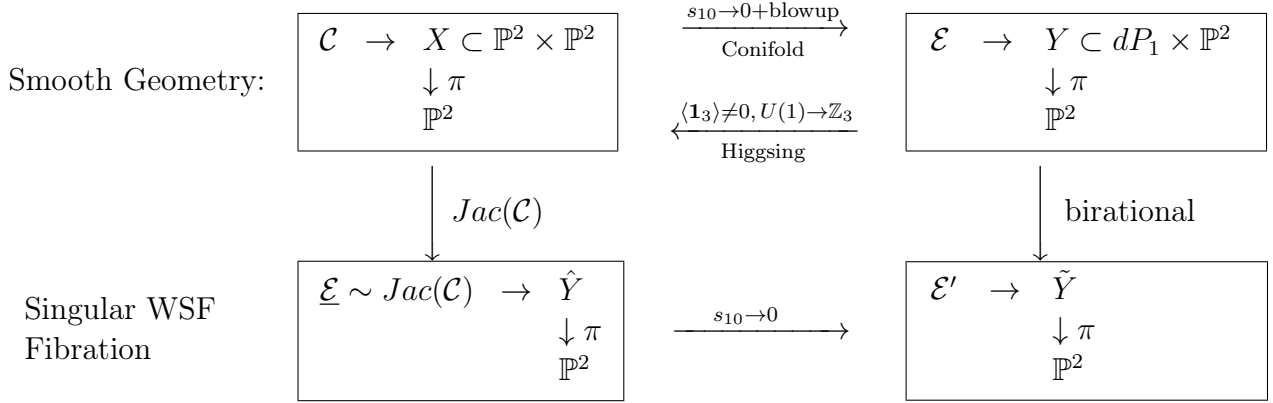

\begin{center}
\begin{tabular}{l     lll}
Smooth Geometry:  &
 
  \framebox{$
\begin{array}{ccl}
\mathcal{C} & \rightarrow & X \subset \mathbb{P}^2 \times \mathbb{P}^2 \\
&&  \downarrow \pi  \\
& & \mathbb{P}^2  
\end{array}
$} 
 
 &$\begin{array}{c}      \xrightarrow[\text{Conifold}]{s_{10}\rightarrow0+\text{blowup}}  \\
 \\
      \xleftarrow[\text{Higgsing }]{\langle \mathbf{1}_3 \rangle \neq 0,\,  U(1) \rightarrow \mathbb{Z}_3  }  
    \end{array} $ &

 \framebox{$
\begin{tabular}{ccl}
$\mathcal{E}$ & $\rightarrow $& $ Y \subset dP_1 \times \mathbb{P}^2  $ \\
&& $ \downarrow \pi $ \\
& & $ \mathbb{P}^2 $
\end{tabular}
 $ } \\
 &\qquad \qquad $ \left\downarrow\rule{0cm}{0.7cm}\right. Jac(\mathcal{C})$ & &\qquad \qquad $ \left\downarrow\rule{0cm}{0.7cm}\right.$ birational \\  
 
 \begin{tabular}{l}Singular WSF \\  Fibration \end{tabular}    & 
 
   \framebox{$
\begin{array}{ccl}
\underline{\mathcal{E}} \sim Jac(\mathcal{C})  & \rightarrow & \hat{Y} \\
&&  \downarrow \pi  \\
& & \mathbb{P}^2  
\end{array}
$} 
 &  \,  $ \xrightarrow{\, \quad s_{10}\rightarrow0 \qquad} $ & 
 
 \framebox{$
\begin{tabular}{ccl}
$\mathcal{E}^\prime $ & $\rightarrow $& $  \tilde{Y}     $ \qquad  \qquad  \, \, \\
&& $ \downarrow \pi $ \\
& & $ \mathbb{P}^2 $
\end{tabular}
 $ } \\

\end{tabular} 
\end{center}
\caption{\label{fig:QuotientDiagramSmall}{\it  The genus one fibered geometry and its un-Higgsing to a U(1) theory from left to right. The first row shows the conifold transition in the smooth CY geometry whereas the second row shows the same procedure in the singular Weierstrass model / Jacobian of the genus one-fibration. }}
 
\end{figure}

The physics of the above geometry is made most clear by un-Higgsing the discrete symmetry to a U(1), realized by a transition to a smooth elliptic fibration $Y$ with enhanced Mordell-Weil rank. In the context of toric geometry this is done by a complex structure deformation  
by tuning:
\begin{align}
s_{10}(y_0,y_1,y_2) \rightarrow 0 \, .
\end{align}
 As $s_{10}$ is a generic cubic by itself, this amounts to setting ten complex structure coefficients to zero. After the deformation the threefold $\tilde{Y}$ admits nodal singularities and is therefore a conifold that can be resolved to another smooth threefold $Y$. Torically this resolution can be performed as a blow-up of the ambient space $Z$ by adding a vertex to the associated polytope $\Delta$ \eqref{eq:DeltaCover}. The threefold $Y$ is now the anti-canonical surface in $dP_1 \times \mathbb{P}^2$ ambient space, with polytope $\Delta$ given as
\begin{align}
\begin{tabular}{cccc|ccc}
$x_0$ & $x_1$ & $x_2$ & $e_1$ & $y_0$ & $y_1$ & $y_2$ \\ \hline
1 & 0 & -1 & 1 & 0 & 0 & 0 \\
0 & 1 & -1 & 1 & 0 & 0 & 0 \\
0 & 0 & 0 & 0 & 1  & 0 & -1 \\
0 & 0 & 0 & 0&0 &1 & -1 \\
\end{tabular} \, ,
\end{align}
and Stanley-Reisner ideal
\begin{align}
SRI: \{ x_0 x_1, x_2 e_1, y_0 y_1 y_2 \} \, .
\end{align}
We compute the Hodge and Euler numbers of this geometry as
\begin{align}
(h^{1,1},h^{2,1})_\chi = (3,75)_{-144} \, .
\end{align}
The smooth elliptic fiber is thus given as the vanishing hypersurface
\begin{align}
\label{eq:dp1}
P = & s_1 e_1^2 x_0^3 + s_2 e_1^2  x_0^2 x_1 + s_3 e_1^2  x_0 x_1^2 + s_4  e_1^2  x_1^3 + s_5 e_1 x_0^2 x_2 + 
 s_6 e_1 x_0 x_1 x_2 + s_7 e_1 x_1^2 x_2 + s_8 x_0 x_2^2 + s_9 x_1 x_2^2  \, .
\end{align}
Indeed, the divisor $ D_{e_1}$ intersects the fiber exactly once  $D_{e_1} \cdot \mathcal{E}=1$ which yields a zero-section. In addition, another non-toric section can be constructed which generates a non-trivial MW group \cite{Klevers:2014bqa}.
We summarize the full matter spectrum in the following table 
\begin{align}
\label{eq:spectrum}
\begin{array}{c|c|c}
$$6$-d Rep.$& $Base Intersection$ & $Multiplicity$ \\ \hline 
\mathbf{1}_{1}  & \begin{array}{c} 12[K_b^{-1} ]^2 + [K_b^{-1}](8 \mathcal{S}_7-\mathcal{S}_9) \\ -4\mathcal{S}_7^2
 + \mathcal{S}_7\mathcal{S}_9 - \mathcal{S}_9^2 \end{array}
 & 135 \\ \hline 
\mathbf{1}_{2} & \begin{array}{c} 6 [K_B^{-1} ]^2+[K_B^{-1} ](4\mathcal{S}_9-5\mathcal{S}_7)\\
+\mathcal{S}_7^2+ 2 \mathcal{S}_7\mathcal{S}_9-2\mathcal{S}_9^2 \end{array} 
   &  54 \\ \hline
\mathbf{1}_{3}   &  \mathcal{S}_9([K_b^{-1} ] + \mathcal{S}_9-\mathcal{S}_7) &    9\\ \hline
\mathbf{1}_0  & h^{2,1}(X) +1  & 76 \\
\mathbf{V}  &  h^{1,1}(X)-h^{1,1}(B)-1 & 1\\
\mathbf{T}_{(1,0)} &  9-(K_b^{-1})^2  & 0 \\ \hline
\end{array} \, .
\end{align}
The spectrum is again computed by identifying $\mathcal{S}_7 = \mathcal{S}_9 = K_b^{-1}$ and using self intersection $(K_b^{-1})^2=9$.
 For this spectrum again all anomalies are canceled and in particular it is free of the gravitational one \eqref{eq:gravanomaly}. That all gauge anomalies are canceled can be seen by using the associated U(1) height pairing
 \begin{align}b_{11} = -2 (\mathcal{S}_7 -2 \mathcal{S}_9 - 3 K_b^{-1}) = 8 K_b^{-1} \, ,
 \end{align}
 and plugging this into equations \eqref{eq:6dAnomalies} in Appendix \ref{app:Anomalies}. 
We remark that the geometrical transition back to the bi-cubic is induced by a vev in the hypermultiplets $\langle \mathbf{1}_{3} \rangle \neq 0$. 
   Upon this breaking we find that the eight D-flat directions appear as new complex structure coefficients whereas the Goldstone mode renders the U(1) generator massive. On the other hand
   the $135$ and $54$ charged states get identified upon the unbroken $\mathbb{Z}_3$ residual symmetry and match the counting for the discrete charged states as given in Table~\eqref{eq:cubicspectrum} for the genus one fibered geometry.
 The transition that we have performed above is summarized in the Figure~\ref{fig:QuotientDiagramSmall}.
    
 \subsubsection{The quotient of the bi-cubic}
 \label{sec:Example1Quotient}
For specific values of the complex structure, the bi-cubic hypersurface considered above admits
a $\mathbb{Z}_3$ symmetry which can be used to take a free quotient. In terms of the coordinates this quotient is given by
 \begin{align}\label{z3act}
(x_i; y_i) \sim (\Gamma_3^i x_i; \Gamma_3^i y_i) \, ,
 \end{align}
 with $\Gamma_3^3=\mathbf{1}$. Thus the quotient is possible when all
 monomials in the bi-cubic equation that do not respect the above action are absent thereby reducing the amount of complex structure moduli.\\
 
From the point of view of the ambient variety $Z$ the points of the dual polytope $\Delta^*$  to the polytope $\Delta$ corresponds to the monomials of the CY hypersurface via the Batyrev prescription.
 Hence  one can view the quotient as a lattice refinement of the dual lattice. This refinement can be rephrased as a basis change \cite{Davies:2011is} of the vertices in $\Delta$, that now have the coordinates:
\begin{align}
\begin{tabular}{ccc|ccc}
$x_0$ & $x_1$ & $x_2$ & $y_0$ & $y_1$ & $y_2$ \\ \hline
1 & 0 & -1 & 0 & 1 & -1 \\
0 & 1 & -1 & 0 & -1 & 1 \\
0 & 0 & 0 & 1  & 1 & -2 \\
0 & 0 & 0 & 0 &3 & -3 \\
\end{tabular} \, .
\end{align}
In the language of \cite{Batyrev:2005jc} the above ambient space geometry is fixed by the following relations of the integral vertices
\begin{align}
&v_{x_0} + v_{x_1} + v_{x_2} = v_{y_0} + v_{y_1} + v_{y_2}  =0 \, ,\\
&v_{\gamma} = \frac13 (v_{x_0} + 2 v_{x_1} + v_{y_1} + 2 v_{y_2}) \, ,
\end{align}
where the first relation is simply the specification of the two $\mathbb{P}^2$'s and the last one is the additional fractional relation that refines the lattice.
Before we turn to the CY hypersurface, it is worth to consider the $\mathbb{C}^*$ scalings
of the above ambient space geometry, that are 
\begin{align}
\phi: \quad \mathbb{C}^6 \rightarrow ( \prod^3_ i x_i^{v^j_{x_i}} y_i^{v^j_{y_i}}) = 
(\frac{x_0 y_1}{x_2 y_2}, \frac{x_1 y_2 }{x_2 y_1}, \frac{y_0 y_1}{y_2^2}, \frac{y_1^3}{y^3_2})\  . 
\end{align}
The scaling relations of this variety are given by the kernel of the map $\phi$ as
\begin{align}
(\lambda_1, \lambda_1, \lambda_1, \lambda_2, \lambda_2, \lambda_2)& \, \text{ with } \lambda_1, \lambda_2 \in \mathbb{C}^* \, ,\\
(\Gamma^0,\Gamma^1 , \Gamma^2, \Gamma^0, \Gamma^1, \Gamma^2)& \, \text{ with } \Gamma^3 =1 \, .
\end{align}
We find that the variety indeed admits the $\Gamma_3$ action as an additional relation on the coordinates and hence we conclude that this is indeed the polytope of  $(\mathbb{P}^2 \times \mathbb{P}^2) / \mathbb{Z}_3$ with the same SRI as in Equation~\ref{eq:SRIquotient}. 
Note that the above geometry is not smooth and admits nine equivalent codimension-four fixed points, that are of the form
\begin{align}
(x_0,x_1,x_2; y_0,y_1,y_2) = (\underline{0,0,1};\underline{0,0,1})\, ,
\end{align}
where the underline indicates permutations. On the ambient space variety intersections are not integer valued but instead fractional 
\begin{align}
D_{x_i} \cdot D_{x_j} \cdot D_{y_i} \cdot D_{y_j} = \frac13 \, .
\end{align}
The associated Calabi-Yau hypersurface is smooth as we will argue in the following and admits the Hodge numbers 
\begin{align}
\label{eq:hodgebicquotient}
(h^{1,1}, h^{2,1})_\chi=(2,29)_{-54} \, .
\end{align}
The dual polyhedron $\Delta^*$ consists of 34 vertices and encodes all monomials 
of the bi-cubic, that are invariant under the $\mathbb{Z}_3$ action. As expected the Euler number gets reduced by $1/3$ upon the quotient.\\
The resulting CY hypersurface in the quotient admits the same structure in terms of a cubic polynomial in the fiber coordinates \eqref{eq:cubic}. This time however we must
 specify base dependent sections $s_i$ that are not generic cubic functions in the $y_i$ anymore but are restricted such that they transform in a well defined way under the $\Gamma_3$ action as we have presented in Section~\ref{section2}. See their explicit form eq. \eqref{eq:siiny} in Appendix~\ref{app:sections}. The general structure of the fiber equation stays invariant: 
\begin{align}
\begin{split}
P = & s^{(0)}_1 x_0^3 + s^{(2)}_2  x_0^2 x_1 + s^{(1)}_3   x_0 x_1^2 + s^{(0)}_4   x_1^3 + s^{(1)}_5  x_0^2 x_2 \\ &+ 
 s^{(0)}_6  x_0 x_1 x_2 + s^{(2)}_7   x_1^2 x_2 + s^{(2)}_8 x_0 x_2^2 + s^{(1)}_9 x_1 x_2^2  + s^{(0)}_{10} x_2^3\, .
 \end{split}
\end{align}
We have added a superscript $s^{(j)}$ that denotes the weight of the base sections $s_i$ under the $\Gamma_{3,b}$ action in the base as 
\begin{align}
s^{(j)}_i \rightarrow (\Gamma_{3,b})^{j} s^{(j)}_i \, .
\end{align}
In order to identify the behavior of the fiber close to the fixed points we choose a coordinate patch including the fixed point by using the $\mathbb{C}^*$ action to fix the coordinate dependence such as
\begin{align}
(y_0,y_1,y_2) = ( 1,u,v)
\end{align}
that are local coordinates on $\mathbb{C}^2/\Gamma_{3,b}$ i.e. we still have the additional phase identification $(1,u,v) \sim (1, \Gamma_3 u, \Gamma_3^2 v )$ with the orbifold singularity at the origin. Choosing a radial coordinate of the form $(1, \Gamma_3^k  z, \Gamma_3^{2k} z)$ the sections $s_i^{(j)}$ factor as:
\begin{align}
\begin{array}{lllll}
s_1^{(0)} \rightarrow \hat{s}_1 \, , &   s_2^{(2)} \rightarrow   \Gamma_3^{2k}  z \hat{s}_2 \, , & s_3^{(1)} \rightarrow  \Gamma_3^{ k} z \hat{s}_3\, , & s_4^{(0)} \rightarrow  \hat{s}_4 \, ,& s_5^{(1)} \rightarrow  \Gamma_3^{ k} z \hat{s}_5\, , \\
s_6^{(0)} \rightarrow \hat{s}_6 \, ,& s_7^{(2)} \rightarrow   \Gamma_3^{2k} z \hat{s}_7\, , & s_8^{(2)} \rightarrow  \Gamma_3^{2k} z \hat{s}_8\, , & s_9^{(1)} \rightarrow  \Gamma_3^{ k} z \hat{s}_9\, , & s_{10}^{0)} \rightarrow  \hat{s}_{10}\, ,
\end{array}
\end{align}  
with $\hat{s}_i$ being $\Gamma_3$ invariant non vanishing functions at $z\rightarrow 0$ for generic complex structures. In this parametrization it is easy to see that that all sections $s_i$ that transform non-trivially under the $\Gamma_3$ action vanish at the fixed point for $z \rightarrow 0$.
Hence the fiber equation over any fixed point in the base becomes
\begin{align}
\label{eq:bicubicfiberfp}
P_{fp_b} =  \hat{s}_1  x_0^3 +\hat{s}_4  x_1^3 +\hat{s}_{10}  x_2^3 + \hat{s}_6 x_0 x_1  x_2 \, .
\end{align}
with $\hat{s}_i$ being generic coefficients.
Moving onto a fixed point in the fiber ambient space  $(x_0,x_1,x_2) = (\underline{0,0,1})$ we indeed find that the coefficients $\hat{s}_i$ prevent the ambient space singularity to hit the CY hypersurface $P=0$ which justifies the computation of Hodge and Euler numbers. However we also observe, that we can tune in those ambient space singularities by choosing one of the sections $\hat{s}_i$ for $i=1,4,10$ to vanish over $z\rightarrow 0$.\\\\
As the CY hypersurface is still a generic cubic in the fiber coordinates $x_i$, this is a genus one fibered smooth CY and therefore F-theory should be well defined.
First we find, that after mapping the $s^{(j)}_i$ into Weierstrass coefficients using eqn. \ref{eq:gcubic} in Appendix~\ref{app:cubicinWSF} that $f$ and $g$ are invariant well defined sections\footnote{Also the Weierstrass coordinates $x,y,z$ of $\mathbb{P}^{2,3,1}$ in the Jacobian are $\Gamma_3$ invariant.} under the $\Gamma_{3}$ action.

The fibers over the fixed points in the base are \emph{multiple} in the sense that they are non-reduced copies $n\mathcal{E}$ of a smooth genus one curve, $\mathcal{E}$. Intuitively the multiple fibers arise from the fact that away from the fixed point, the group action in \eqref{z3act} maps three distinct torus fibers into one another, while over the fixed points, a single torus is mapped to itself three times, as illustrated in Figure \ref{fig:MultipleFiberAC}. This action locally behaves as a translation along an elliptic fiber (locally the tri-section is identical to three honest sections since the fixed points are generically far away from any branch loci in the multi-section), a classic origin of multiple fibers in algebraic geometry \cite{hulek}. More explicitly, the multiple nature of the fiber can be seen by residual $\mathbb{Z}_3$ scaling freedom in the fiber. As this discussion is rather lengthy, we defer it to Appendix \ref{app:GIODescription} where it is described in detail. It should be noted that the techniques used to verify the existence of the multiple fibers in Appendix \ref{app:GIODescription} can also be applied to the standard $\mathbb{Z}_2$ quotient of a $K3$ surface which leads to an Enriques surface, where we can also reproduce the standard result of two multiple fibers.

We should also note that the multiple fiber is not visible from the Jacobian. There the fiber itself is smooth over the fixed points in the base where the fiber obtains the form \eqref{eq:bicubicfiberfp}. We find the Weierstrass coefficients to be
\begin{align}
\begin{split}
f&= \frac{1}{48} \hat{s}_{6}(216  \hat{s}_1 \hat{s}_4 \hat{s}_{10}-\hat{s}_{6}^3    )\, , \\
g&=\frac{1}{864} (\hat{s}_{6}^6 + 540\hat{s}_1  \hat{s}_4 \hat{s}_{10}  \hat{s}_{6}^3 - 5832 \hat{s}_{10}^2 \hat{s}_1^2 \hat{s}_4^2)\, ,\\
\Delta &=\frac{1}{16}\hat{s}_1 \hat{s}_4  \hat{s}_{10} (\hat{s}_{6}^3 + 27 \hat{s}_1 \hat{s}_4 \hat{s}_{10} )^3 \, .
\end{split}
\end{align}
and hence non-vanishing. We find that one obtain an I$_1$ fiber  when one tunes one of the ambient space fixed points onto the CY by requiring $\hat{s}_i \rightarrow 0$ for $ i=\{ 1,4,10\}$.
\subsubsection{Quotient action on the multi-section}
Let us consider at this point the explicit form of the three-section and its behavior when we go from the covering to the quotient geometry and discuss the action on the fiber in some more detail. \\ From the cubic equation of the fiber of the covering space in \eqref{eq:cubic} we pick the multi section $x_1 = 0$ with equation:
\begin{align}
\label{eq:threesection}
P_{x_1=0} = s_1 x_0^3 + s_{10} x_2^2  s_5 x_0^2 x_2 + s_8 x_0 x_2^2 \, ,
\end{align}
  \begin{figure}
\begin{center}
\begin{picture}(0,160)
\put(-200,0){\includegraphics[scale=1.6]{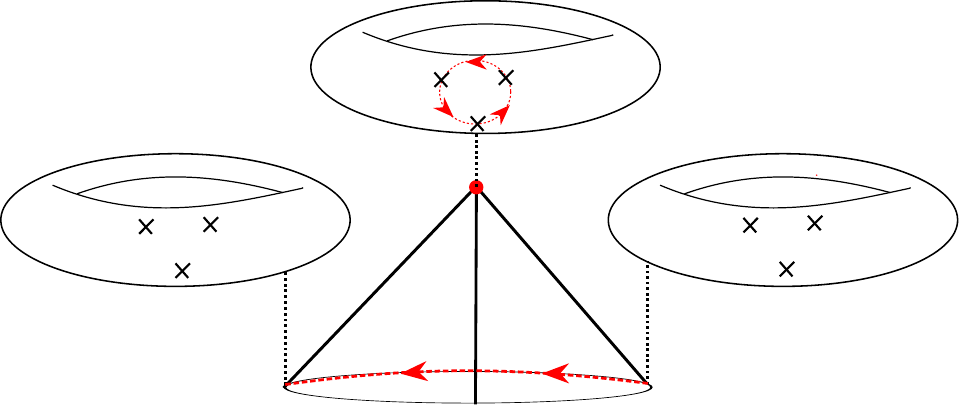}}
\put(-105,60){$x_0^{(3),1}$}
\put(-95,80){$x_0^{(3),2}$}
\put(-160,80){$x_0^{(3),1}$}

\put(170,60){$x_0^{(3),1}$}
\put(180,80){$x_0^{(3),2}$}
\put(120,80){$x_0^{(3),1}$}

\put(40,130){$x_0^{(3),1}$}
\put(40,150){$x_0^{(3),2}$}
\put(-15,150){$x_0^{(3),1}$}
\end{picture}
\end{center}
\caption{\label{fig:BicubicMultipleFibers}{\it Behavior of the multi-sections on the covering geometry with the $\mathbb{Z}_3$ symmetry. On what will become the fixed point, denoted as a red dot, the three-section is mapped into itself by the $\Gamma_{3,f}$ rotation. At a generic point
on the covering base, this translation does not persist.}}
\end{figure}
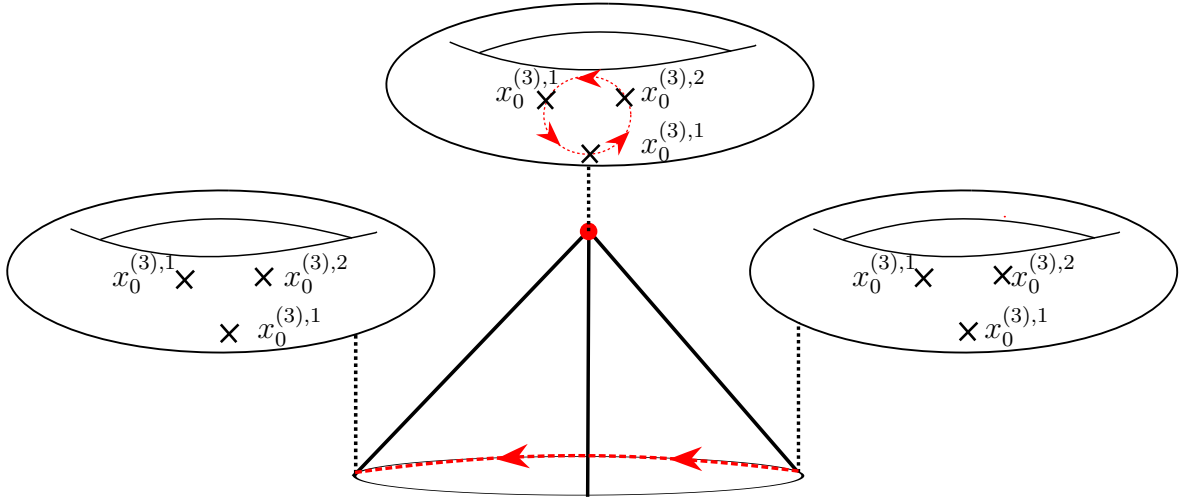
which admits three roots if we want to solve the system, say in $x_0$. These three roots generically get interchanged by moving around the base. Moving onto a $\Gamma_{3,b}$ fixed point in the base, the sections $s_i$ become constant enforcing a non trivial  $\Gamma_{3,f}$ action on the fiber in order to avoid fixed fibers. This action acts as a translation on the fiber as 
\begin{align}
x_i \rightarrow x_i \Gamma_{3,f}^i\, \text{ with } \, \Gamma_{3,f}=e^{\frac{2 \pi i}{3}} \, .
\end{align}
The translation becomes a symmetry precisely when $s_5=s_8 = 0$ which is the behavior we obtained and here Equation~\eqref{eq:threesection} becomes
\begin{align}
P = \hat{s}_4 x_0^3 + \hat{s}_{10} x_2^3 \, ,
\end{align}
and thus the three solutions
\begin{align}
x_0^{(3),r} =  \{ (\frac{\hat{s}_4}{\hat{s}_{10}}         )^{1/3}\, \Gamma_{3,f}^{r} \, x_2   \} \, ,
\end{align}
labeled by $r=0,1,2$ get related by the action of $\Gamma_{3,f}$. Let us now consider the action away from the fixed points. First there, the sections transforms non-trivially under $\Gamma_{3,b}$ in order to obtain an invariant hypersurface \eqref{eq:cubic}. Thus on a generic point on the $\mathbb{P}^2$ covering space we rotate by $\Gamma_{3,b}$ and find that the accompanied $\Gamma_{3,f}$ action indeed preserves the three-section of Equation~\ref{eq:threesection}. We have depicted the geometry of the fibration from the perspective of the covering geometry, with the $\mathbb{Z}_3$ symmetry of the multi-section in Figure~\ref{fig:BicubicMultipleFibers}.

\begin{figure}
 \begin{center}
 \includegraphics[scale=1.2]{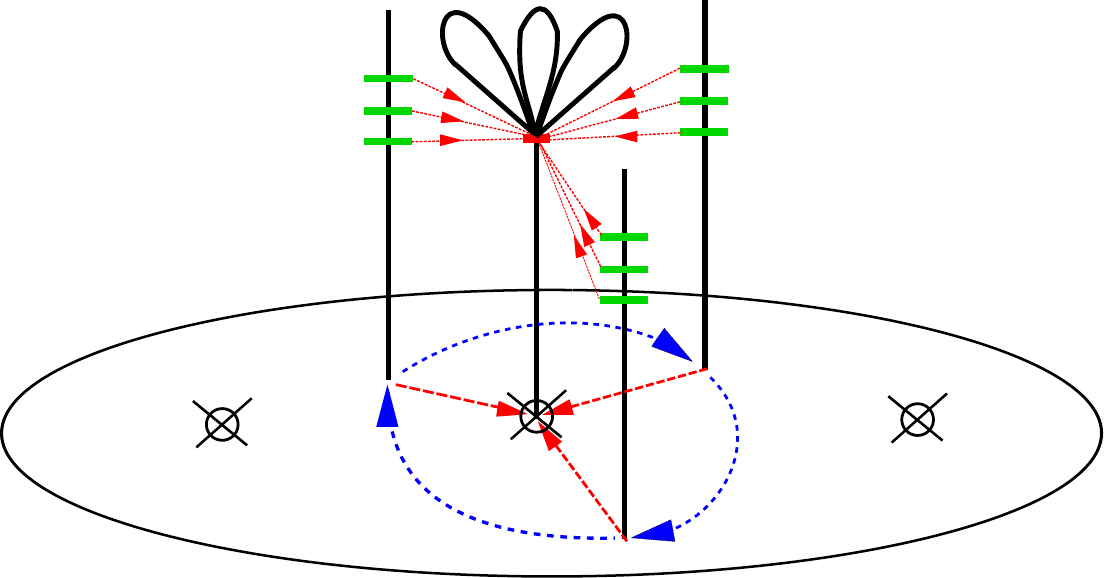}
 \caption{\label{fig:MultipleFiberAC}{\it Depiction of the genus one fiber in the quotient geometry. Moving onto the fixed point in the base, brings the thee three-section together and produces a multiple fiber such that the total space is smooth.}
 }
 \end{center}
\end{figure}

When taking the $\Gamma_3$ quotient, the $\Gamma_3$ symmetry of the three-sections becomes an identification. 
Hence over a fixed point in the base the three-sections come together to form a three multiple fiber as depicted in Figure~\ref{fig:MultipleFiberAC}.

\subsubsection{Spectrum of the quotient geometry}
After having discussed the geometry of the genus one fiber over the fixed points, we turn to the associated spectrum. As we showed before, the spectrum is fully fixed once we know the classes of the base line bundles  $s_7$ and $s_9$ and $K_b^{-1}$ and insert them into the respective formulas, given in \eqref{eq:spectrum}. This time however, $s_7^{(2)}$ and $s_9^{(1)}$ although being degree three polynomials in the $y_i$ are not $\Gamma_{3,b}$ invariant sections. This in particular implies, that they vanish over the fixed points and are therefore not Cartier divisors anymore as we have argued already in Section~\ref{sec:A2DiscreteCoupled}. However as they are both degree three polynomials in the base, we denote their class as $\mathcal{S}_7 = \mathcal{S}_9 = K^{-1}_{B}$ by abuse of notation. However we should keep in mind that their class is actually non-Cartier unlike the anticanonical class of the base. Using these classes we can compute the multiplicity of discrete charged matter as \eqref{eq:spectrum} using the intersection numbers $(K_b^{-1})^2 = 3$ resulting in the following spectrum
\begin{align}
\begin{tabular}{c|c|c}
$6$-d Rep.& Geometric Intersection & Multiplicity \\ \hline 
$\mathbf{1}_{1}$ & $ 21(K_b^{-1})^2  $ & 63 \\ \hline
$\mathbf{1}_0$ & $h^{2,1}(X) +1$ & 30 \\
$\mathbf{V}$ & $h^{1,1}(X)-h^{1,1}(B)-1$& 0\\
$\mathbf{T}_{(2,0)}$ & $ 9-(K_b^{-1})^2 $& 6 \\  
$\mathbf{T}_{(1,0)}$ & $ h^{1,1}(B)-1  $& 0 \\  
\end{tabular} \, .
\end{align}
The number of discrete charged states thus gets divided by three, which is intuitively clear as none of them resides on a fixed point and the fundamental domain of the $\mathbb{P}^2$ gets reduced.
Again, we check for the consistency by checking the gravitational anomalies:
\begin{align*}
H-V+29T -273 -30 T_{(2,0)} =0  \, , \nonumber
\end{align*} 
As expected all anomalies cancel due to the contribution of the three fixed points that each support two (2,0) tensors at $(y_0, y_1, y_2)=(\underline{0,0,1})$:
\begin{align}
X/\Gamma_3 \, \xrightarrow{\pi} \,  \mathbb{P}^2/\Gamma_{3,B} \, .
\end{align}
 Again we note,  that the whole fibration $X/\Gamma$ is smooth, while the base (which is the physical space of F-theory) is not. Hence these singularities signal the presence of additional light string states from M5 brane stacks that support (2,0) superconformal tensor multiplets \cite{DelZotto:2014fia}. 
In the following sections we consider various phases of the quotient fibration that are connected by conifold transitions i.e. by tuning of complex structure coefficients and subsequent toric resolutions. Note that every resolution breaks the $\pi_1$ to a trivial group  but the additional matter we find will give a hint of the symmetry of the quotient geometry. 
\subsubsection{Hyperconifold resolution of the fixed points}
\label{sec:Example1Resolution}
We fix the fiber coordinates of an ambient space fixed point to $(x_0,x_1,x_2) = (0,0,1)$ using the residual $\mathbb{C}^*$ action and obtain for the CY hyper surface:
\begin{align}
P=y_0^3 a_0 + y_1^3 a_3 + y_2^3 a_6 + y_0 y_1 y_2 a_{26}\, . 
\end{align}
Each of the three coefficients $a_0, a_3$ and $a_6$ should be non-vanishing in order that the hypersurface does not intersect the $(y_0,y_1,y_2) = (\underline{0,0,1})$ fixed points.   By tuning $a_0\rightarrow 0 $ the CY becomes singular and we reach a hyperconifold point which can be resolved by two blow-up divisors $e_{1,1},e_{1,2}$ leading to a smooth CY with Hodge numbers
\begin{align}
(h^{1,1},h^{2,1})_\chi = (4,28)_{-48} \, ,
\end{align}
This threefold has reduced first fundamental group $\pi_1(X) = \mathbf{1}$. 

Similarly we can tune the other two ambient space singularities to coincide with the CY hypersurface and resolve with two additional divisors for each. Luckily there exists a nice toric description of these blow-ups directly in the the ambient space parametrized by the polytope $\Delta$ spanned by the vertices: 
 \begin{align}
 \label{eq:Example1Tensor}
\begin{tabular}{ccc|ccc|cc|cc|cc}
$x_0$ & $x_1$ & $x_2$ & $y_0$ & $y_1$ & $y_2$ & $e_{1,1}$ & $e_{1,2}$ & $e_{2,1}$ & $a_{2,2}$ &  $e_{3,1}$ & $a_{3,2}$\\ \hline
1 & 0 & -1 & 0 & 1 & -1&1&0 & 0 & 0 & 1 & 1    \\
0 & 1 & -1 & 0 & -1 & 1 &0&1 & 1 & 1 & 0 & 0 \\
0 & 0 & 0 & 1  & 1 & -2 &0&-1& -1 & 0 &1 & 1\\
0 & 0 & 0 & 0 &3 & -3 &1&-1&-2 & -1 &1&2\\
\end{tabular} \, .
\end{align}
From 3295 triangulations we chose one with the following Stanley-Reisner ideal
\begin{align}
\begin{split}
SRI: \{&  x_0 e_{32}, x_2 e_{11}, y_0 e_{11}, e_{11} e_{21}, e_{11} e_{22}, x_2 e_{12}, y_0 e_{12}, e_{12} e_{22}, e_{12} e_{31}, e_{12} e_{32}, x_2 e_{31}, y_2 e_{31}, \\ & e_{21} e_{31}, x_2 e_{21}, y_1 e_{21}, e_{21} e_{32}, x_2 e_{22}, y_1 e_{22}, e_{22} e_{32}, x_2 e_{32}, y_2 e_{32}, x_0 x_1 x_2, x_0 x_1 y_0, x_0 x_1 y_1, \\ &x_0 x_1 y_2, x_0 y_1 e_{12}, x_1 y_2 e_{11}, x_1 y_1 e_{31}, x_1 y_0 e_{21}, x_0 y_2 e_{22}, y_0 y_1 y_2                    \} \, .
\end{split}
\end{align}
The CY hypersurface constructed from $\Delta$ admits the Hodge numbers:
\begin{align}
( h^{1,1}, h^{2,1})_{\chi} = (8,26)_{-36} \, .
\end{align}
This geometry is now simply connected and still genus one fibered, albeit over a different base. Hence we still expect to have a $\mathbb{Z}_3$ discrete symmetry. The fiber admits the following expression
\begin{align}
\label{eq:blowupbase}
p=&e_{1,1} e_{2,1} e_{3,1} d_1   x_0^3 + e_{1,1} e_{1,2} e_{2,1} e_{2,2} e_{3,1} e_{3,2} d_2   x_0^2 x_1 +
 e_{1,1} e_{1,2} e_{2,1} e_{2,2} e_{3,1} e_{3,2} d_3   x_0 x_1^2 \nonumber \\&+ e_{1,2} e_{2,2} e_{3,2} d_4   x_1^3  + e_{1,1} e_{2,1} e_{3,1} d_5   x_0^2 x_2 + d_6   x_0 x_1 x_2 + e_{1,2} e_{2,2} e_{3,2} d_7   x_1^2 x_2 + 
 d_8 x_0 x_2^2 + d_9 x_1 x_2^2  + x_2^3 d_{10}   \, .
\end{align} 
In particular, we observe a factorization of the sections of the base $s_i$ that factor out  resolution divisors $E_{i,j}$. Note that in particular the last coefficient 
\begin{align}
d_{10} = y_0 y_1 y_2 a_{26} \, ,
\end{align}
is non vanishing and we therefore sill preserve the form of a generic cubic without a section. Moreover the projection to the base is given by the toric morphism inherited from the ambient space $\pi_B$ that projects the vertices $v_i \in \Delta$ onto their last two coordinates. Thus, we find the ambient space to 
be dP$_6$ consistent with the six blow-ups we performed. A depiction of the 2d polytope of the base is given in Figure~\ref{fig:dP6ResBase}.
\begin{figure}
 \hspace{1cm}
\begin{picture}(0,160)
\put(150,10){\includegraphics[scale=0.25]{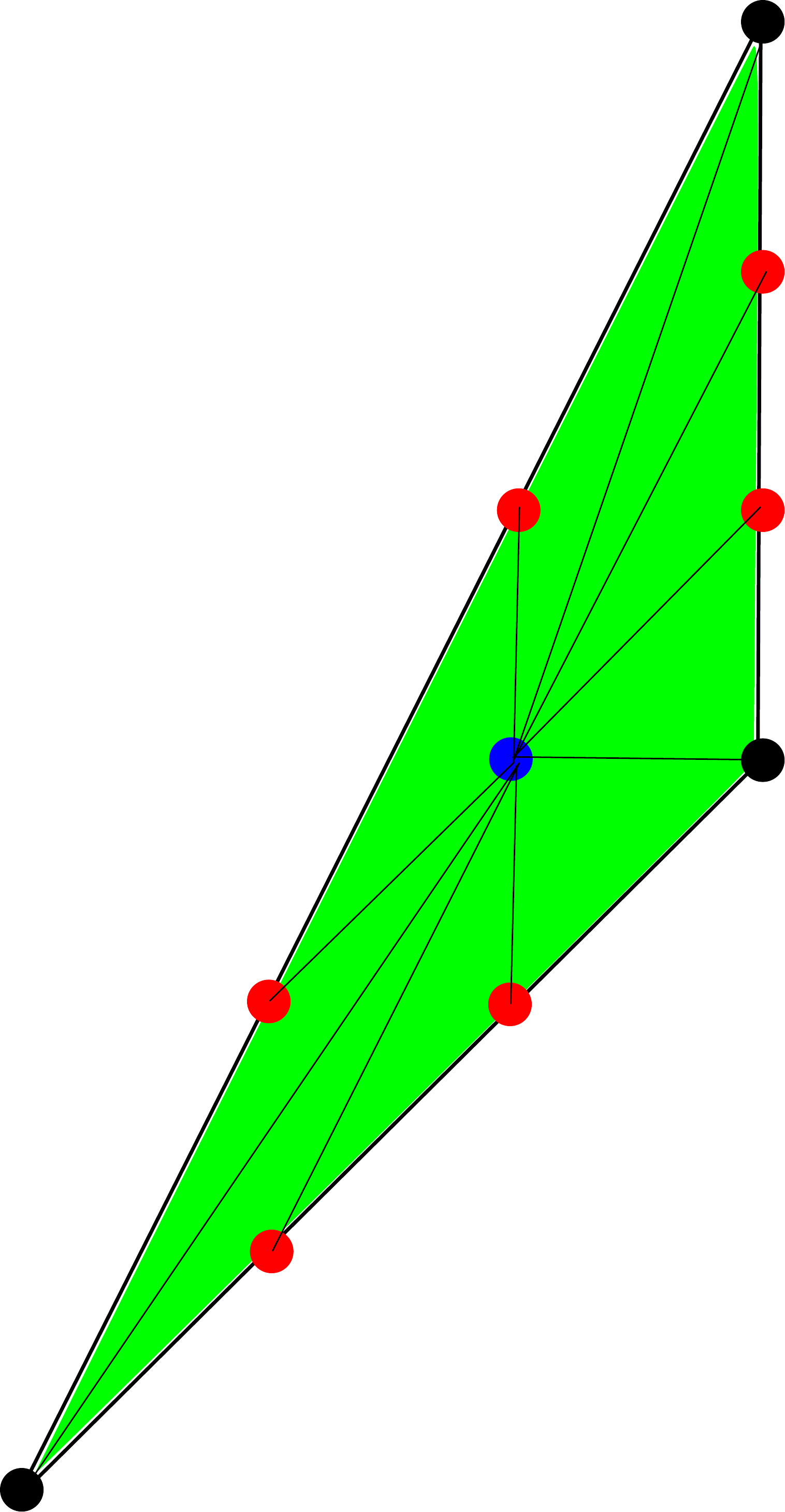}}
\put(265,110){$  y_0 $}
\put(160,10){$  y_2 $}
\put(265,215){$  y_1 $}
 
 \put(265,180){$  e_{3,2} $}
 \put(265,145){$  e_{3,1} $}
 
  \put(235,80){$  e_{2,2} $}
   \put(200,45){$  e_{2,1} $}
 
    \put(163,80){$  e_{1,2} $}
    \put(200,145){$ e_{1,1}  $}
    
\end{picture}
\caption{\label{fig:dP6ResBase}{\it The polytope of the resolved dP$_6$ base as given in \eqref{eq:Example1Tensor}. The blow-up vertices of the hyperconifold are red dotted.
}}
 
\end{figure}
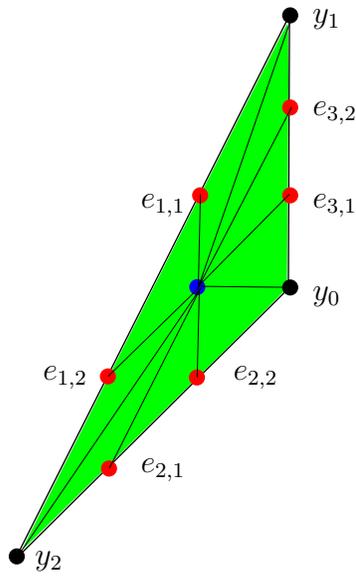
For convenience we repeat the computation of the base cohomology and intersections of Section~\ref{section2}.
From the polytope we calculate the full cohomology generated by $y_2$ and the  $e_{i,j}$ where $y_0$ and $y_1$ are linear equivalent to
\begin{align}
\begin{split}
[y_0]& \sim[y_2 + 1/3 e_{1,1} + 2/3 e_{1,2} + 1/3 e_{2,1} - 1/3 e_{2,2} - 2/3 e_{3,1} - 1/3e_{3,2}] \\
[y_1]& \sim[y_2 - 1/3 e_{1,1} + 1/3 e_{1,2} + 2/3 e_{2,1} + 1/3e_{2,2} - 1/3e_{3,1} - 2/3e_{3,2}]\, .
\end{split}
\end{align}
With this information it is easy to see that the curves in the base  have genus given as 
\begin{align}
D_{y_i}:\quad g =  1\, \qquad E_{i,j}:\quad g=  0 \, ,
\end{align}
where we have used that the anticanonical class of the base is equivalent to:
\begin{align}
\label{eq:dp6class}
K_{B}^{-1} = [3 y_2 + e_{1,1} + 2 e_{1,2} + 2 e_{3,1}  + e_{3,2}] \, ,
\end{align}
with $(K_B^{-1})^2=3$ which has unchanged intersection numbers, as it is a Cartier divisor.
This blow-up changes the base dependency and therefore also the spectrum of the theory. Making use  of the general formulas \eqref{eq:cubicspectrum} we can compute the full spectrum together with an identification of the base classes in front of the $x_1^2 x_2$ and $x_1 x_2^2$ monomials that are the sections $s_7$ and $s_9$ that we identify according to the conventions in \cite{Klevers:2014bqa} as the base classes
\begin{align}
\label{eq:BlowUpBaseclasses}
\mathcal{S}_9 = &[3y_2 + 1/3 e_{1,1}+ 5/3e_{1,2}+ 4/3 e_{2,1} + 2/3e_{2,2} - 2/3e_{3,1}- 1/3e_{3,2}]  \, ,\\
\mathcal{S}_7 =& [3y_2 + 2/3e_{1,1} + 7/3e_{1,2} + 5/3e_{2,1}+ 4/3e_{2,2} - 1/3e_{3,1} + 1/3e_{3,2}]  \, .
 \end{align}
 Those classes admit the linear equivalences and intersections:
 \begin{align}
 \label{eq:Example1Intersections}
 \begin{split}
 2 K_b^{-1} - \mathcal{S}_7 - \mathcal{S}_9 &\sim     [e_{1,1}+e_{2,1}+e_{3,1}]\, , \\
 2 \mathcal{S}_7 - \mathcal{S}_9-K_b^{-1}& \sim   [e_{1,2}+e_{2,2}+e_{3,2}]\, , \\
 2 \mathcal{S}_9 - \mathcal{S}_7-K_b^{-1} &\sim   [+e_{1,1}+e_{2,1}+e_{3,1}e_{1,2}+e_{2,2}+e_{3,2}]\, , \\  
 \mathcal{S}_7 K_b^{-1}&=  \mathcal{S}_9 K_b^{-1} = (K_b^{-1})^2=3 \, , \\
 \mathcal{S}_7 \mathcal{S}_7& =  \mathcal{S}_9 \mathcal{S}_9 = 1 \, , \\
 \mathcal{S}_7 \mathcal{S}_9 &= 2 \, .
 \end{split}
 \end{align}
 We remark, that the classes $\mathcal{S}_7$ and $\mathcal{S}_9$ have different intersection numbers now, which is consistent with the fact that they are  non-Cartier divisors on the quotient geometry.
Those classes, together with \eqref{eq:dp6class} can be plugged into \eqref{eq:cubicspectrum} which yields the spectrum:
\begin{align}
 \begin{tabular}{l|l}
 Multiplet & Multiplicity \\ \hline
      $\mathbf{1}_1$ & $72$  \\ \hline
      $\mathbf{1}_0$   & $27$ \\
      $\mathbf{V}$ & $0$ \\
      $\mathbf{T}_{(1,0)}$ & $6 $
   \end{tabular}
\end{align}
Indeed, the above spectrum cancels the gravitational anomaly and therefore captures all  massless degrees of freedom. Some comments are in order concerning the form of the blow-up divisors in the hypersurface equation \eqref{eq:blowupbase}. Indeed, by plugging this form into the equations for the associated Weierstrass form, we obtain the following dependencies on the discriminant to leading order in the blow-up divisors
\begin{align}
\Delta = e_{1,1} e_{1,2} e_{2,1} e_{2,2} e_{3,1} e_{3,1} \left(P_1 + \mathcal{O}((e_{1,1} e_{1,2} e_{2,1} e_{2,2} e_{3,1} e_{3,1} )^2) \right) \, ,
\end{align}
with the polynomial
\begin{align}
P_1 = (-d_{10} d_6^3 -   d_6 d_7 d_8^2 +  d_4 d_8^3 + d_6^2 d_8 d_9) \, .
\end{align}
 In particular we find an $A_1$ locus at the collision points of two blow-up divisors, such as $e_{i,1} = e_{i,2} = 0$. Hence over these toric loci we expect charged matter which can be confirmed by imposing the same locus in equation \eqref{eq:blowupbase}, say $e_{1,1} = e_{1,2} = 0$, which yields a factorized fiber equation 
 \begin{align}
p= x_2 (\hat{d}_6 x_0 x_1 + \hat{d}_8 x_0 x_2 + \hat{d}_9 x_1 x_2 + d_{10} x_2^2) \, .
 \end{align}
The multi-section equips us with a charge generator, in analogy to  the Shioda-map $\sigma(s_i)$, which for this case is given\footnote{Note that we have left out any potential base divisor parts.} as
\begin{align}
\sigma_{\mathbb{Z}_3} = [x_0] \, .
\end{align}
Intersecting the reducible curves of the fiber with $\sigma_{\mathbb{Z}_3}$ computes the discrete $6$-dimensional charge of the associated hypermultiplets, which yields the equivalent degrees of freedom of charge one and two, which is also the only non-trivial charge possible.
 Besides those toric loci, we also find discrete charged states over each of the resolution $\mathbb{P}^1$'s of the fixed points  which is depicted in Figure~\ref{fig:dP6ResBase} which can be found at the vanishing of $e_{i,j}= P_1 = 0$. Solving $P_1=0$ for $d_9$ and inserting those constraints into the cubic yields the desired factorization of the genus one curve into two $\mathbb{P}^1$'s 
 \begin{align}
 p_{|e_{i.j} = P_1 =0 } =\frac{(d_6 x_1 + d_8 x_2) (d_4 d_6 d_8 x_1^2 + d_6^2 d_8 x_0 x_2 + d_6 d_7 d_8 x_1 x_2 - 
   d_4 d_8^2 x_1 x_2 + d_{10} d_6^2 x_2^2)}{d_6^2 d_8} \, .
 \end{align} 
 Again the two fibral curves result in hypers of charge one and two as before.
 \begin{figure}
\begin{picture}(50,140)
 \put(75,130){$D_{y_1}$}
 \put(40,80){$D_{y_2}$}
 \put(115,80){$D_{y_0}$}
\put(0,20){\includegraphics[scale=0.6]{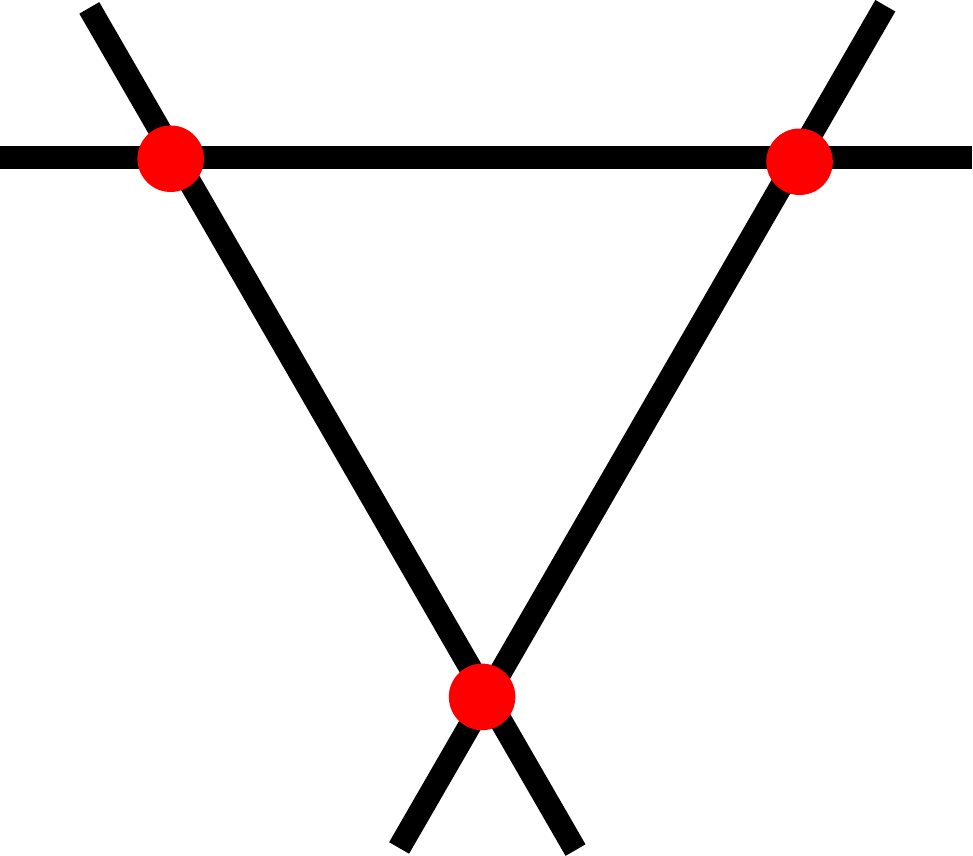}}

\put(250,20){\includegraphics[scale=0.6]{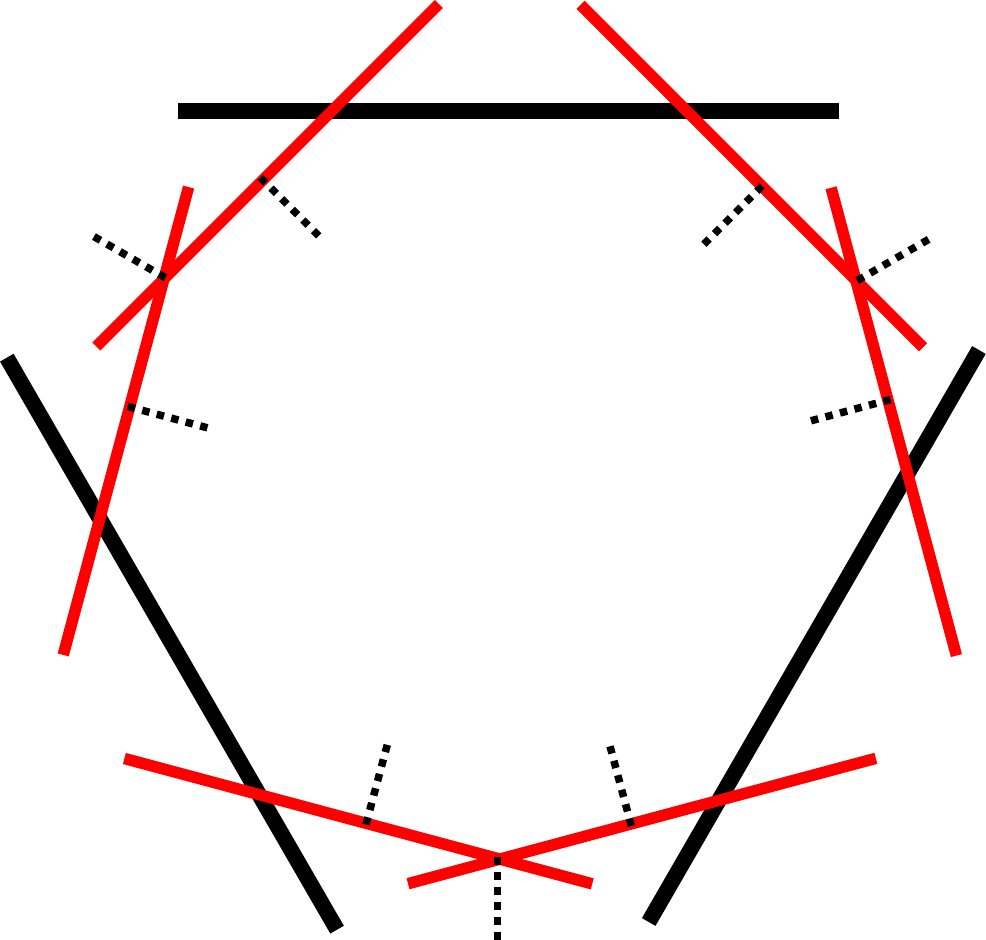}}

\put(248,139){\framebox{$\mathbf{1}_1$}}
\put(305,139){\framebox{$\mathbf{1}_1$}}
\put(286,105){\framebox{$\mathbf{1}_1$}}

\put(371,107){\framebox{$\mathbf{1}_1$}}
\put(352,137){\framebox{$\mathbf{1}_1$}}
\put(411,139){\framebox{$\mathbf{1}_1$}}

\put(327,8){\framebox{$\mathbf{1}_1$}}
\put(308,60){\framebox{$\mathbf{1}_1$}}
\put(347,60){\framebox{$\mathbf{1}_1$}}

\end{picture}

\caption{\label{fig:dP6ResBase}{\it Intersections of toric Base divisors before and after the blow-up of the $\mathbb{P}^2/\mathbb{Z}_3$ base. Each tensor branch consists of three $\mathbb{Z}_3$ charged singlets.}}
\end{figure} 
We summarize the change of the total spectrum that we have induced by the hyperconifold transition again
\begin{align}
 \begin{tabular}{ccc}
\begin{tabular}{c|c}
$6$-d Rep.&  Multi. \\ \hline 
$\mathbf{1}_{1}$      & 63 \\ \hline
$\mathbf{1}_0$ &     30 \\ 
$\mathbf{V} $ &  0 \\  
$\mathbf{T}_{(1,0)}$ &  0 \\  
$\mathbf{T}_{(2,0)}$ & 6 
\end{tabular}
& $\xrightarrow{\text{Hyperconifold}} $&
 \begin{tabular}{l}
 Multi.  \\ \hline
 $72$  \\  \hline
    $27$ \\  
    $0$ \\
      $6 $ \\
   $0$
   \end{tabular}
\end{tabular}\, .
\end{align}
Before we deform the theory further to a U(1) gauge theory, we summarize the F-theory picture we obtain when deforming back to the deformation phase of the hyperconifold.
In the F-theory picture we take the limit to a singular base which however admits a smooth CY {\it resolution} by multiple fibers over the fixed points. As there have been discrete charged states over the resolution divisors, those form new states with the collapsing tensor multiplets. The change in the spectrum suggests that one linear combination of the discrete charged hypermultiplets forms a new neutral hyper whereas two others combine with the tensors to $\mathcal{A}_2$ superconformal matter. 

\subsubsection{Un-Higgsing  to an elliptic fibration}
In the next step we want to further deform the above geometry  to an elliptic fibration that admits a section as well as a non-trivial Mordell-Weil group. For this we can build upon the configuration that we had before and tune the coefficient $d_{10} = y_0 y_1 y_2 b_{24} $ to zero 
which can be achieved by a single complex structure deformation.
The blown-up ambient space is given by the reflexive hull of the polytope spanned by the following vertices
 \begin{align}
\begin{tabular}{ccc|ccc|cc|cc|cc|c}
$x_0$ & $x_1$ & $x_2$ & $y_0$ & $y_1$ & $y_2$ & $e_{1,1}$ & $e_{1,2}$ & $e_{2,1}$ & $a_{2,2}$ &  $e_{3,1}$ & $a_{3,2}$ & $e_1$\\ \hline
1 & 0 & -1 & 0 & 1 & -1&1&0 & 0 & 0 & 1 & 1   & 1  \\
0 & 1 & -1 & 0 & -1 & 1 &0&1 & 1 & 1 & 0 & 0 & 1\\
0 & 0 & 0 & 1  & 1 & -2 &0&-1& -1 & 0 &1 & 1& 0\\
0 & 0 & 0 & 0 &3 & -3 &1&-1&-2 & -1 &1&2 & 0\\
\end{tabular} \, ,
\end{align}
with a choice of a triangulation resulting in the Stanley-Reisner ideal
\begin{align}
\begin{split}
SRI: \{& 
x_0 x_1, x_0 e_{32}, x_2 e_{11}, x_2 e_{12}, x_2 e_{21}, x_2 e_{22}, x_2 e_{31}, x_2 e_{32}, x_2 e_1, y_0 e_{11}, y_0 e_{12}, y_1 e_{21}, y_1 e_{22},\\&  y_1 e_1, y_2 e_{31}, y_2 e_{32}, e_{12} e_{31}, e_{21} e_{31}, e_{22} e_{31}, e_{11} e_{21}, e_{12} e_{21}, e_{21} e_{32}, e_{11} e_{22}, e_{12} e_{22}, e_{22} e_{32},\\& e_{12} e_{32}, x_0 y_1 e_{12}, y_0 y_1 y_2, x_0 y_2 e_{22}, y_0 y_2 e_1, x_1 y_1 e_{31}, x_1 y_0 e_{21}, y_0 e_{21} e_1, y_2 e_{22} e_1\\ &, y_0 e_{32} e_1, x_1 y_2 e_{11}, y_2 e_{11} e_1, x_1 e_{11} e_{31}
\} \;.
\end{split}
\end{align}
The additional divisor $e_1=0$ is a rational section of the elliptic fibration, that admits the Hodge numbers
\begin{align}
( h^{1,1}, h^{2,1})_{\chi} = (9,25)_{-32} \,,
\end{align}
as expected. The fiber equation becomes a restricted cubic in the $x_i$ given as
\begin{align}
\label{eq:Dp1Dp6fiber}
p=&e_{1,1} e_{2,1} e_{3,1} d_1 e_1^2 x_0^3 + e_{1,1} e_{1,2} e_{2,1} e_{2,2} e_{3,1} e_{3,2} d_2 e_1^2 x_0^2 x_1 \nonumber \\&+
 e_{1,1} e_{1,2} e_{2,1} e_{2,2} e_{3,1} e_{3,2} d_3 e_1^2 x_0 x_1^2 + e_{1,2} e_{2,2} e_{3,2} d_4 e_1^2 x_1^3 \\
 &+ e_{1,1} e_{2,1} e_{3,1} d_5 e_1 x_0^2 x_2 + d_6 e_1 x_0 x_1 x_2 + e_{1,2} e_{2,2} e_{3,2} d_7 e_1 x_1^2 x_2 + 
 d_8 x_0 x_2^2 + d_9 x_1 x_2^2 \, . \nonumber 
\end{align}
We compute the change in spectrum by the identification of the divisor classes as in \eqref{eq:BlowUpBaseclasses} and insert them in the general expressions \eqref{eq:spectrum} to obtain the spectrum
\begin{align}
 \begin{tabular}{l|l}
 $6$-d Reps & Multi. \\ \hline
   $\mathbf{1}_3$  & $2$ \\ 
      $\mathbf{1}_2$ & $18$ \\
      $\mathbf{1}_1$ & $54$ \\   \hline
      $\mathbf{1}_0 $ &$ 26$ \\
      $\mathbf{V}$ & $1 $\\
      $\mathbf{T}_{(1,0)}$ & $6 $
   \end{tabular}
\end{align}
consistent with all anomalies. The spectrum above is consistent with the Higgsing back to the genus one fibration  induced by $\langle \mathbf{1}_3 \rangle \neq 0$. The two singlets then become the Goldstone mode for the massive U(1) vector and the additional neutral singlet in the genus one geometry. Also the multiplicity of the discrete charged singlets is matched with those in the genus one geometry. Again, singlets of U(1) charge one are located at the intersection of the $\mathbb{Z}_3$ resolution divisors. However, now we have an elliptic fibration with a non-trivial Mordell-Weil rank for which we redo the computation of the matter on the resolution divisors  as a consistency check. 
\subsubsection{Location of charged matter}
\begin{figure}[h]
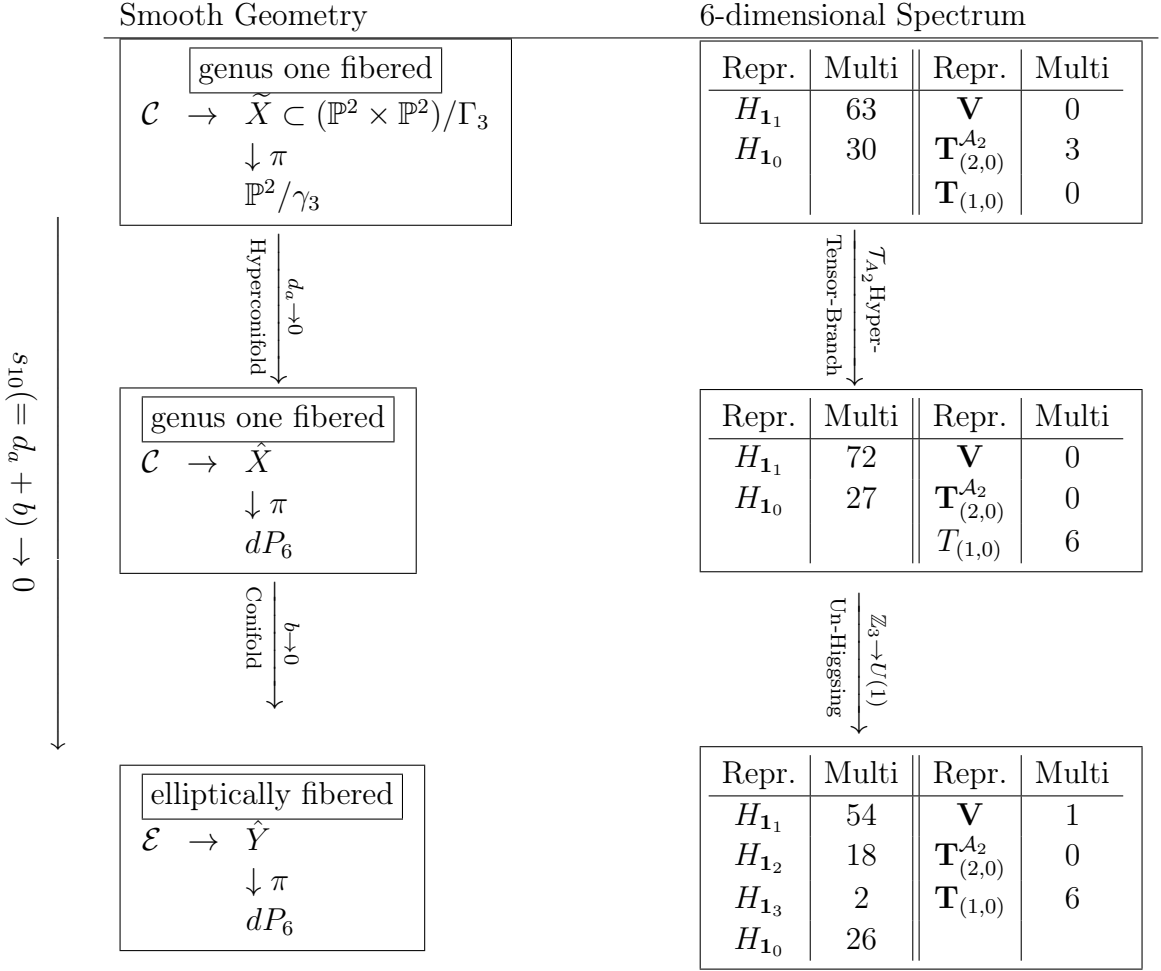

\begin{center}
\begin{tabular}{ll}    \rotatebox{270}{ $\hspace{-2cm} s_{10}( = d_a + b )\rightarrow 0 $} $ \left\downarrow\rule{0cm}{3.7cm}\right.$   &

   \begin{tabular}{ l  c    l}
Smooth Geometry  & & $6$-dimensional Spectrum \\ \hline
 
  \framebox{$
\begin{array}{ccl}
\multicolumn{3}{c}{\framebox{genus one fibered}} \\
\mathcal{C} & \rightarrow & \widetilde{X} \subset (\mathbb{P}^2 \times \mathbb{P}^2) /\Gamma_3 \\
&&  \downarrow \pi  \\
& & \mathbb{P}^2 /\gamma_3 
\end{array}
$} 
 
 & \qquad 
 &

 \framebox{$
\begin{array}{c|c||c|c}
\text{Repr.} & \text{Multi} &\text{Repr.} & \text{Multi}  \\ \hline
H_{\mathbf{1}_1} & 63   &  \mathbf{V}  & 0 \\
 
H_{\mathbf{1}_0} & 30 &  \mathbf{T}^{\mathcal{A}_2}_{(2,0)}  & 3 \\       
& &  \mathbf{T}_{(1,0)} & 0  
\end{array}
 $ } \\

\qquad \qquad \rotatebox{270}{$\hspace{-0.3cm}    \xrightarrow[\text{Hyperconifold}]{d_a \rightarrow  0          } $}   & & \qquad \qquad \rotatebox{270}{ $ \hspace{-0.5cm}    \xrightarrow[\text{Tensor-Branch}]{ \mathcal{T}_{A_2} \text{Hyper-\, }}$} \\
  
   \framebox{$
\begin{array}{ccl}
\multicolumn{3}{c}{\framebox{genus one fibered}} \\
\mathcal{C} & \rightarrow & \hat{X} \\
&&  \downarrow \pi  \\
& & dP_6
\end{array}
$}

 & \qquad \qquad \qquad
 &

 \framebox{$
\begin{array}{c|c||c|c}
\text{Repr.} & \text{Multi} &\text{Repr.} & \text{Multi}  \\ \hline
H_{\mathbf{1}_1} & 72   &  \mathbf{V} & 0 \\
 
H_{\mathbf{1}_0} & 27 & \mathbf{T}^{ \mathcal{A}_{2}}_{(2,0)} & 0 \\       
& & T_{(1,0)} & 6 
\end{array}
 $ } \\

  \qquad \qquad \rotatebox{270}{$ \hspace{-0.3cm}     \xrightarrow[\text{Conifold}\quad]{b \rightarrow  0          } $}   & & \qquad \qquad \rotatebox{270}{ $  \hspace{-0.3cm}   \xrightarrow[\text{Un-Higgsing }]{\mathbb{Z}_3 \rightarrow U(1) } $   } \\

   \framebox{$
\begin{array}{ccl}
\multicolumn{3}{c}{\framebox{elliptically fibered}} \\
\mathcal{E} & \rightarrow & \hat{Y} \\
&&  \downarrow \pi  \\
& & dP_6
\end{array}
$} 
 
 & \qquad \qquad \qquad
 &

 \framebox{$
\begin{array}{c|c||c|c}
\text{Repr.} & \text{Multi} &\text{Repr.} & \text{Multi}  \\ \hline
H_{\mathbf{1}_1} & 54   & \mathbf{V} & 1 \\
H_{\mathbf{1}_2} & 18 & \mathbf{T}^{ \mathcal{A}_{2}}_{(2,0)} & 0 \\   
H_{\mathbf{1}_3} & 2 &     \mathbf{T}_{(1,0)} & 6 \\ 
 H_{\mathbf{1}_0}&26  & &
\end{array}
 $ } \\

\end{tabular} 
\end{tabular}
\end{center}
\caption{\label{fig:QuotientDiagram}{\it The genus one fibered geometry of the bicubic-quotient and its transition to an elliptic fibration from top to bottom crossing three hyperconifold transitions. The $6$-dimensional spectrum is highlighted in every step.}}
\end{figure}

In the following we want to re-compute the matter loci over the orbifold resolution divisors that are affected when going back to the $ \mathcal{A}_2 $ tensor branch. In the case of the above mentioned U(1) theory, those loci have been analyzed in \cite{Klevers:2014bqa} by a prime ideal decomposition of the rational sections in Weierstrass form.\\\\
We start by listing those loci for the charge one, two and three singlets whose general multiplicity we have computed in the section before . They are summarized in the following table:
\begin{align}
\begin{array}{c|l}\label{ideals1}
\text{singlet} & \qquad\qquad\text{constraint}  \\ \hline
\mathbf{1}_3 & V(I_3): \{ s_8 = s_9 = 0 \}  \\ 
\mathbf{1}_2 & V(I_2): 
\{ s_4 s_8^3 -s_3 s_8^2 s_9 + s_2 s_8 s_9^2 -s_1 s_9^3 \\ 
&\qquad\qquad =s_7 s_8^2 + s_5 s_9^2 - s_6 s_8 s_9=0 \}  \\
&\qquad\qquad (s_8,s_9) \neq (0,0) \\
\mathbf{1}_1 & V(I_1): \{ y_1 = f z_1^4 + 3 x_1^2 =0\} / ( (V(I_1 ) \& V(I_2))     \\ 
\end{array}
\end{align}
where the Weierstrass coordinates of the rational section $(y_1, x_1, z_1)$ are given in the Appendix B of \cite{Klevers:2014bqa}. In the following we discuss the three ideals in more detail and determine whether their associated matter is located over the $A_2$ resolution divisors or not. 
\begin{itemize}
\item $\mathbf{I_3}$ {\bf locus:} Imposing the $e_{i,j}= 0$ on the $I_3$ locus and imposing the SRI results in two constant  non-vanishing functions for a generic complex structure. Hence there is no charge three matter found over these loci and hence those states are located far away from the $A_2$ singularity and its resolution.
\item $\mathbf{I_2}$ {\bf locus: } For the charge two matter the situation is very similar and 
we find i.e. for $e_{1,1} = 0$ and using the SRI, the two functions to be of the form:
    \begin{align}
I_{2|e_{1,1}=0} = \{ e_{1,2} a_1 a_{21}^3\, ,\, a_{21} (-y_1 a_9 a_{13} +e_{1,2} a_{20} a_{21} -e_{1,2} a_9 a_{28}) \} \, ,
\end{align}
which also admits no solution that is codimension two in the dP$_6$ coordinates. 
\item $\mathbf{I_1}$ {\bf locus: } Here we do find a solution, which can be seen by imposing again $e_{1,1} = 0$ where the ideal becomes of the form:
\begin{align}
I_{1|e_{1,1} = 0=0} =\{ e_{1,2}  a_{1} a_{
  21}^5 Q_1(y_1,e_{1,2})Q_2(y_1, e_{1,2}  ) \,
, \, 
    e_{1,2}  a_{1} a_{ 
  21}^6 Q_1(y_1,e_{1,2})Q_3 (y_1, e_{1,2}   \} 
\end{align}
with $Q_i(y_1, e_{1,2}  )$ being degree $i$ polynomials in $y_1$ and $e_{1,2} $. The two solutions are
\begin{align}
Q_1(y_1,e_{1,2}  )=(y_1 a_{9}^2 a_{13} - e^{a_0}_2 a_{9} a_{20} a_{21} + e_{1,2} a_{1} a_{21}^2 + 
  e_{1,2}  a_{9}^2 a_{28}) =0 \, .
\end{align}
Hence we find one charge state at the intersection of the two $\mathbb{P}^1$'s and another one over a non-toric locus just like in the higgsed case.
 
 \end{itemize}
The above calculations show  that each resolved $A_2$ singularity in the base actually carries three charged singlet states with minimal charge: One located over each $\mathbb{P}^1$ and another one at their intersection just as in the $\mathbb{Z}_3$ case.
We summarize the whole flow of geometries and their respective $6$-dimensional F-theory spectra in Figure~\ref{fig:QuotientDiagram}.\\
Finally we note that  in the U(1) theory we considered the height pairing, given as
\begin{align*}
b_{11}= -2 (3 K_b^{-1} + \mathcal{S}_7 - 2 \mathcal{S}_9) \,.
 \end{align*} 
 is actually not a Cartier divisor when taking the singular limit of the base.
 This is because $\mathcal{S}_7 - 2 \mathcal{S}_9$ is exactly the sum of the classes of the $\mathbb{Z}_3$ resolution divisors $b_{11}$  as can be seen from Equation~\ref{eq:Example1Intersections}. Thus if we do not Higgs the above theory  we can couple the U(1) theory to the $A_2$ (2,0) points.
 \newpage 
 \subsection{Example 2: Threefold in $(\mathbb{P}^{1,1,2}\times F_0)/\mathbb{Z}_2 $}
 \label{sec:example2}
The second example we chose admits a $\mathbb{Z}_2$ gauge symmetry and four $\mathcal{A}_1$ (2,0) points in the base  but in addition it also admits a non-Abelian SU(2) gauge symmetry which gives it more structure than the example we have studied before. In the following we go again through the explicit construction of the quotient geometry and  follow the change of the spectrum. Finally we perform  the hyperconifolds and check that the tensor branch of the (2,0) theories admits additional purely discrete charged hypermultiplets consistent with anomaly cancellation. 

 \subsubsection{The Geometric setup}
 For the sake of keeping the discussion short we go to the quotient geometry $\widehat{X}$ straight away\footnote{The polytope of the covering  CY,  $X$  can be obtained by giving the base coordinates of $F_0$ trivial legs in the fiber such that the polytope coordinates become block diagonal.} which is given by the toric hypersurface in the $(\mathbb{P}^{1,1,2}\times \mathbb{F}_0)/\mathbb{Z}_2$ ambient space that is encoded in the polytope generated by the following vertices
 \begin{align}
\begin{tabular}{cccc|cc|cc}
$X$ & $Y$ & $Z$ &  $e_1$ &  $x$ & $t$ & $y$ & $s$  \\ \hline
-1 & -1 & 1 & -1 & 0 & 0 & 0 & 0 \\ 
1 & -1 & 0 & 0 & 1 & -1 & 0 & 0 \\ 
0 & 0 & 0 & 0 & -1 & 1 & -1 & 1 \\ 
0 & 0 & 0 & 0 & 0 & 0 & 2 & -2 
\end{tabular} \, .
\end{align}
The first four coordinates are those of $\mathbb{P}^{1,1,2}$ whereas the second four parametrize $\mathbb{F}_0$. From the ambient space we find the toric morphism $\phi$ 
\begin{align}
 \phi: \qquad \mathbb{C}^8 \rightarrow \{\frac{Z}{X Y e_1}, \frac{X  x}{Y t}, \frac{t s}{x y}, \frac{y^2}{s^2}          \} \, ,
 \end{align}
 whose kernel generates the usual four $\mathbb{C}^*$ identifications of $\mathbb{P}^{1,1,2}$ and $\mathbb{F}_0$. However, in addition, we also find the discrete $\Gamma_2$ identification:
 \begin{align}
 \Gamma_2: \quad  (X,Y,Z,e_1; x,t,y,s) \sim (     \Gamma_2 X, Y, \Gamma_2 Z,e_1; \Gamma_2 x, t,\Gamma_2 y ,s       ) \, ,
 \end{align}
with $\gamma_2^2=1$. This geometry admits the same standard Stanley Reisner ideal as we would have for the direct product manifold
\begin{align}
SRI := \{X Y, Z e_1; xt, ys      \} \, ,
\end{align}
 but admits 16 fixed points in total that come as the combinations:
\begin{align}
(X,Z,Y,e_1; x,t,y,s) = (\underline{0,1},\underline{0,1};\underline{0,1},\underline{0,1}) \, ,
\end{align}
using the SRI and the $\mathbb{C}^*$ transformations to set the residual coordinates to one.
The CY hypersurface $\widehat{X} \subset (\mathbb{P}^{1,1,2} \times F_0)/\Gamma_2$ misses those fixed points  as we will discuss momentarily and the Hodge numbers are given by
\begin{align}
\label{eq:HodgeEx2}
(h^{(1,1)}(\widehat{X}),h^{(2,1)}(\widehat{X}))_{\chi} = (4,36)_{-64} \, .
\end{align}
These are indeed the expected Hodge numbers  when we compared to the the covering CY $X \subset (\mathbb{P}^{1,1,2} \times F_0)$  geometry
\begin{align}
(h^{(1,1)}(X),h^{(2,1)}(X))_{\chi} = (4,68)_{-128} \, .
\end{align}
Due to the two different ambient factors before quotienting, we actually have two choices to pick a genus one fibration. We  start by picking the $\mathbb{P}^{1,1,2}$ as the fiber ambient space whereas the second one is presented in Section~\ref{sec:example3}.
 The CY hypersurface in terms of fiber coordinates has been discussed several times in the literature already \cite{Braun:2014oya, Anderson:2014yva, Morrison:2014era,Klevers:2014bqa} and is given by
 \begin{align}
\begin{split}\label{eq:quartichypersurface}
p&= d_1^{(+)} e_1^2 X^4 + d_2^{(-)}  e_1^2 X^3 Y + d_3^{(+)}  e_1^2 X^2 Y^2 + d_4^{(-)}  e_1^2 X Y^3 + d_5^{(+)}  e_1^2 Y^4 + d_6^{(-)}  e_1 X^2 Z \\
&\phantom{=} + d_7^{(+)}  e_1 X Y Z + d_8^{(-)}  e_1 Y^2 Z + d_9^{(+)}  Z^2\,,
\end{split}
\end{align}
 where the $d_i^{(\pm)}$ are sections in the anticanonical class of the base that transform even or odd under the $\Gamma_{2,b}$ action on the base. Similarly as in the first example all odd sections $d_i^{(-)}$ vanish over a fixed point in the base where the fiber attains the form
 \begin{align}
 \label{eq:quartic}
\begin{split} 
p = \hat{d}_1  e_1^2 X^4 +  \hat{d}_3  e_1^2 X^2 Y^2 +  \hat{d}_5  e_1^2 Y^4  
  + \hat{d}_7   e_1 X Y Z   + \hat{d}_9  Z^2\,,
\end{split}
\end{align}
 while the $\hat{d}_i$ are non-vanishing for generic complex structures. Hence we find that the sections $\hat{d}_i$ with $i=1,3,5,9$  prevent the singularities to lie on the hypersurface.   Smoothness of the fiber is readily checked by the following generically non-vanishing discriminant 
\begin{align}
 \Delta= -(1/16) \hat{d}_1 \hat{d}_5 \hat{d}_9^2 (-\hat{d}_7^4 + 8 \hat{d}_3 \hat{d}_7^2 \hat{d}_9 - 16 \hat{d}_3^2 \hat{d}_9^2 + 
   64 \hat{d}_1 \hat{d}_5 \hat{d}_9^2)^2 \, .
 \end{align} 
 \subsubsection{Spectrum and hyperconifold tensor branch}
 \label{sec:Example2Conifold}
 The spectrum associated to the geometry described in the previous Subsection can be computed by using the general results given in \cite{Klevers:2014bqa}. The gauge group of this genus one geometry is given as $SU(2) \times \mathbb{Z}_2$. Indeed, the locus $d_9^{(+)}=0$ gives an SU(2) singularity that is resolved by $e_1$ in the fiber but does not cross any of the base fixed points as it transforms as an even section under the base $\Gamma_2$ action. In addition we find several discrete charged singlet states located away from the fixed points. The general formulas for the spectrum computation and four different geometries are summarized in Table~\ref{eq:Example2Spectrum}. We would like to contrast the tensor branch of the $\mathcal{A}_1$ theories which we obtained by the hyperconifolds, with that of a direct product manifold listed as the last row in Table~\ref{eq:Example2Spectrum} that lacks the additional discrete charged states.
 \begin{table}
  \begin{tabular}{|c|c||c|c|c|c|}
   \multicolumn{6}{c}{Multiplicities} \\ \cline{2-6}  
 \multicolumn{1}{c|}{}  & \multirow{2}{*}{ Generic Base}& \multicolumn{4}{c|}{Ambient Space Geometry} \\ \cline{3-6}  
 \multicolumn{1}{c|}{State}  &  &   $\mathbb{P}^{1,1,2} \times \mathbb{F}_0$  &$\frac{(\mathbb{P}^{1,1,2} \times \mathbb{F}_0)}{\mathbb{Z}_{2}}  $  &$ \frac{(\mathbb{P}^{1,1,2} \times \mathbb{F}_0)}{\mathbb{Z}_{2}}^{\text{HC}}$   &    {\footnotesize $(\mathbb{P}^{1,1,2} \times BL_4 \mathbb{F}_0)$}   \\ \hline 
  $ H_{\mathbf{2}_{1}}$  & $\begin{array}{l}6(K_b^{-1} + 2 \mathcal{S}_7-2 \mathcal{S}_9)\\ 
\times(K_b^{-1} - \mathcal{S}_7 + \mathcal{S}_9)\end{array}$ &  48 & 24&24 & 24 \\ \hline
$H_{1_{2}}$& $\begin{array}{l}  6(K_b^{-1})^2 + 13 K_b^{-1} \mathcal{S}_7 - 3\mathcal{S}_7^2 \\
-5K_b^{-1} \mathcal{S}_9-2\mathcal{S}_7 \mathcal{S}_9 + \mathcal{S}_9^2 \end{array}$
& 80 & 40 & 48 & 40 \\ \hline
$H_{3_{0}}$ & $1 + \frac12 (K_b^{-1}-\mathcal{S}_7 + \mathcal{S}_9)
(\mathcal{S}_9-\mathcal{S}_7)$ & 1 & 1&1 & 1  \\ \hline
$H_{1_{0}}$ & - & 69 & 37 & 33 & 41 \\ \hline
V & 3 & 3 & 3 & 3  & 3 \\
$T_{(1,0)}$   & $h^{1,1}(B)-1$               & 1 & 1 & 5 & 5 \\
$T_{(2,0)} $  & $10-h^{1,1}(B)-K_b^{2}$ & 0 & 4 & 0  & 0  \\ \hline
 \end{tabular} 
 \caption{
 \label{eq:Example2Spectrum}{\it  Spectra of four genus one fibrations with $(SU(2) \times \mathbb{Z}_4  )/\mathbb{Z}_2$ gauge group and their ambient spaces. We compare covering geometry, quotient, hyperconifold tensor branch and highlight the change in spectrum. This is contrasted to the spectrum of a regular $A_1$ tensor branch theories given in the last column.}} 
 \end{table}
In the computation of the charged spectrum we used $\mathcal{S}_7 = \mathcal{S}_9 = K_b^{-1}$ for the (un-)quotiented geometry with the self intersection $(K_b^{-1})^2 = (8)4$.
We note again, that we used the general formulas of the discrete charged matter spectrum, obtained in \cite{Klevers:2014bqa}, with the identification
\begin{align}
[d^{(+)}_7] \sim K_b^{-1}\, , \qquad [d^{(-)}_8]\sim \mathcal{S}_7 \, , \qquad [d^{(-)}_2] \sim 2 K_b^{-1} - \mathcal{S}_9 \, .  
\end{align}
Similar to what we have described in the case of the bicubic quotient, we find that $\mathcal{S}_7$ and $\mathcal{S}_9$ are degree $(2,2)$ non-Cartier divisors in the base, unlike $K_b^{-1}$. However, by abuse of notation we set their classes to be equal when computing their intersections in Table \ref{eq:Example2Spectrum}.

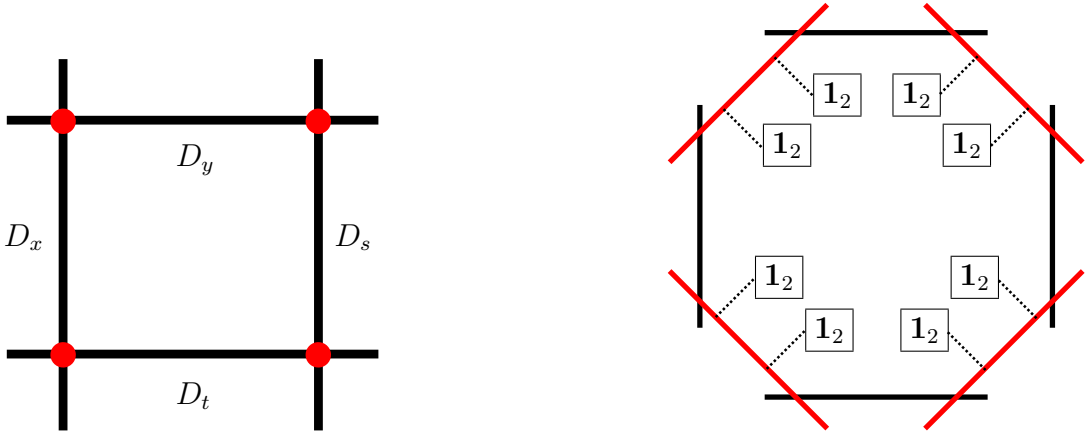
\begin{figure}[h!]
 \hspace{1.5cm}
\begin{picture}(130,170) 
\put(0,20){\includegraphics[scale=0.5]{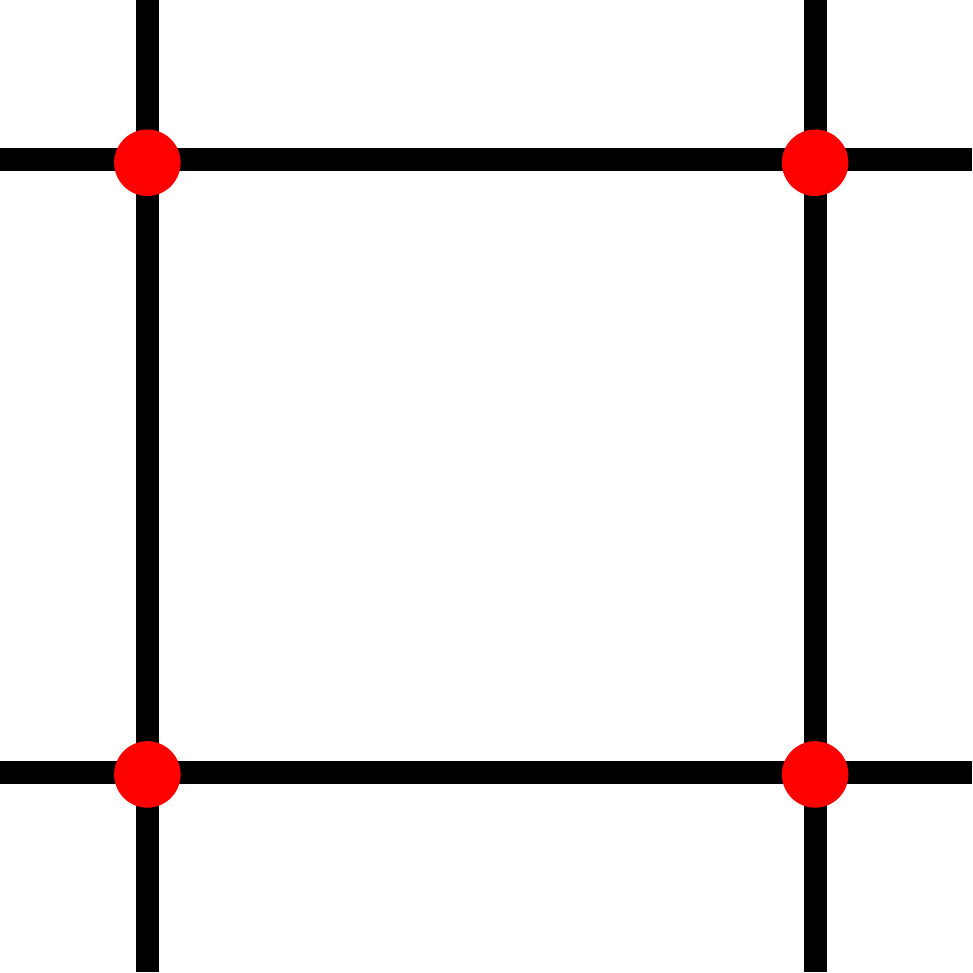}}

\put(250,20){\includegraphics[scale=0.3]{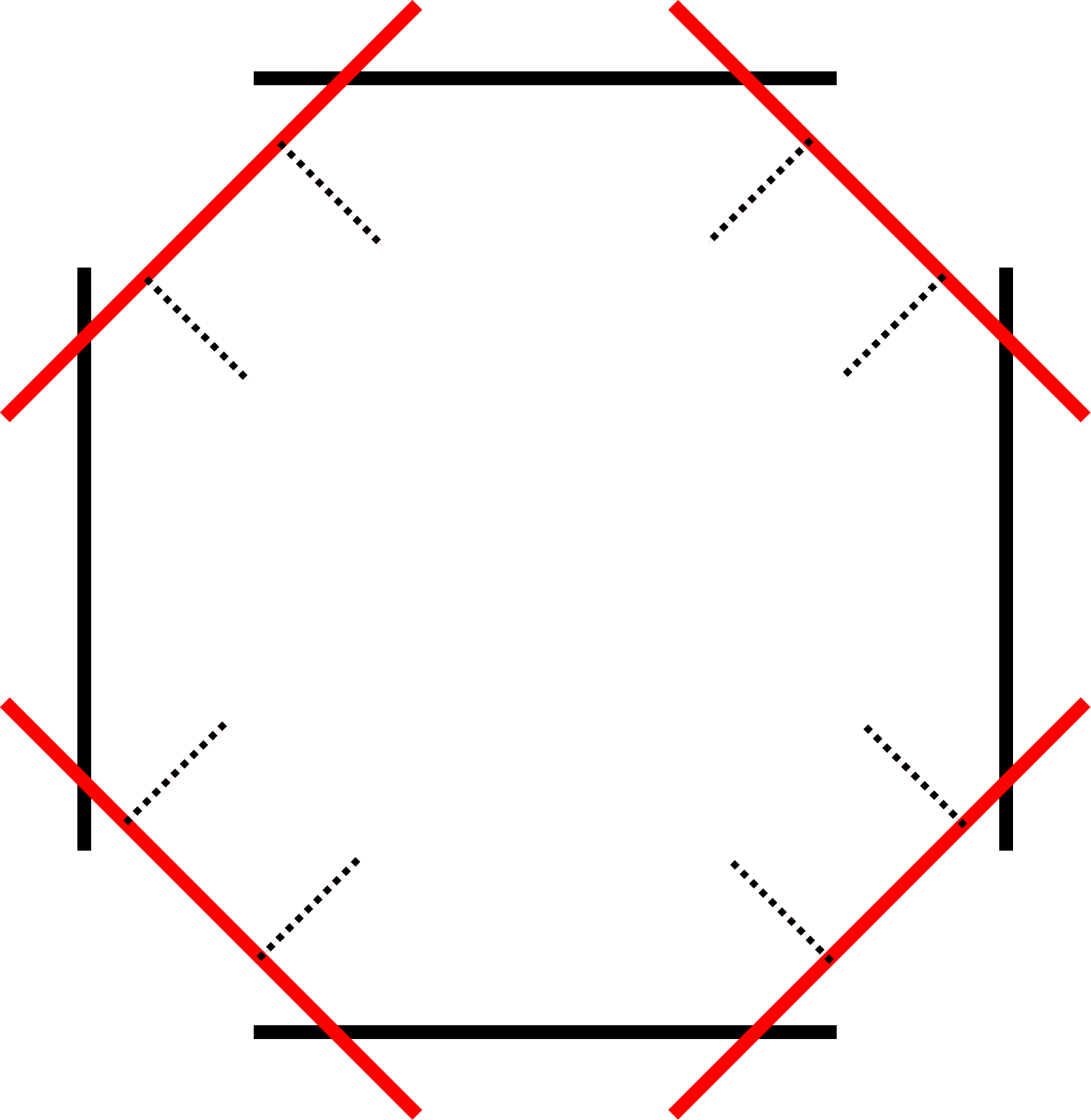}}

\put(305,144){\framebox{$\mathbf{1}_2$}}
\put(286,125){\framebox{$\mathbf{1}_2$}}

\put(335,144){\framebox{$\mathbf{1}_2$}}
 \put(354,125){\framebox{$\mathbf{1}_2$}}

\put(-5,90){ $D_x$ }
\put(120,90){ $D_s$ }
\put(60,120){ $D_y$ }
\put(60,30){ $D_t$ }
 
\put(302,55){\framebox{$\mathbf{1}_2$}}
\put(283,75){\framebox{$\mathbf{1}_2$}}

\put(338,55){\framebox{$\mathbf{1}_2$}}
\put(357,75){\framebox{$\mathbf{1}_2$}}

\end{picture}

\caption{\label{fig:dP4ResBaseB}{\it Intersections of toric divisors for singular and resolved $\mathbb{F}_0/\mathbb{Z}_2$ base. In red we denote the orbifold fixed points and their resolution divisors. After resolution we find two discrete charged matter states per $-2$ curve.}}
 
\end{figure}

Note that all spectra satisfy all gauge and gravitational anomalies \eqref{eq:6dAnomalies} listed in Appendix~\ref{app:Anomalies}.\\\\
In the following we want to comment on the resolution of the $(2,0)$ subsectors and the location of the newly appearing discretely charged matter states over the blow-ups, which is the main difference to ordinary $A_1$ (2,0) superconformal points.
For this we tune the following four ambient space fixed points 
\begin{align}
(X,Y,Z,e_1; x,y,s,t) = (0,1,0,1 ; \underline{0,1 }, \underline{0,1}); 
\end{align}
 onto the CY hypersurface which amounts to tune the complex structure coefficients $a_i$ in
 \begin{align}
 \label{eq:U1CoeffExample2}
 d_5^{+} = x^2 t^2 a_1 + y^2 x^2 a_2 + s^2 t^2 a_3 + y^2 s^2 a_4 + 
  s t x y b \, ,
 \end{align}
to zero and then resolving\footnote{Tuning the $b$ coefficient to zero as well creates a section, and the blow-up of the the singular model results in the familiar $Bl_1 \mathbb{P}^{1,1,2}$ model of  \cite{Morrison:2012ei} with $U(1) \times SU(2)$ gauge group.} the singular CY. 
 This choice leads to a new polytope spanned by the following vertices
 \begin{align}
\begin{tabular}{cccc|cc|cc|cccc}
$X$ & $Y$ & $Z$ &  $e_1$ &  $x$ & $t$ & $y$ & $s$ & $e_{1,1}$ & $e_{2,1}$ & $e_{3,1} $ & $e_{4,1} $  \\ \hline
-1 & -1 & 1 & -1 & 0 & 0 & 0 & 0 &  0  &    0  &  0  &  0                            \\ 
1 & -1 & 0 & 0 & 1 & -1 & 0 & 0   &  1   &    1  &  0  &  0      \\ 
0 & 0 & 0 & 0 & -1 & 1 & -1 & 1  &  0   &    -1 &  0  &  1                     \\ 
0 & 0 & 0 & 0 & 0 & 0 & 2 & -2   &  -1  &    1  &  1  &  -1
\end{tabular} \, .
\end{align}
A choice of some triangulation yields a Stanley-Reisner ideal of the form
\begin{align}
\begin{split}
SRI: \{ &
X Y, X Z, Z e_1, e_1 e_{1,1}, e_1 e_{2,1}, e_1 e_{3,1}, e_1 e_{4,1}, x t, x e_{3,1}, x e_{4,1}, y s, y e_{1,1}, \\ &y e_{4,1}, s e_{2,1}, s e_{3,1}, t e_{1,1}, t e_{2,1}, Y e_{2,1}, e_{2,1} e_{4,1}, Y e_{1,1}, Y e_{3,1}, Y e_{4,1}
\} \, ,
\end{split}
\end{align}
and the Hodge numbers
\begin{align}
(h^{(1,1)},h^{(2,1)})_{\chi} = (8,32)_{-48} \, .
\end{align} 
The base can be identified via the projection of the polytope onto the last two coordinates  which gives the toric diagram of a resolved $\mathbb{F}_0 /\mathbb{Z}_2$ as shown in Figure~\ref{fig:dP4ResBaseA}.
\begin{center}
\begin{figure}
\vspace{1cm}
\begin{picture}(0,110)
\put(150,10){\includegraphics[scale=0.25]{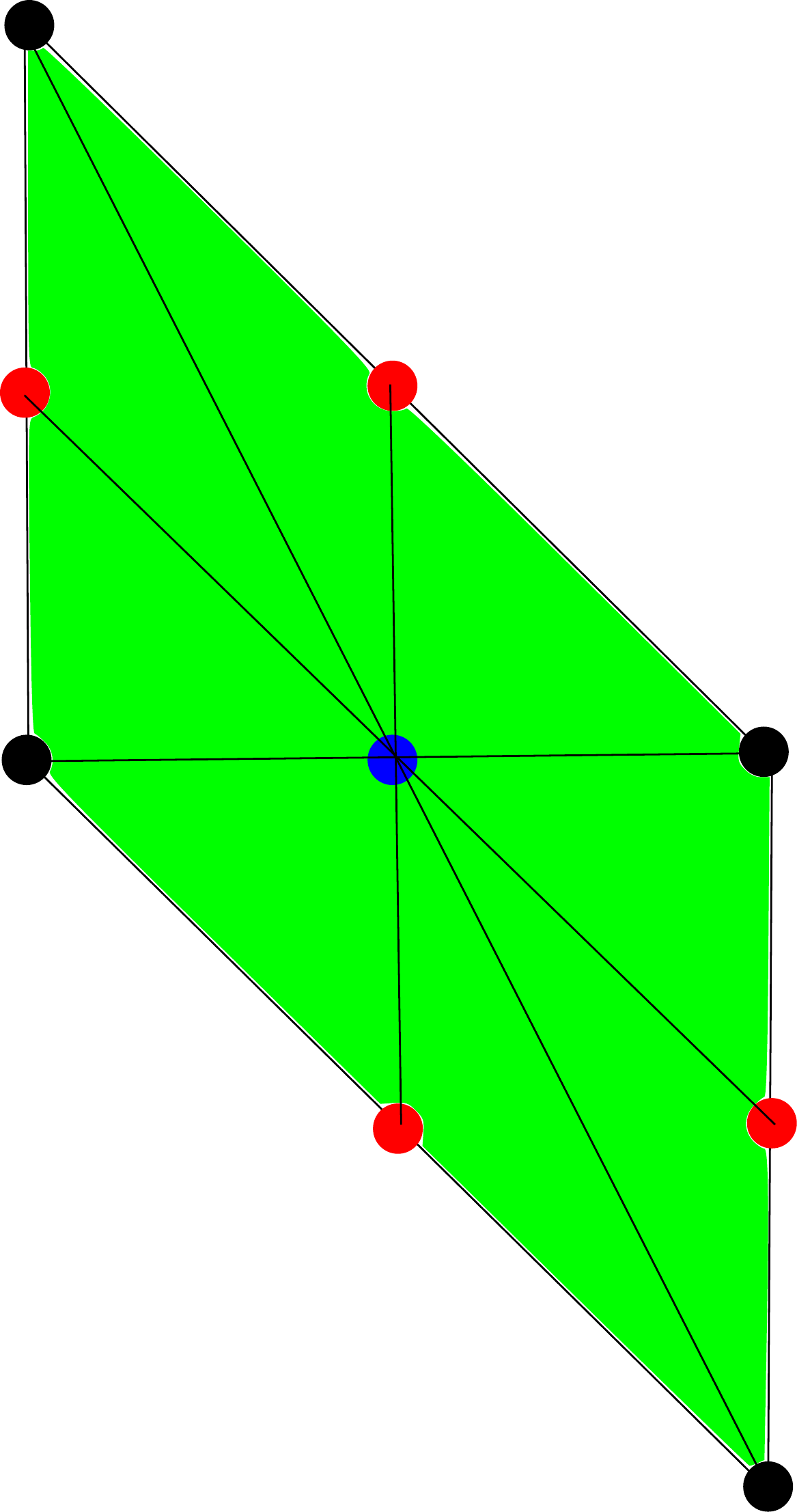}}

  \put(161,166){$y  $}
 
 \put(200,130){$ e_{3,1}  $}

  \put(240,85){$  t $}
  
    \put(240,50){$  e_{4,1} $}
    
      \put(240,10){$  s $}

   \put(170,45){$  e_{1,1} $}

        \put(130,125){$  e_{2,1} $}
            \put(135,85){$  x $}

\end{picture}
\caption{\label{fig:dP4ResBaseA}{\it The polytope of the resolved $\mathbb{F}_0/\mathbb{Z}_2$ base. Resolution divisors are highlighted by red dotted vertices.}}
\end{figure}
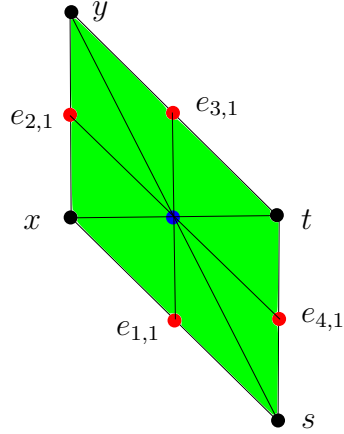
\end{center}
Upon the shown resolution,  the sections $d_i$ of the genus one curve factor out $e_{i,1}$ coordinates in the following way
\begin{align}
\begin{array}{lll}
d_1 \rightarrow e_{1,1} e_{2,1} e_{3,1} e_{4,1} \hat{d}_1,& d_2 \rightarrow e_{1,1} e_{2,1} e_{3,1} e_{4,1} \hat{d}_2, &d_3 \rightarrow \hat{d}_3
, \\ 
 d_4 \rightarrow \hat{d}_4,& d_5 \rightarrow 
 \hat{d}_5  &    d_6 \rightarrow e_{1,1} e_{2,1} e_{3,1} e_{4,1} \hat{d}_6,  \\  d_7 \rightarrow \hat{d}_7,& d_8 \rightarrow \hat{d}_8,& d_9 \rightarrow \hat{d}_9
\end{array}
\end{align}
Again we want to compute the full spectrum on this geometry but we are particular interested if there are new multiplets over the resolution divisors which we can find by inserting the above factorization into the Jacobian of $\mathbb{P}^{1,1,2}$ given in Appendix~\ref{app:cubicinWSF}. The discriminant then obtains the following form
introducing the collective notation $\mathcal{D} = \sum_i e_{i,1}$
\begin{align}
\Delta =\mathcal{D}  \,  \left(R \, \hat{d}_9 \, Q^3 + \mathcal{O}(\mathcal{D}^2)                  \right) \, ,
\end{align}
whereas $R$ and $Q$ are polynomials in the $d_i$ and $d_9 =0$ is the locus of the aforementioned SU(2) gauge symmetry.
The singlets are found\footnote{The $\mathcal{A}=Q=0$ is a $(2,3,4)$ point and carries no matter.}, where the I$_1$ fiber enhances to  I$_2$ which exactly happens for $\mathcal{D} = R = 0$.
Thus the singlets reside where also the polynomial $R$ vanishes which is explicitly given as
\begin{align}
R = -\hat{d}_4 \hat{d}_7 \hat{d}_8 + \hat{d}_3 \hat{d}_8^2 + \hat{d}_4^2 \hat{d}_9 +  \hat{d}_5 \hat{d}_7^2   - 4  \hat{d}_3 \hat{d}_5 \hat{d}_9  \;.
\end{align}
 As the resolution divisors within $\mathcal{D}$ do not intersect the SU(2) divisor in the base
 $\hat{d}_9 = 0$ we can solve the above locus $R=0$ for $\hat{d}_9$ and insert this solution over the blow up divisors $\mathcal{D}=0$ into the fiber equation \eqref{eq:quartic} using the factorization of the blow-up divisors which results in a fiber of form:
 \begin{align}
 p_{|\mathcal{A}=R=0} = 	\frac{e_1^2 Y^2 (\hat{d}_3 X^2 + Y (\hat{d}_4 X + \hat{d}_5 Y)) + 
 e_1 Y (\hat{d}_7 X + \hat{d}_8 Y) Z + ((-\hat{d}_5 \hat{d}_7^2 + \hat{d}_8 (\hat{d}_4 \hat{d}_7 - \hat{d}_3 \hat{d}_8)) Z^2)}{(
 \hat{d}_4^2 - 4 \hat{d}_3 \hat{d}_5)} \;,
 \end{align}
which indeed can be represented a reducible polynomial of the form
\begin{align}
\label{eq:singletSplit}
\hat{P} = (Z + e_1 Y(\beta_1 Y + \beta_2 X)      ) (\beta_3 Z+e_1 Y(\beta_4 Y + \beta_5 X ))
\end{align}
which admits solutions for the $\beta_i$ in terms of $\hat{d}_i$ that have a $\mathbb{Z}_2$ monodromy that interchanges the two $\mathbb{P}^1$'s around the locus $d=0$ with
\begin{align}
d= (\hat{d}_4^2 - 4 \hat{d}_3 \hat{d}_5) (-2 \hat{d}_5 \hat{d}_7 + \hat{d}_4 \hat{d}_8)^2.
\end{align}
Hence as expected  the fiber in \eqref{eq:singletSplit} splits into two $\mathbb{P}^1_{1/2}$ that are both in the same fibral homology class
\begin{align}
\mathbb{P}^1_{1/2} \in [Z] \, .
\end{align}
The multi-section generator, which can be written as
\begin{align}
\sigma_{\mathbb{Z}_4} = [Z] \, ,
\end{align}
intersects the two matter $\mathbb{P}^1_{1/2}$ curves indeed $2$ times and hence the discrete charged singlets\footnote{To be precise we have not a $\mathbb{Z}_4$ discrete symmetry but  an $(SU(2) \times \mathbb{Z}_4)/\mathbb{Z}_2$ symmetry due to an non-trivial SU(2) center \cite{ Grimm:2015wda,Cvetic:2017epq, Buchmuller:2017wpe} that mixes with the two-section.} have charge 2.
The multiplicities can again be obtained by reading off $\mathcal{S}_7$ and $\mathcal{S}_9$ 
along the conventions of \cite{Klevers:2014bqa} that are given as
\begin{align}
\mathcal{S}_7 \sim \mathcal{S}_9  =& [2 t + y + e_{3,1} + e_{4,1} +s]  \, , \ 
\end{align}
using linear equivalence. From the SRI that is easily read off from the toric diagram in Figure~ \ref{fig:dP4ResBaseA} we deduce the relevant intersections:
\begin{align}
\label{eq:example2intersections}
\begin{split}
(K_b)^2 =& K_b^{-1} \mathcal{S}_9  = 4 \, , \quad \mathcal{S}_7^2 = 2\ , \quad [e_{i,1}] \mathcal{S}_7 =1 \, ,
\end{split}
\end{align}
which is enough to compute the spectrum given in Table~\ref{eq:Example2Spectrum}. Again we remark, that the change in the intersection numbers results from the fact that $\mathcal{S}_7$ and $\mathcal{S}_9$ were non-Cartier on the orbifold base.
Let us finally return to the  I$_2$ loci of the discrete charged matter. Here we have found the polynomial $R_1$ which can be written to be in the class
\begin{align}
[R_1]  \sim [3 \mathcal{S}_7 - \mathcal{S}_9 + K_b^{-1}] \, .
\end{align}
Hence, using the intersections in \eqref{eq:example2intersections} we find 
\begin{align}
[e_{i,1}] [R_1] = 2 \, ,
\end{align}
 and thus exactly two discrete charged matter states over each of the four resolution divisors as depicted in Figure~\ref{fig:dP4ResBaseB}.

\subsubsection{Tuning an SU(2) collision}
\begin{figure}
 
\begin{picture}(0,190)
 
\put(140,20){\includegraphics[scale=0.3]{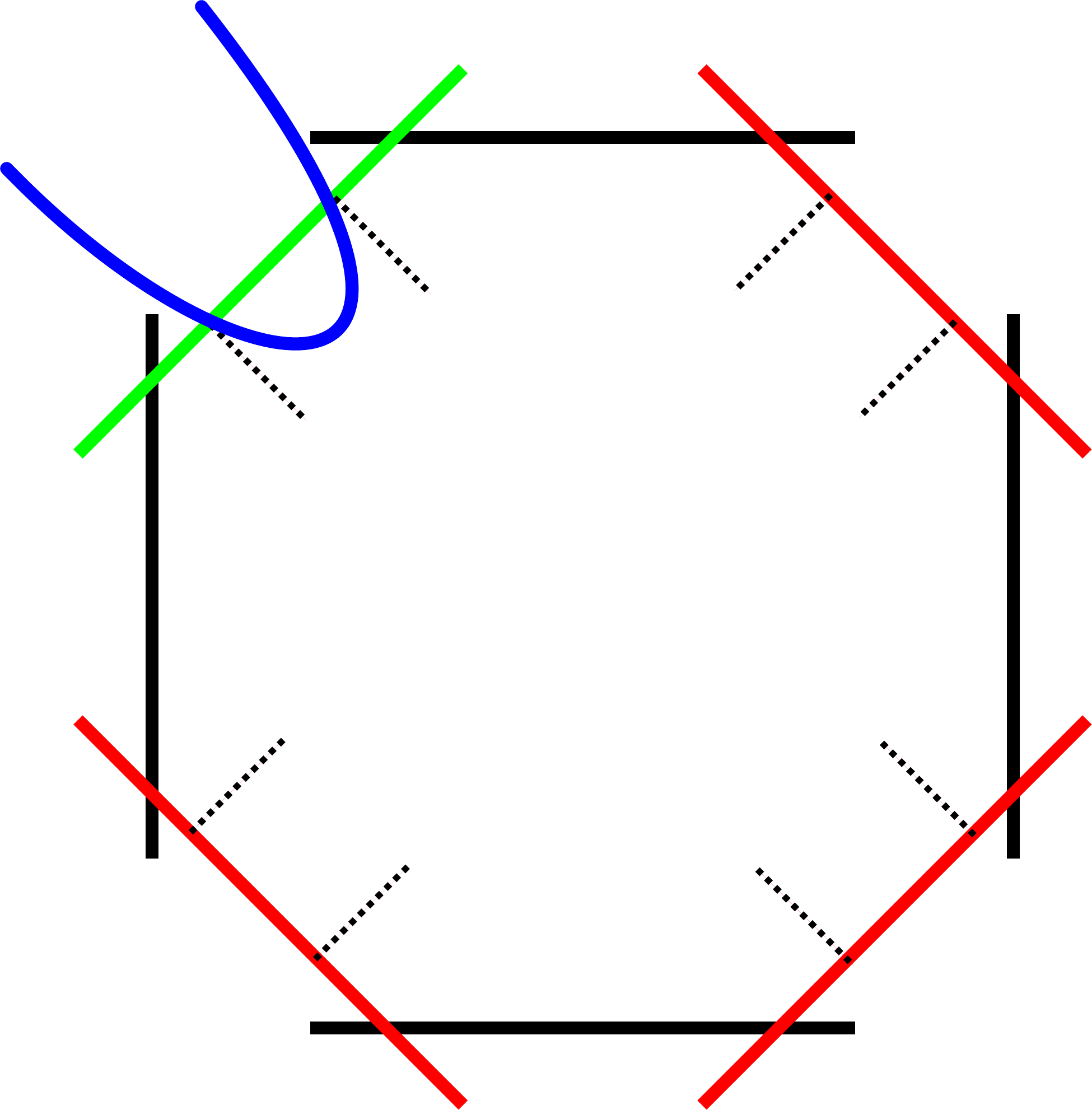}}

\put(190,128){\framebox{$(\mathbf{2},\mathbf{2})_1$}}
\put(110,120){ $SU(2)$ }
\put(110,170){ $SU(2) $ }
\put(235,144){\framebox{$\mathbf{1}_2$}}
 \put(254,125){\framebox{$\mathbf{1}_2$}}

\put(202,55){\framebox{$\mathbf{1}_2$}}
\put(183,75){\framebox{$\mathbf{1}_2$}}

\put(238,55){\framebox{$\mathbf{1}_2$}}
\put(257,75){\framebox{$\mathbf{1}_2$}}
 
\end{picture}

\caption{\label{fig:dP4ResBaseC}{\it Matter locations after the blue SU(2) curve on the green resolution $-2$ curve, which gets gauge enhanced to SU(2) as well. The former discrete charged singlet states enhance to bifundamental matter at the intersection. }
}
 
\end{figure}

In the following we want to tune in some additional singularities onto the fixed points, which we do in the fully resolved CY. Our first example is to tune the SU(2) divisor $d_9^+ = 0$ given as
\begin{align}
d_9^{+} = x^2 y^2 a_{9,1} + t^2 y^2 a_{9,2} + s^2 x^2 a_{9,3} + t^2 s^2 a_{9,4} + 
 s t x y b_9 \, ,
\end{align}
onto the $x=y=0$ fixed point by tuning $a_{4} \rightarrow 0$ in addition to the hyperconifolds that we considered above. The resulting SU(2) singularity over the -2 curve in the base is, as expected, 
over the $e_{2,1}=0$ divisor in the base and can be resolved by adding the vertex
\begin{align}
v_{e_{2,2}}=(-1, 1, -1, 1)\, .
\end{align}
Actually we can also understand this deformation as another hyperconifold, where the resolution divisors in the base do not subdivide a cone but restrict onto a divisor that was already present before as discussed in Section~\ref{sec:LensReview}. This deformation changes the geometry such that $d_9=0$ becomes a genus 0 curve of self intersection $-2$. Moreover also the two discrete charged singlets on $e_{3,1}=0$ are now gauge enhanced to bifundamentals, as $d_9=0$ intersects $e_{3,1}=0$ two times. The full spectrum
is given in the following table and is fully consistent with all anomalies and depicted in Figure~\ref{fig:dP4ResBaseC}.
\begin{align}
\begin{array}{c|c}
$ $6$-d Rep. $& $Multi.$ \\ \hline
(\mathbf{2},\mathbf{1})_1 & 24 \\
(\mathbf{2},\mathbf{2})_1 & 2 \\
(\mathbf{1},\mathbf{1})_2 & 46 \\ \hline
(\mathbf{1},\mathbf{1})_0 & 32 \\
\mathbf{V} & 6 \\
\mathbf{T}_{(1,0)} & 5 
\end{array} \, .
 \end{align}

\subsubsection{The U(1) un-Higgsed theory}
Another phase can be obtained by tuning in a section which un-Higgses the $\mathbb{Z}$ to a U(1). This can be achieved tuning $d_5 \rightarrow 0$ which amounts to set the residual $b$ coefficient in Equation~\eqref{eq:U1CoeffExample2} and resolve. In terms of the ambient space we add the vertex 
$v = ( 0, 1,  0,  0)$ which blows-up the fiber ambient space to $BL_1 \mathbb{P}^{1,1,2}$ which is the prototype of an elliptic fibration with Mordell-Weil rank one. The full CY hypersurface $\widehat{Y}$ has the following Hodge numbers
\begin{align}
(h^{1,1}(\widehat{Y}), h^{2,1}(\widehat{Y}))_\chi = (9,31)_{-44} \, ,
\end{align}
as expected. Again, this theory admits two hypers that have charge $q=4$ under the U(1) whose VEV triggers the Higgsing to the discrete symmetry when performing the conifold. The spectrum is free of all anomalies which can be checked by incorporating the new height pairing  $b_{11}$ with the class
\begin{align}
\begin{split}
b_{11} =& \frac32 K_b^{-1} + \frac52 \mathcal{S}_7 - \frac12 \mathcal{S}_9 \, ,\\ 
=& [7 x + 2 t + \frac72 e_{3,1} + 5 e_{4,1} + 5 s + \frac32 e_{1,1}] \, .
\end{split}
 \end{align}
 Indeed we find that the height pairing contains fractional parts of resolution divisors and is not Cartier when we go back to the singular base. Hence again this model opens up the possibility to construct a strongly coupled sector with U(1) charged superconformal matter. 

 \subsection{Example 3: Threefold  in $(F_0 \times \mathbb{P}^{1,1,2})/\mathbb{Z}_2$ }
 \label{sec:example3}
The final example is the same geometry as we discussed in the proceeding section, but this time we have switched the fiber and base ambient spaces\footnote{From the ingredients above, one could also easily have considered the non-simply connected threefolds in $(F_0 \times F_0)/\mathbb{Z}_2$ and $(P^{1,1,2} \times \mathbb{P}^{1,1,2})/\mathbb{Z}_2$ \cite{Batyrev:2005jc}.}. This time we have a $\mathbb{Z}_2 \times U(1)$ gauge symmetry coupled to four $\mathcal{A}_1$ (2,0) theories. In this model there are four types of hypermultiplets in the spectrum distinguished by their $U(1) \times \mathbb{Z}_2$ charge where only the $\mathbb{Z}_2$ charged hypers  appear on the tensor branch as expected. Moreover we find, that a collection of three $-2$ curves can actually be shrunken to an $\mathcal{A}_3$ (2,0) theory where exactly four discrete charged singlets disappear  consistent with the general picture. Finally we show that in order to tune in a section, one must also necessarily enhance the gauge symmetry by another SU(2) over one of the $-2$ curves.
 
 \subsubsection{The geometric setup}
 The geometry we consider is actually the same as in Section~\ref{sec:example2} however we consider a different $GL(4,\mathbb{Z})$ frame of the polytope $\Delta$ to make the projection to the base, that is $\mathbb{P}^{1,1,2}/\gamma_2$, more evident. In the new frame the polytope is given by the vertices:
  \begin{align}
\begin{tabular}{cccc|cc cc}
       $x$ & $t$ & $y$ & $s$  & $Z$   & $e_1$  &  $X$ & $Y$       \\ \hline
 -1 & 1 & 0 & 0 & 1 & -1 & 0 & -2 \\
 0 & 0 & 1 & -1 & 1 & -1 & -2 & 0 \\
 0 & 0 & 0 & 0  & 1 & -1 & -1 & -1 \\
 0 & 0 & 0 & 0  & 0 & 0  & 2  & -2 \\ 
\end{tabular} \, .
\end{align}
Here we have the same  $\mathbb{C}^*$ and $\gamma_2$ identifications as before with the same fixed points and Hodge numbers as given in Equation \eqref{eq:HodgeEx2}. The genus one fiber is now described as the vanishing of a biquadric equation
 \begin{align}
 \label{eq:biquadric}
p=  \left( b^{(+)}_1 y^2 + b^{(-)}_2 sy + b^{(+)}_3 s^2  \right)x^2 
+ \left( b^{(-)}_5 y^2 + b^{(+)}_6 sy + b^{(-)}_7 s^2  \right)t x 
+ \left( b^{(+)}_8 y^2 + b^{(-)}_9 sy + b^{(+)}_{10} s^2  \right)t^2 
 \end{align}
 with the $b_i$ being non generic sections in the anticanonical class of the base such that they transform under the $\Gamma_{2,b}$ action as highlighted by their superscript. The explicit expressions can be found in  Appendix~\ref{app:SectionsExample3} which shows that all odd sections vanish over any base fixed point. The base $\mathbb{P}^{1,1,2}/\mathbb{Z}_2$ is again identified by the projection $\pi$ onto the last two coordinates of the ambient space polytope $\Delta$ given above. 
 \subsubsection{Spectrum and hyperconifold tensor branch}
 \label{sec:Example3Conifold}
The presented model admits a $U(1) \times \mathbb{Z}_2$ gauge symmetry \cite{Klevers:2014bqa} as well as four $\mathcal{A}_1$ (2,0) points coupled to the discrete symmetry. The spectrum admits three kinds of fiber degenerations, corresponding to singlets of charges $\mathbf{1}_{(1,+)}, \mathbf{1}_{(1,-)}$ and in particular purely discrete charged singlets $\mathbf{1}_{(0,-)}$.
 \begin{table}[h]
 \begin{center}
 \begin{tabular}{|c|c||c|c|c|c|}
   \multicolumn{6}{c}{Multiplicities} \\ \cline{2-6}  
 \multicolumn{1}{c|}{}  &\multirow{2}{*}{ Generic Base}& \multicolumn{4}{c|}{Ambient Space Geometry} \\ \cline{3-6}  
 \multicolumn{1}{c|}{State}  & &   $  \mathbb{F}_0 \times \mathbb{P}^{1,1,2} $  &$\frac{( \mathbb{F}_0 \times \mathbb{P}^{1,1,2})}{\mathbb{Z}_{2}}  $  &$ \frac{( \mathbb{F}_0 \times \mathbb{P}^{1,1,2}  )}{\mathbb{Z}_{2}}^{\text{HC}}$   &    {\footnotesize $(\mathbb{F}_0 \times F_{13})$}   \\ \hline 
  $ H_{\mathbf{1}_{(0,-)}}$  & $\begin{array}{l} 4 K_b^{-1}( \mathcal{S}_7+  \mathcal{S}_9)+\\ 
  6 (K_b^{-1})^2 -   (\mathcal{S}_9+ \mathcal{S}_7 ) \end{array}$ &  80 & 40&48 & 40  \\ \hline
$H_{1_{(1,-)}}$& $\begin{array}{l}  4K_b^{-1} (\mathcal{S}_9 -\mathcal{S}_7)+   \\6(K_b^{-1})^2  +2 \mathcal{S}_7^2 -2 \mathcal{S}_9^2 \end{array}$
& 48 & 24 & 24 & 24\\ \hline
$H_{1_{(1,+)}}$ & $\begin{array}{l}  4K_b^{-1} (\mathcal{S}_9 -\mathcal{S}_7)  + \\6(K_b^{-1})^2  -2 \mathcal{S}_7^2 +2 \mathcal{S}_9^2 \end{array}$ & 48 & 24 & 24& 24\\ \hline
$H_{1_{0}}$ & - & 69 & 37 & 33 & 41 \\ \hline
V & 1 & 1 & 1 & 1 & 1 \\
$T_{(1,0)}$   & $h^{1,1}(B)-1$               & 1 & 1 & 5 & 5 \\
$T_{(2,0)} $  & $10-h^{1,1}(B)-K_b^{2}$ & 0 & 4 & 0  & 0 \\ \hline
 \end{tabular}
 \caption{
 \label{tab:Example3Spectrum}{\it Spectra of four genus one fibrations with $ U(1) \times \mathbb{Z}_2$ gauge group and their ambient spaces. We compare covering geometry, quotient, hyperconifold tensor branch and highlight the change in spectrum. This is contrasted to the spectrum of the tensor branch of a regular $A_1$ theory in the last column. }
 }
 \end{center}
 \end{table}
 \\
In Table~\ref{tab:Example3Spectrum} we summarize the general formulas for the spectrum computation as well as the concrete values for four CY threefold ambient spaces and their F-theory spectra. We list the covering CY, the quotient geometry and its hyperconifold resolution. The last
column shows the tensor branch spectrum of a trivial fibration, where the discrete symmetry is not coupled to the (2,0) points and where eight discrete charged hypers are missing.\\
 It is readily checked that for all theories above all anomalies are Green-Schwarz canceled. In addition to the gravitational anomalies,
 we repeat the U(1) anomalies here
 \begin{align}
 \begin{split}
 \begin{array}{rll}
\text{grav}^2 U(1)^2&:\qquad  -\frac16  \sum_i H_{\mathbf{1}_q} q^2 &=  a \cdot b_{11}  \\  
  U(1)^4&:\qquad \phantom{-} \frac13  \sum_i H_{\mathbf{1}_q} q^4  &=   b_{11} \cdot b_{11} 
  \end{array}
  \end{split}  
\end{align}
 where the anomaly coefficient on the right hand side can be deduced from the anticanonical class of the base
 and the U(1) height pairing \cite{Klevers:2014bqa}  that are
 \begin{align}
 a = K_b \, , \qquad b_{11} = 2 K_b^{-1} \, .
 \end{align}
   The multiplicities of the charged matter states can again be computed using the formulas in \cite{Klevers:2014bqa} and the identification of the classes $\mathcal{S}_7, \mathcal{S}_9$ and $ K_b^{-1}$. For any base those classes can be taken from the line bundle classes of the genus one fiber \eqref{eq:biquadric} as
   \begin{align}
   [b_{6}^{(+)}] \sim K_b^{-1} \, , \qquad [b_7^{(-)}] \sim \mathcal{S}_7\, , \qquad [b_9^{(-)}] \sim \mathcal{S}_9 \, .
   \end{align}
The $\mathcal{S}_{7}$ and $\mathcal{S}_9$ do not descent from $\Gamma_{2,b}$ invariant classes of $B_\text{cov}$ but covariant ones, with the same degree as $K_b^{-1}$. Thus, by abuse of notation we consider them as the same, keeping the difference in mind and use the self intersection $(K_b)^{-1}=(8)4$ for the (un-)quotiented case. Using the formulas in the first column of Table~\ref{tab:Example3Spectrum}, the full charged spectrum of the covering, quotient and direct product CY can easily be computed.
   
   \begin{figure}[h!]
 
\hspace{2cm} \begin{picture}(0,180)
\put(150,10){\includegraphics[scale=0.25]{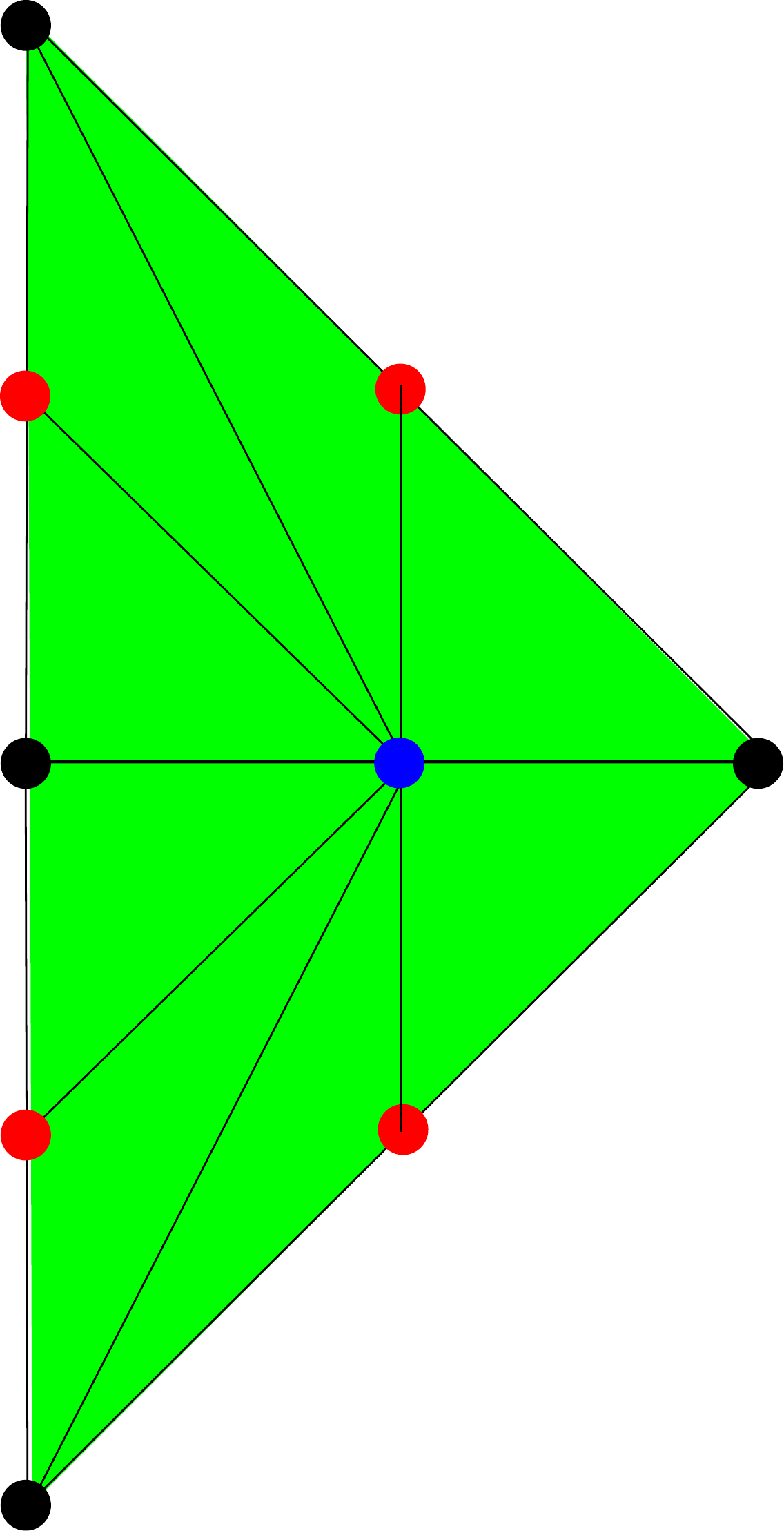}}

  \put(135,165){$X$}
 
 \put(200,130){$ B_x  $}

  \put(240,85){$  Z $}
  
    \put(200,50){$  B_y $}
    
      \put(135,10){$Y$}

   \put(132,45){$  A_y $}

        \put(132,125){$  A_x $}
            \put(135,85){$ e_1 $}

\end{picture}
 
\caption{ \label{fig:p112ResBase}{\it  The polytope of the resolved $\mathbb{P}^{1,1,2}/\mathbb{Z}_2$ base. Red doted vertices represent the additional resolution divisors.}}
 \end{figure}
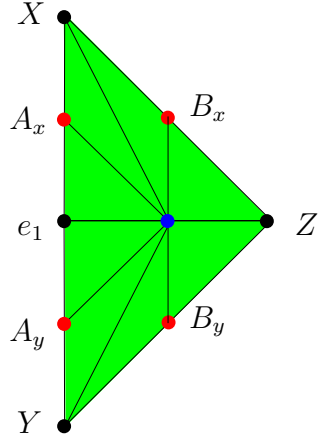 
   
To obtain the spectrum of the hyperconifold tensor branch of the quotient geometry, we proceed by tuning ambient space fixed points onto the CY and resolving. We choose to fix a fiber fixed point and tune in the four base fixed points that we resolve by adding the four additional coordinates $A_x, A_y, B_x,B_y$ which amounts to four blow-ups of the ambient variety, with polytope $\Delta$ given as
   \begin{align}
\begin{tabular}{cccc|cc cc| cccc}
       $x$ & $t$ & $y$ & $s$  & $Z$   & $e_1$  &  $X$ & $Y$ &$A_x$ & $A_y$ & $B_x$ & $B_y$       \\ \hline
 -1 & 1 & 0 & 0 & 1 & -1 & 0 & -2  & -1 & -2  & -1 &0     \\
 0 & 0 & 1 & -1 & 1 & -1 & -2 & 0 & -1 & 0 & 1 &0\\
 0 & 0 & 0 & 0  & 1 & -1 & -1 & -1 & -1 & -1 & 0& 0 \\
 0 & 0 & 0 & 0  & 0 & 0  & 2  & -2 &  1& -1 & -1 & 1 \\ 
\end{tabular} \, .
\end{align}
Choosing a triangulation yields a Stanley-Reisner ideal 
\begin{align}
\begin{split}
SRI: \{& 
t x, t A_x, t A_y, t B_y, t B_x, y s, Z e_1, Z A_x, Z A_y, X Y, X A_y, X B_y, Y B_x, e_1 A_y,\\ & e_1 B_y, e_1 B_x, s A_x, s A_y, s B_y, s B_x, A_x B_y, B_y B_x, A_y B_x, x y X, x y Y, x y e_1
\} \, ,
\end{split}
\end{align}
and a CY hypersurface with Hodge numbers
\begin{align}
(h^{(1,1)},h^{(2,1)})_{\chi} = (8,32)_{-48} \, .
\end{align}
 After the standard projection down onto the last two coordinates of $\Delta$, we find indeed the base to be that of polytope $F_{13}$ as shown in Figure~\ref{fig:p112ResBase}, the resolved orbifold of $\mathbb{P}^{1,1,2}$. 
 After performing the above steps, we find a factorized biquadric as in Equation~\eqref{eq:biquadric}
 with base sections given in Equation~\eqref{eq:sectionsExample3b} of Appendix~\ref{app:SectionsExample3}.
The hyperconifold resolution leads to factorized base sections $b_i$ that are of the form
\begin{align}
\label{eq:factorblexample3}
\begin{array}{lllll}
b_1 = A_x A_y B_x B_y  \hat{b}_1 \, ,    & b_3 = \hat{b}_3              \, ,                  &b_6 = \hat{b}_6\, , &b_8 = \hat{b}_8 \, , &b_{10} = e_1 X Y \hat{b}_{10} \, , \\ 
b_2= A_x A_y B_x B_y e_1 \hat{b}_2 \, ,&  b_5 = A_x A_y B_x B_y e_1 \hat{b}_5\,  , &b_7 = e_1        \hat{b}_7\, , &b_9 = e_1  \hat{b}_9\, ,  &  
\end{array}
\end{align}
with the $\hat{b}_i$ being some residual polynomials, spelled out in detail in Equation~\eqref{eq:sectionsExample3b}. 

In the convention of \cite{Klevers:2014bqa}  the base classes $\mathcal{S}_7$ and $\mathcal{S}_9$ are given by the base classes of the polynomials $b_7$ and $b_9$ respectively that are read off to be linear equivalent to
\begin{align}
\label{eq:secExample3}
\begin{split}
\mathcal{S}_7 \sim \mathcal{S}_9 \sim [2 Z + Y - A_x - X + B_y]  \, ,\\
K_{b}^{-1} \sim [2 Z + B_x + B_y] \, .
\end{split}
\end{align}
These last quantities have intersections
 \begin{align}
 (K_b^{-1})^2 = \mathcal{S}_{7} K_b^{-1} = 4\, , \qquad \mathcal{S}_7^2 = 2 \, ,
 \end{align}
 which can be checked by using the intersection relations as read off from the toric diagram in Figure~\ref{fig:p112ResBase}. Thus we have all information needed in order to compute the multiplicity of all states by inserting them into the general formulas, see Table~\ref{tab:Example3Spectrum}. 
 
  \begin{figure} 
  \hspace{0.6cm}
\begin{picture}(50,170)
 
\put(0,20){\includegraphics[scale=0.55]{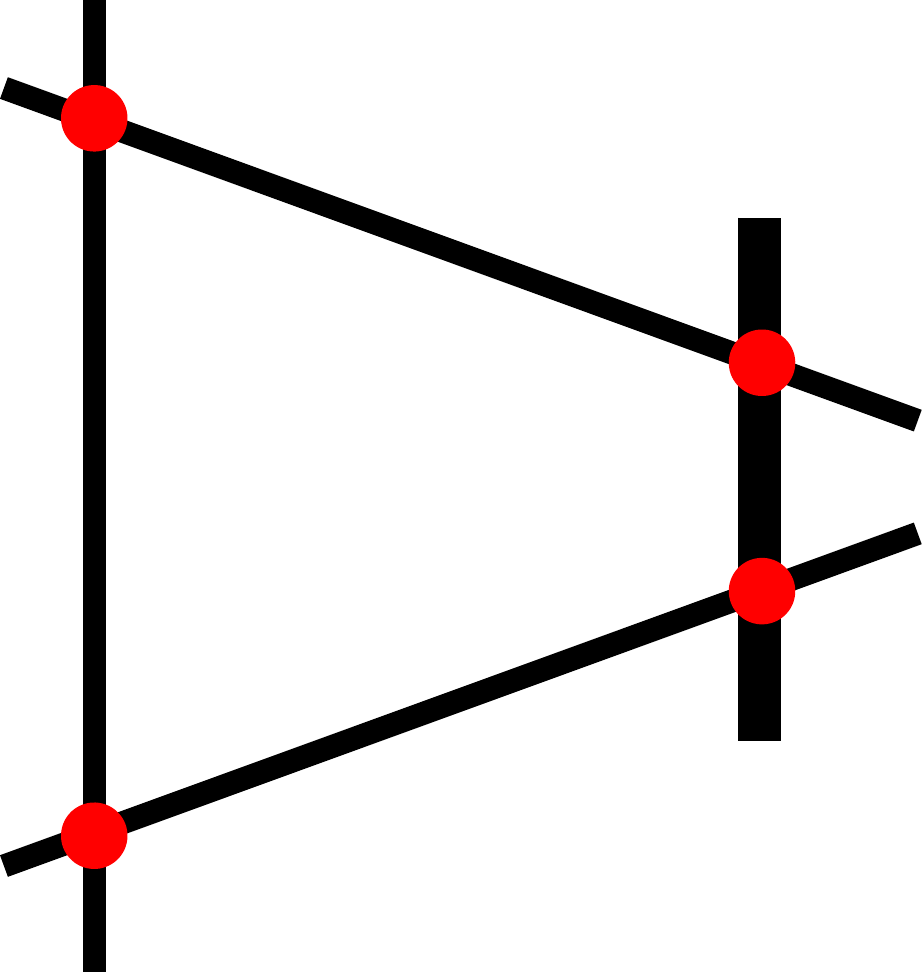}}

\put(240,20){\includegraphics[scale=0.4]{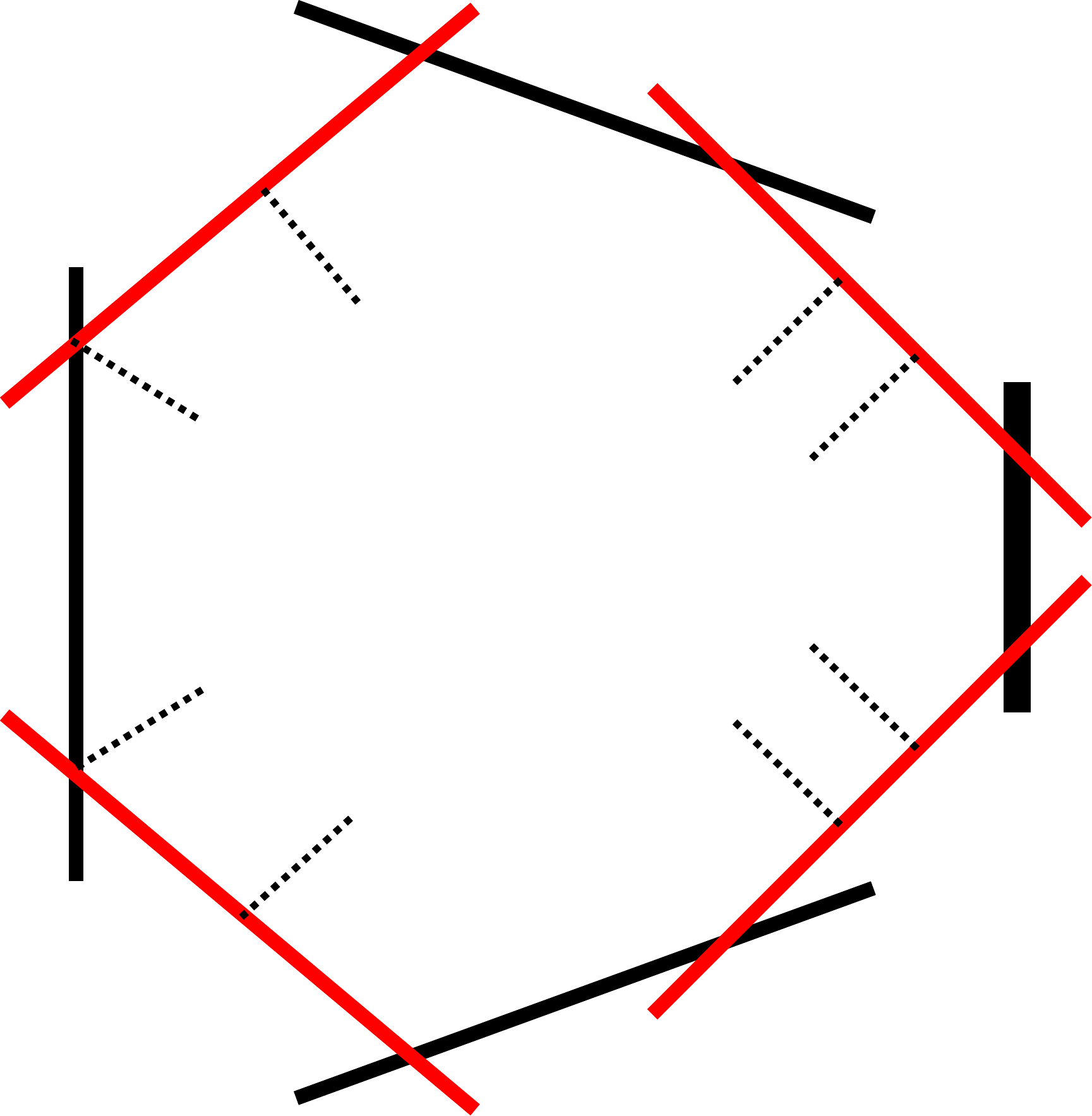}}

\put(305,165){\framebox{$\mathbf{1}_{(0,-)}$}}
\put(276,145){\framebox{$\mathbf{1}_{(0,-)}$}}

\put(340,154){\framebox{$\mathbf{1}_{(0,-)}$}}
 \put(354,135){\framebox{$\mathbf{1}_{(0,-)}$}}
 
\put(-10,90){ $D_{e_1}$ }
\put(122,95){ $D_Z$ }
\put(60,120){ $D_X$ }
\put(60,45){ $D_Y$ }

\put(276,96){\framebox{$\mathbf{1}_{(0,-)}$}}

 \put(354,106){\framebox{$\mathbf{1}_{(0,-)}$}}
 \put(340,87){\framebox{$\mathbf{1}_{(0,-)}$}}
 
 \put(305,75){\framebox{$\mathbf{1}_{(0,-)}$}}

\end{picture}

\caption{\label{fig:dP4ResBase}{\it Intersections of toric divisors for singular and resolved $\mathbb{P}^{1,1,2}/\mathbb{Z}_2$ base. In red we denote the orbifold fixed points and their resolution divisors. Over every resolution divisor, additional discrete charge states appear.}}
\end{figure}
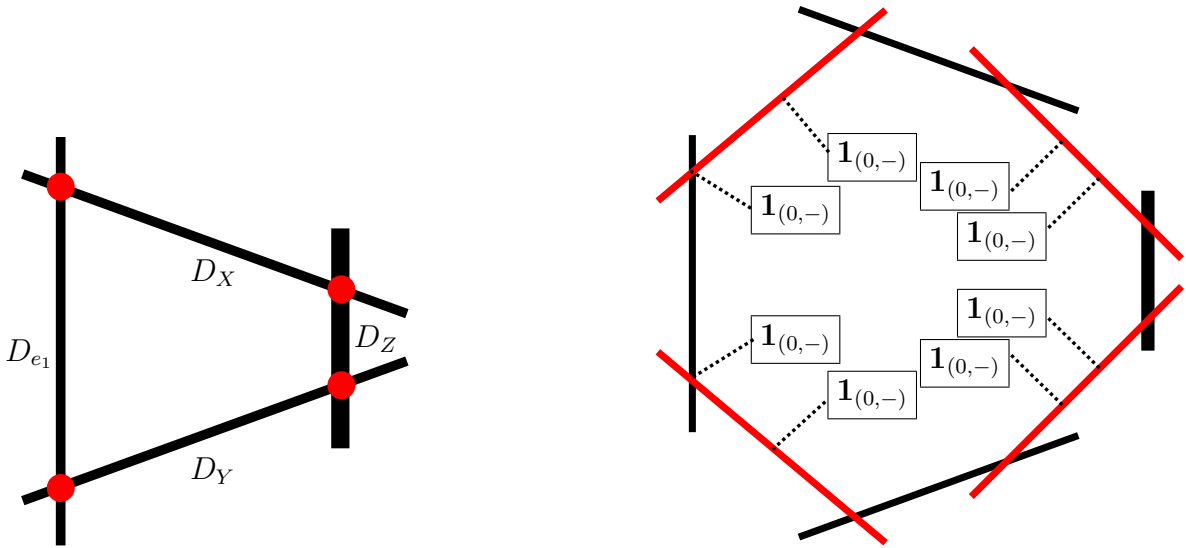 
 As a last point we want to consider the states that appear over the four resolution divisors. In order to do so, we 
 consider the Jacobian of the biquadric given in Appendix~\ref{app:cubicinWSF} and insert the factorization  \eqref{eq:factorblexample3}. 
We factor the discriminant with respect to $\mathcal{A} = A_x A_y$ and $\mathcal{B} = B_x B_y$ as 
 \begin{align}
\label{eq:DiscExample3}
 \Delta =\mathcal{A} \mathcal{B} e_1 \left(\hat{b}_1 \hat{b}_3 \hat{b}_8 P_1 Q_1^3 \mathcal{B} + \mathcal{O}((\mathcal{A} \mathcal{B})^2...   \right) \, .
 \end{align}
 $Q_1$ and $P_1$ define polynomials that signal enhanced codimension two loci where matter resides, whereas $Q_1 = \mathcal{A} \mathcal{B}=0$ denotes a $(2,3,4)$ locus which does not lead to additional matter degrees of freedom. The polynomial $P_1$ is given as
  \begin{align}
 \label{eq:P1Example3}
 P_1= -\hat{b}_7^2 \hat{b}_8 e_1 + \hat{b}_6 \hat{b}_7 \hat{b}_9 e_1 - \hat{b}_3 \hat{b}_9^2 e_1 - \hat{b}_{10} \hat{b}_6^2 X Y + 
 4 \hat{b}_{10} d_3 \hat{b}_8 X Y\, ,
 \end{align}
 which defines an I$_2$ fiber together with $\mathcal{A},\mathcal{B}=0$ in addition to the toric locus $\mathcal{A}=e_1 =0$.
 For completion we confirm the factorization of the fiber into two degree (1,1) curves in terms of the $\mathbb{F}_0$ ambient space classes of the fiber. Indeed over those loci, the fiber reduces to
 \begin{align}
 \begin{split}
 p_{\mathcal{A} = e_1 =0} =& \hat{d}_3 s^2 x^2 + \hat{d}_6 s t x y + \hat{d}_8 t^2 y^2 \\
=& (\beta_1 sx +  \beta_2 ty)(\beta_3 sx + \beta_4 ty) \, ,
\end{split}
  \end{align}
where the last factorization admits solutions of the $\beta_i$ in terms of the $d_i$. Hence we find that the fiber reduces to two $\mathbb{P}^1$'s that get interchanged by a monodromy around the $Q_1 = 0$ locus.  The charges of the state under $U(1) \times \mathbb{Z}_2$ we compute by intersecting one choice of component of the reducible fiber with the U(1) and $\mathbb{Z}_2$ generators that are given \cite{Klevers:2014bqa} as
  \begin{align}
  \sigma(s_1) = [y]-[x] \, , \qquad \sigma_{\mathbb{Z}_2} = [x] \, .  
    \end{align}
Using the familiar intersection relations of $\mathbb{F}_0$ we find for each choice of component of the reducible fiber a state with charge $\mathbf{1}_{(0,-)}$ where the $-$ denotes charge 1 under the $\mathbb{Z}_2$ symmetry. The states at the loci $\mathcal{A},\mathcal{B} = P_1 =0$ are exactly of the same type which, however, is harder to see. Things however get more transparent when we tune in a section which amounts to an un-Higgsing of $\mathbb{Z}_2 \rightarrow U(1)$ which can be achieved by tuning $\hat{b}_{10}\rightarrow 0$. In that case we have an elliptic fibration with Mordell-Weil rank two \cite{Cvetic:2013nia,Borchmann:2013hta,Braun:2013nqa}. We consider the geometry in more detail in the next section however now we simply use that a singlet of of charge $\mathbf{1}_{(0,1)}$ is found at the vanishing locus $V(I_s)$
\begin{align}
\begin{split}
I_s = \{ &b_1 b_9^4 b_7^2 + (b_3 b_9^2 + b_7 (-b_6 b_9 + b_8 b_7)) (b_3 b_8 b_9^2 + 
     b_7 (-b_6 b_8 b_9 + b_8^2 b_7 + b_9^2 b_5)), \\
& b_2 b_9^3 b_7^2 + 
  b_3^2 b_9^4- b_3 b_6 b_9^3 b_7-   b_7^3 (-b_6 b_8 b_9 + b_8^2 b_7 + b_9^2 b_5) \} 
  \end{split}
 \end{align} 
 which vanishes precisely for $\mathcal{A},\mathcal{B}=P_1=0$ after inserting the factorization \eqref{eq:factorblexample3}. Hence after Higgsing the second U(1) factor those become the desired discrete charged singlets. The multiplicities we compute by using the homology class of $P_1$, which is
 \begin{align}
 [P_1] \sim [2 K_b^{-2} + X +Y +Z  ] \, , 
 \end{align}
 which admits the following intersections
 \begin{align}
 \label{eq:Example3P1Intersection}
[ \mathcal{A}] [e_1] = [\mathcal{A}] [P_1] = 1 \, ,\qquad [\mathcal{B}] [P_1] = 2 \, , \qquad [e_1] [P_1] = 0 \, .
 \end{align}
 This results in exactly two discrete charged singlets over each$-2$ curves as depicted in Figure~\ref{fig:dP4ResBase}.
\subsubsection{The $U(1)^2$ un-Higgsed theory}
We conclude this section with some final remarks on the un-Higgsing of the $\mathbb{Z}_2$ by tuning $\hat{b}_{10} \rightarrow 0$. In the geometry above, the polynomial $\hat{b}_{10}$ admits the form
\begin{align}
\label{eq:b10Example3}
\hat{b}_{10} =  \left( A_x A_y e_1 X Y b  + Z c \right) \, ,
\end{align}
with $b,c$ being the complex coefficients that we need to tune to zero. 
Tuning $c \rightarrow 0$ leads to another factorization in $P_1 \rightarrow e_1 \hat{P_1}$ of  Equation~\eqref{eq:P1Example3} and therefore to another factorization of the discriminant \eqref{eq:DiscExample3} as 
\begin{align}
\Delta =\mathcal{A} \mathcal{B} e_1^2 \left (\hat{b}_1 \hat{b}_3 \hat{b}_8 \hat{P}_1 Q_1^3 \mathcal{B} + \mathcal{O}(\mathcal{A} \mathcal{B})^2...   \right) \, ,
\end{align}
with the polynomial
\begin{align}
\hat{P}_1 = \hat{b}_7^2 \hat{b}_8 - \hat{b}_6 \hat{b}_7 \hat{b}_9 + \hat{b}_3 \hat{b}_9^2\,. 
\end{align}
 This is an SU(2) singularity over $e_1 = 0$, which can be resolved torically by adding the vertex $v_{1,1}=(-2,0,-1,0)$ with corresponding coordinate $e_{1,1}$ to the polytope inducing another K\"ahler parameter and reducing one complex structure degree of freedom\footnote{This is an example of another hyperconifold resolution whose new divisor restricts onto a divisor in the base that was already present.}. Consequently the former discrete charged singlet loci get enhanced from I$_2$ to I$_3$ fibers that are present over
 \begin{align}
 e_1 = \{ A_x =0, A_y = 0, \hat{P}_1 = 0 \} \, .
 \end{align}   
 Over the $A_{x/y}=0$ loci, we find two neutral doublets, whereas over $\hat{P}_1=0$ we find
discrete charged doublets as we will describe in detail in the following. 

\begin{figure}
   \hspace{-3cm}
\begin{picture}(50,170) 
\put(240,20){\includegraphics[scale=0.4]{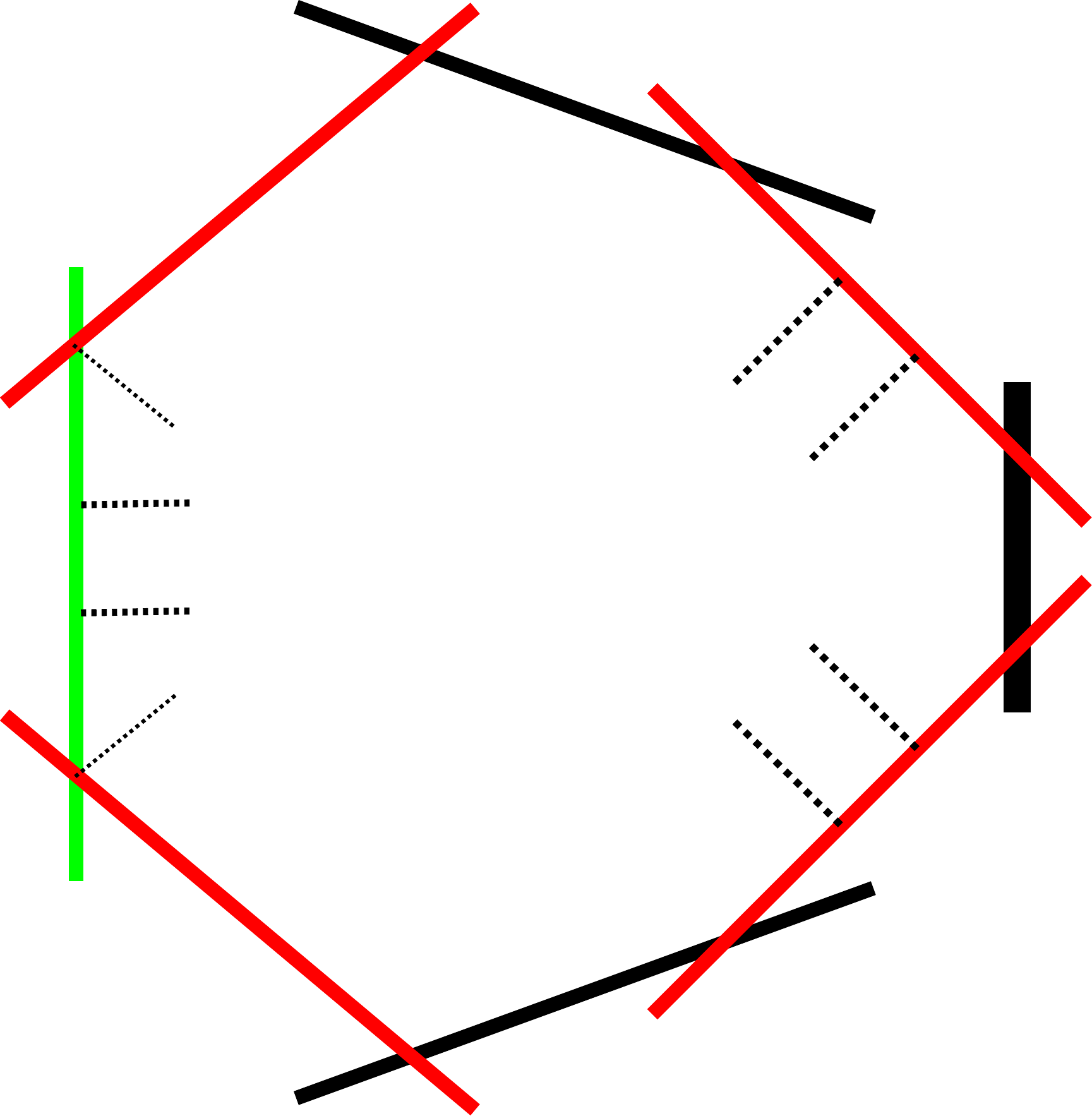}}

 
\put(340,154){\framebox{$\mathbf{1}_{(0,-)}$}}
 \put(354,135){\framebox{$\mathbf{1}_{(0,-)}$}}
 
\put(272,145){\framebox{$\mathbf{2}_{(0,+)}$}}
\put(272,95){\framebox{$\mathbf{2}_{(0,+)}$}}

\put(276,120){\framebox{$\mathbf{2}_{(0,-)}$}}

\put(210,120){$SU(2)$}

 \put(354,106){\framebox{$\mathbf{1}_{(0,-)}$}}
 \put(340,87){\framebox{$\mathbf{1}_{(0,-)}$}}
 

\end{picture}
\caption{\label{fig:tunedp112base}Matter locations over the resolved $\mathbb{P}^{1,1,2}/\mathbb{Z}_2$ base after a partial tuning. The green $-2$ curve admits an enhanced SU(2) symmetry and the former discrete charged singlets have been enhanced to fundamentals.     }
 \end{figure}
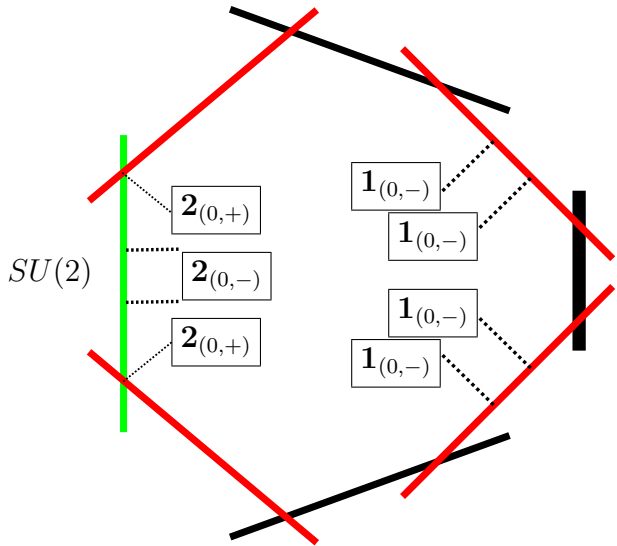

 Due to the factorization, the $\hat{P}_1$ divisor is linear equivalent to
\begin{align}
[\hat{P}_1] \sim [P_1 - e_1] \, ,  
\end{align}
which, after using \eqref{eq:Example3P1Intersection} and $[e_1][e_1]=-2$ leads to the following multiplicities
\begin{align}
\label{eq:SU2U1Z2spectrum}
\begin{array}{c|c}
 $ $6$-d Rep. $& $Multiplicity$ \\ \hline
\mathbf{2}_{(0,-)} & 2 \\
\mathbf{2}_{(0,+)} & 2 \\ \hline
\mathbf{1}_{(0,-)} & 44 \\
\mathbf{1}_{(1,+)} & 24 \\
\mathbf{1}_{(1,-)} & 24 \\ \hline
\mathbf{1}_{(0,+)} & 32 \\  
\mathbf{V} & 4 \\
\mathbf{T}_{(1,0)}& 5  
\end{array} \, ,
 \end{align}  
 which is again consistent with all anomalies. The location of the matter is depicted in Figure~\ref{fig:tunedp112base}.
 
 As a final step, we can also perform the conifold $b\rightarrow 0$ in \eqref{eq:b10Example3} which is resolved by adding the vertex $v_{2,1} = (-1,1,0,0)$ with corresponding coordinate $x_{2,1}$. As this vertex lies in the pre-image over a generic point on the base, the ambient space of the generic fiber can be modified to that of dP$_2$. The generic elliptic fiber in $dP_2$ \cite{Borchmann:2013hta, Braun:2013nqa,Cvetic:2013nia} admits a MW group of rank two and therefore corresponds to the un-Higgsing of the $\mathbb{Z}_2$ symmetry. In addition, we can use the corresponding divisor of $x_{2,1}=0$ to construct a new U(1) divisor as
\begin{align}
   \sigma(s_2) =[x_{2,1}] -  [x] \, .
\end{align}    
  The reduction in the complex structure by one and the additional U(1) vector is compensated in the gravitational anomaly by two singlets in the representation $\mathbf{1}_{(0,2)}$ whose vevs trigger the Higgsing. As a consistency check, we recompute the full massless spectrum in the $SU(2) \times U(1)^2$ theory which is given as
\begin{align}
\label{eq:SU(2)U12spectrum}
\begin{array}{c|c}
 $ $6$-d Rep. $& $Multiplicity$ \\ \hline
\mathbf{2}_{(0,0)} & 2 \\
\mathbf{2}_{(0,1)} & 2 \\ \hline
\mathbf{1}_{(1,-1)} & 4 \\
\mathbf{1}_{(1,1)} & 20 \\
\mathbf{1}_{(-1,-2)} & 4 \\  
\mathbf{1}_{(1,0)} & 20 \\  
  \mathbf{1}_{(0,2)} & 2 \\  
  \mathbf{1}_{(0,1)} & 44 \\  \hline
\mathbf{1}_{(0,0)} & 31 \\  
\mathbf{V} & 5 \\
\mathbf{T}_{(1,0)}& 5  
\end{array} \, ,
\end{align} 
and is fully consistent with the Higgsed multiplicities in \eqref{eq:SU2U1Z2spectrum}. Over the
loci  $A_{x/y} = e_1 = 0$ we find the fibral divisor $e_{1,1}$ of the SU(2) to split as a perfect square
\begin{align}
p = \hat{b}_3 x^2 s^2 + \hat{b}_6 y x s t + \hat{b}_8 y^2 t^2 \equiv (\beta_1 xs+ \beta_2 y t )(\beta_3 xs+ \beta_4 y t ) \, ,
\end{align}
where the coefficients $\beta_i$ can be expressed in terms of the $\hat{b}_i$ involving square root factors. Both $\mathbb{P}^1$ factors are homological equivalent and get exchanged by monodromies in the base around the square root factors in the $\beta_i$. Therefore both states must have vanishing $U(1)^2$ quantum numbers. For the $e_1 = \hat{P}_1$ locus on the other hand we do not find a solution which interchanges the monodromy in a similar way, such that we conclude the charges as given in \eqref{eq:SU(2)U12spectrum}.
 
Finally we turn to the anomaly coefficients of the U(1) generators that have the form \cite{Klevers:2014bqa} 
\begin{align}
b_{mn}= \left(\begin{array}{cc}      2 K_b^{-1} & K_b^{-1} + \mathcal{S}_9 - \mathcal{S}_7         \\
K_b^{-1} + \mathcal{S}_9 - \mathcal{S}_7 & 2 (K_B^{-1} +  \mathcal{S}_9 -  [e_1]) \end{array}
   \right)_{mn} \, .
\end{align}
In order to be consistent with all anomalies, the $b_{22}$ height pairing of the Unhiggsed U(1)  we had to shift $b_{22}$ by $-2 [e_1]$ . From \eqref{eq:secExample3} we find that the height pairing of the second U(1) is equivalent to
\begin{align}
b_{22} \sim [2 (3 Z + A_x + 2 X + 2B_x + B_y)] \, .
\end{align}
Thus, in distinction to the other entries, that are proportional to $K_b^{-1}$ and are thus Cartier when taking the limit to a singular base, the $b_{22}$ coefficient is non-Cartier. \\\\
In summary we find various interesting effects in this example that are worth studying further: First we find exactly eight purely discrete charged singlets appearing on the tensor branch of four $A_1$ singularities where two of them are connected by another $-2$ curve. Thus we have a chain of three $-2$ curves that can be collapsed\footnote{Note that the resulting threefold is not smooth.} to an $\mathbb{Z}_4$ singularity. The resulting superconformal matter is again an unconventional $\mathcal{A}_3$ theory as the collapse involves four discrete charged hypers consistent with the overall picture. Second, the un-Higgsing of the discrete symmetry admits non-trivial structure, as the involved tuning necessarily introduces an SU(2) divisor over one $-2$ curve and introduces further superconformal matter. 
\section{Summary and Outlook}
\label{sec:Summary}

In this work we have taken a preliminary look at \emph{smooth} Calabi-Yau threefold quotient geometries and their consequences for M-/F-theory compactifications in 5- and 6-dimensions. We find that quotients of CY threefolds can lead to smooth, non-simply connected genus one fibrations with singular base geometries, multiple fibers and a discrete gauged symmetry. These orbifold singularities in the base lead to $(2,0)$ SCFT sectors in the associated 6-dimensional physics. The non-simply connected quotiented threefold can be mapped to a geometry with section via its Jacobian. Once in that Jacobian geometry we have the tools to easily read off the effective physics of F-theory compactifications in 6-dimensions. There we find our primary result -- an F-theory vacuum with discretely charged  superconformal matter charged under the discrete symmetry. Within the context of M-theory in 5-dimensions The genus one fibered CY quotient manifolds can be connected to geometries with section via hyperconifold transitions that represent the tensor branch of the superconformal sector with an U(1) gauge symmetry. In addition, we find a number of new results linking Abelian gauge symmetries to superconformal sectors. There are a number of open questions that would be interesting to explore both in the context of the physics and geometry described in this work and we will address each in turn briefly here. 

First, from a geometrical perspective we have found that CY quotient geometries generically lead to theories with superconformal loci (i.e. orbifold fixed points) in the base geometry. The CY quotients remain globally smooth by the addition of multiple fibers over the singular points in the base. In future work it would be intriguing to consider such multiple fibers in more generality. It is possible to ask for example whether the number of such multiple fibers (and relatedly, orbifold fixed points in the base) can be bounded in a CY threefold. If such a bound existed, it could provide intriguing physical constraints on the order of discrete symmetries appearing in such F-theory compactifications.

In addition to being interesting in their own right, the presence of such multiple fibers can effect the form of CY torsors (genus one fibered geometries sharing a common Jacobian). This difference may be manifest in that the full Weil-Ch$\hat{\rm{a}}$telet group could differ from its subgroup, the Tate-Shafarevich group (which has to-date been commonly employed in the F-theory literature). It would be very interesting to fully explore the Weil-Ch$\hat{\rm{a}}$telet group for the quotient manifolds considered here and understand its physical relevance.

Turning next to the associated physics, it is clear that the non-trivial torsion in homology can generate discrete fluxes in Type IIA/IIB or M-theory vacua, but it is unclear whether such fluxes tied to the torsional cycles considered here uplift into F-theory vacua. In future work it would be interesting to investigate more fully the role played by the non-trivial first fundamental group. In addition, completing a careful analysis of the 5-dimensional to 6-dimensional M-theory to F-theory uplift (in the spirit of \cite{Anderson:2014yva}) would be fruitful. This analysis in principle should be straightforward since the CY quotient geometries studied here are smooth, and hence the light states in M-theory are well understood. In a similar spirit, one could also ask whether such CY quotient geometries in F-theory could give rise to interesting dual 6-dimensional heterotic theories. It remains to be seen if $K3$ fibers can survive the quotienting procedures described here as the elliptic fibers have. If so, the multi-section geometry must be understood in the context of heterotic/F-theory duality, generalizing previous efforts in this regard \cite{Cvetic:2016ner}.

Finally, the coupling of discretely charged matter to superconformal points studied here seems to give rise to a potentially novel form of 6-dimensional SCFT. The discrete symmetries studied here arise from the Higgsing of a U(1) theory in the F-theory geometry. The fact that the U(1) symmetries are not localized in F-theory (in the neighborhood of the $\mathbb{C}^2/\Gamma$ singularity) but rather a feature of the global, compact threefold, leads to the interesting question of what happens to the structure of the superconformal sector in a decompactification limit leading to an SCFT or Little String Theory. How are such theories related to the classification of \cite{Heckman:2015bfa, DelZotto:2014hpa,Heckman:2013pva}? Do the discrete charges persist or are they really intrinsic to the compact geometry (i.e. the strongly coupled theory linked to gravity)? We hope to turn to some of these fruitful questions in future work.

\section*{Acknowledgments}
We thank Fabio Apruzzi, Michele del Zotto, I\~naki García-Etxebarria, Markus Dierigl, 
Fabian Ruehle and Sav Sethi for valuable discussions. L. A., A. G.  and J. G,  are grateful to the Aspen Center for Physics for hospitality during the beginning part of this project. We also thank the Banff Research Station for hosting during the final stage of the work. 
 The work of L.A. and J.G. is supported in part by NSF
grant PHY-1720321 and is part of the working group activities of the the 4-VA initiative ``A Synthesis of Two Approaches to String Phenomenology". A.G. was supported in part by a University of Pennsylvania Faculty Working Group. The work of P.K.O. is supported by an individual DFG grant OE 657/1-1.

\appendix
\section{ Threefold in $\mathbb{P}^2 \times$ dP$_6$ fibration}
\label{app:P2Dp6Direct}
In this appendix we want to contrast the F-theory physics of the bi-cubic quotient we have considered in Section~\ref{sec:example1} with 
a geometry that admits also a $\mathbb{Z}_3$ discrete symmetry as well as three $A_2$ points
which however is not coupled to the discrete symmetry. This model represents the smooth CY realization of the first model presented as a Weierstrass model in Section~\ref{sec:Sec2Discrete} and therefore corresponds to the tensor branch of the $A_2$ points.
The geometric setup is very similar as the one we considered in Section~\ref{sec:example1} for the bicubic and the CY threefold is given by the hypersurface in the following polytope $\Delta$
\begin{align}
\begin{tabular}{ccc||ccc}
$x_0$ & $x_1$ & $x_2$ & $y_0$ & $y_1$ & $y_2$ \\ \hline
1 & 0 & -1 & 0 & 0 & 0 \\
0 & 1 & -1 & 0 & 0 & 0 \\
0 & 0 & 0 & 1  & 1 & -2 \\
0 & 0 & 0 & 0 &3 & -3 \\
\end{tabular} \, .
\end{align} 
It is important to note that the base coordinates $y_i$ do not have a 'leg' in the fiber. The above 
threefold is singular and corresponds to the genus one fibration over a singular $\mathbb{P}^2 /\mathbb{Z}_3$ base. The resolved geometry that corresponds to the tensor branch is then given by the following polytope
 \begin{align}
\begin{tabular}{ccc||ccc|cc|cc|cc}
$x_0$ & $x_1$ & $x_2$ & $y_0$ & $y_1$ & $y_2$ & $e_{1,1}$ & $e_{1,2}$ & $e_{2,1}$ & $e_{2,2}$ &  $e_{3,1}$ & $a_{3,2}$\\ \hline
1 & 0 & -1 & 0 & 0 & 0&0&0 & 0 & 0 & 0 & 0    \\
0 & 1 & -1 & 0 & 0 & 0 &0&0 & 0 & 0 & 0 & 0 \\
0 & 0 & 0 & 1  & 1 & -2 &0&-1& -1 & 0 &1 & 1\\
0 & 0 & 0 & 0 &3 & -3 &1&-1&-2 & -1 &1&2\\
\end{tabular} \, .
\end{align}
This ambient space is a direct product manifold $\mathbb{P}^2 \times dP_6$ whereas the CY hypersurface admits Hodge numbers:
\begin{align}
(h^{1,1},h^{2,1})_\chi = (8, 35) \, .
\end{align}
The triangulation is unique and the composition of the two ambient spaces
\begin{align}
\begin{split}
SRI: \{& y_0 y_1, y_0 y_2, y_0 e_{1,1}, y_0 e_{1,2}, y_0 e_{2,1}, y_0 e_{3,2}, y_1 e_{2,2}, y_2 e_{2,2}, e_{1,1} e_{2,2}, e_{1,2} e_{2,2}, e_{2,2} e_{3,1}, e_{2,2} e_{3,2}, \\ & y_1 e_{3,1}, y_2 e_{3,1}, e_{1,1} e_{3,1}, e_{1,2} e_{3,1}, e_{2,1} e_{3,1}, y_1 y_2, y_1 e_{1,2}, y_1 e_{2,1}, y_2 e_{1,1}, e_{1,1} e_{2,1}, e_{1,1} e_{3,2},\\ & y_2 e_{3,2}, e_{1,2} e_{3,2}, e_{2,1} e_{3,2}, e_{1,2} e_{2,1}, x_0 x_1 x_2 \} \, .
\end{split} 
\end{align}
The mayor distinction to the hyperconifold resolution is that all dP$_6$ divisors have no 'leg' in the fiber and hence the ambient space is simply a direct product of $\mathbb{P}^2$ and dP$_6$. The genus one fiber is again described as a generic cubic hypersurface \eqref{eq:cubic} where the ten sections $s_i$ 
all transform in the anticanonical class of the dP$_6$ base and do not factorize further. 
Thus by symmetry we have  $\mathcal{S}_7 = \mathcal{S}_9 = K_b^{-1}$ which we can use to compute the spectrum using the formulas given in \eqref{eq:cubicspectrum}. We compute the spectrum again in the following table and compare the spectrum with that of the hyperconifold resolved quotient of Example~1 of Section~\ref{sec:Example1Resolution} to clarify the distinction.
\begin{center}
 \begin{tabular}{|c |c|c|} \hline
\begin{tabular}{c||c} 
& Bicubic-Quotient  \\ \hline
6d Rep.&  Multi   \\ \hline 
$\mathbf{1}_{1}$      & 63 \\
$\mathbf{1}_0$ &     30 \\ \hline
$\mathbf{T}(1,0)$ &  0 \\  
$\mathbf{T}(2,0)$ & 6 
\end{tabular}
&  
 
 \begin{tabular}{l}
 Fixed point resolved \\ \hline
 Multi  \\ \hline
 $72$  \\  
    $27$ \\ \hline
      $6 $ \\  
   $0$  
   \end{tabular}
   & 
   \begin{tabular}{l} 
   dP6 direct fibration \\ \hline
 Multi  \\ \hline
 $63$  \\  
    $36$ \\ \hline
      $6 $ \\
   $0$
   \end{tabular}\\ \hline
\end{tabular}
\end{center} 
Note that we have performed similar computations for all other examples to illustrate the difference to regular $A_n$ theories that are not coupled to a discrete symmetry.

\section{Specializations of Base Sections}
\label{app:sections}
In this appendix we give the full explicit base dependence we of the sections of the three examples we present in the sections \ref{sec:example1}-\ref{sec:example3} as they
appear in the hypersurface equations.
\subsection{Base sections of Example~1}
The full sections $s_i$ of the bicubic quotient which we use in Weierstrass model of Section~\ref{sec:A2DiscreteCoupled} and are obtained from the smooth geometry of a cubic Equation~\eqref{eq:cubic} in Subsection~\ref{sec:Example1Quotient} are given as
 \begin{align}
\label{eq:siiny}
\begin{split}
s^{(0)}_1 =& a_1 y_0^3 + a_4 y_1^3 + a_{17} y_0 y_1 y_2 + a_7 y_2^3\, , \\
s^{(2)}_2 =& a_9 y_0 y_1^2 + a_{18} y_0^2 y_2 + a_{27} y_1 y_2^2\, , \\
s^{(1)}_3 =& a_{10} y_0^2 y_1 + a_{19} y_1^2 y_2 + a_{28} y_0 y_2^2\, , \\
s^{(0)}_4 =& a_2 y_0^3 + a_5 y_1^3 + a_{20} y_0 y_1 y_2 + a_8 y_2^3\, , \\
s^{(1)}_5 =& a_{11} y_0^2 y_1 + a_{21} y_1^2 y_2 + a_{29} y_0 y_2^2\, , \\
s^{(0)}_6 =& a_{12} y_0^3 + a_{13} y_1^3 + a_{22} y_0 y_1 y_2 + a_{33} y_2^3\, , \\
s^{(2)}_7 =&a_{14} y_0 y_1^2 + a_{23} y_0^2 y_2 + a_{30} y_1 y_2^2\, , \\
s^{(2)}_8 =&a_{15} y_0 y_1^2 + a_{24} y_0^2 y_2 + a_{31} y_1 y_2^2 \, , \\
s^{(1)}_9 =& a_{16} y_0^2 y_1 + a_{25} y_1^2 y_2 + a_{32} y_0 y_2^2\, , \\
s^{(0)}_{10} =& a_0 y_0^3 + a_3 y_1^3 + a_{26} y_0 y_1 y_2 + a_6 y_2^3\, .
\end{split}
\end{align}
After performing three hyperconifold transitions we have the following sections with the additional blow-up coordinates $e_{i,j}$, i=1..3, $j=1,2$, and generic complex coefficients $a_i$:
\begin{align}
\label{eq:sectionsEx1}
\begin{array}{rlrl}
s_ 1 &=&e_{1,1} e_{2,1} e_{3,1} &( e_{2,1} e_{2,2}^2  e_{3,1}^2 e_{3,2} y_0^3 a_ {1} + 
   e_{1,1}^2 e_{1,2} e_{3,1} e_{3,2}^2 y_1^3 a_ {3}   \\&& & + e_{1,1} e_{1,2}^2 e_{2,1}^2 e_{2,2} y_2^3 a_ {5}+ 
    e_{1,1} e_{1,2} e_{2,1} e_{2,2} e_{3,1} e_{3,2} y_0 y_1 y_2 a_ {15} )\, , \\ 
 s_ 2 &=&e_{1,1} e_{1,2} e_{2,1} e_{2,2} e_{3,1} e_{3,2}  &( e_{1,1} e_{3,1} e_{3,2} y_0 y_1^2 a_ {7}  + e_{2,1} e_{2,2} e_{3,1} y_0^2 y_2 a_ {16} + 
   e_{1,1} e_{1,2} e_{2,1} y_1 y_2^2 a_ {25})\, , \\ 
 s_ 3 &=& e_{1,1} e_{1,2} e_{2,1} e_{2,2} e_{3,1} e_{3,2} &(e_{2,2} e_{3,1} e_{3,2} y_0^2 y_1 a_ {8} + e_{1,1} e_{1,2} e_{3,2} y_1^2 y_2 a_ {17} + 
   e_{1,2} e_{2,1} e_{2,2} y_0 y_2^2 a_ {26})\, , \\ 
 s_ 4 &=&e_{1,2} e_{2,2} e_{3,2}&( e_{2,1} e_{2,2}^2 e_{3,1}^2 e_{3,2} y_0^3 a_ {2}  + 
   e_{1,1}^2 e_{1,2} e_{3,1} e_{3,2}^2 y_1^3 a_ {4}\\& && + e_{1,1} e_{1,2}^2 e_{2,1}^2 e_{2,2} y_2^3 a_ {6} +
    e_{1,1} e_{1,2} e_{2,1} e_{2,2} e_{3,1} e_{3,2} y_0 y_1 y_2 a_ {18})\, , \\ 
 s_ 5 &=& e_{1,1} e_{2,1} e_{3,1} &( e_{2,2} e_{3,1} e_{3,2} y_0^2 y_1 a_ {9} + e_{1,1} e_{1,2} e_{3,2} y_1^2 y_2 a_ {19}  + 
   e_{1,2} e_{2,1} e_{2,2} y_0 y_2^2 a_ {27})\, , \\ 
 s_ 6 &=&& e_{2,1} e_{2,2}^2 e_{3,1}^2 e_{3,2} y_0^3 a_ {10} + 
   e_{1,1}^2 e_{1,2} e_{3,1} e_{3,2}^2 y_1^3 a_ {11}  \\& &&+ 
   e_{1,1} e_{1,2} e_{2,1} e_{2,2} e_{3,1} e_{3,2} y_0 y_1 y_2 a_ {20} + 
   e_{1,1} e_{1,2}^2 e_{2,1}^2 e_{2,2} y_2^3 a_ {31}\, ,  \\ 
 s_ 7 &=& e_{1,2} e_{2,2} e_{3,2}&(e_{1,1} e_{3,1} e_{3,2} y_0 y_1^2 a_ {12} + e_{2,1} e_{2,2} e_{3,1} y_0^2 y_2 a_ {21} + 
   e_{1,1} e_{1,2} e_{2,1} y_1 y_2^2 a_ {28})\, , \\ 
 s_ 8 &=&& e_{1,1} e_{3,1} e_{3,2} y_0 y_1^2 a_ {13} + e_{2,1} e_{2,2} e_{3,1} y_0^2 y_2 a_ {22} + 
   e_{1,1} e_{1,2} e_{2,1} y_1 y_2^2 a_ {29}\, ,  \\ 
 s_ 9 &=&& e_{2,2} e_{3,1} e_{3,2} y_0^2 y_1 a_ {14} + e_{1,1} e_{1,2} e_{3,2} y_1^2 y_2 a_ {23} + 
   e_{1,2} e_{2,1} e_{2,2} y_0 y_2^2 a_ {30}\, ,   \\ s_ {10} &=& y_0 y_1 y_2 a_ {24} \, .
\end{array}
\end{align}
\subsection{Base sections of Example~2}
\label{app:Example2}
In the second example, presented in Section~\ref{sec:example2} the fiber is represented as the vanishing of a quartic polynomial in $\mathbb{P}^{1,1,2}$, given in Equation~\eqref{eq:quartic} with nine base sections $d_i$ depending on the $\mathbb{F}_0/\mathbb{Z}_2$ base as
\begin{align}
\begin{split}
 d^{(+)}_1 =& x^2 t^2 a_{5} + y^2 x^2 a_{6} + s^2 t^2 a_{11} + y^2 s^2 a_{12} + 
   y s x t a_{21}\, , \\ 
 d^{(-)}_2 =& s x t^2 a_{15} + y s^2 t a_{18} + y x^2 t a_{23} + y^2 s x a_{26}\, , \\ 
 d^{(+)}_3 =& s^2 t^2 a_{13} + x^2 t^2 a_{16} + y s x t a_{20} + y^2 s^2 a_{24} +
    y^2 x^2 a_{27}\, , \\ 
 d^{(-)}_4 =& s x t^2 a_{14} + y s^2 t a_{17} + y x^2 t a_{22} + y^2 s x a_{25}\, , \\ 
 d^{(+)}_5 =& x^2 t^2 a_{1} + y^2 x^2 a_{2} + s^2 t^2 a_{7} + y^2 s^2 a_{8} + 
   y s x t a_{19}\, , \\ 
 d^{(-)}_6 =& s x t^2 a_{30} + y s^2 t a_{33} + y x^2 t a_{36} + y^2 s x a_{39}\, , \\ 
 d^{(+)}_7 =& s^2 t^2 a_{28} + x^2 t^2 a_{31} + y s x t a_{34} + y^2 s^2 a_{37} +
    y^2 x^2 a_{40}\, , \\ 
 d^{(-)}_8 =& s x t^2 a_{29} + y s^2 t a_{32} + y x^2 t a_{35} + y^2 s x a_{38}\, , \\ 
 d^{(+)}_9 =& x^2 t^2 a_{3} + y^2 x^2 a_{4} + s^2 t^2 a_{9} + y^2 s^2 a_{10} + 
   y s x t a_{41} \, ,
   \end{split}
   \end{align}
   with $s,t,x,y$ being the base coordinates and $a_i$ generic complex coefficients. The superscript of the polynomials denotes their weight under the $\Gamma_{2,b}$ transformation. After performing the four hyperconifold transitions described in Section~\ref{sec:Example2Conifold} we introduce four additional blow-up coordinates $e_i$ and the sections $d_i$ attain the following, partially factorized form
  \begin{align}
  \label{eq:example2Base}
  \begin{array}{lrl}
   d_1 =& e_{1,1} e_{2,1} e_{3,1} e_{4,1} & \left(e_{1,1} e_{2,1}^2 e_{3,1} y^2 x^2 a_{3} + 
     e_{2,1} e_{3,1}^2 e_{4,1} t^2 x^2 a_{4}  \right.\\ && \left. + e_{1,1}^2 e_{2,1} e_{4,1} s^2 y^2 a_{7} + 
     e_{1,1} e_{3,1} e_{4,1}^2 t^2 s^2 a_{8} + e_{1,1} e_{2,1} e_{3,1} e_{4,1} t s y x a_{17}\right)\, , \\ 
 d_2 =&  e_{1,1} e_{2,1} e_{3,1} e_{4,1} &\left(e_{1,1} e_{2,1} s y^2 x a_{11} + e_{1,1} e_{4,1} t s^2 y a_{14} + 
     e_{2,1} e_{3,1} t y x^2 a_{19} + e_{3,1} e_{4,1} t^2 s x a_{22}\right)\, , \\ 
 d_3 =&& e_{1,1}^2 e_{2,1} e_{4,1} s^2 y^2 a_{9} + e_{1,1} e_{2,1}^2 e_{3,1} y^2 x^2 a_{12} + 
   e_{1,1} e_{2,1} e_{3,1} e_{4,1} t s y x a_{16}\\ && + e_{1,1} e_{3,1} e_{4,1}^2 t^2 s^2 a_{20} + 
   e_{2,1} e_{3,1}^2 e_{4,1} t^2 x^2 a_{23}\, , \\ 
 d_4 =&& e_{1,1} e_{2,1} s y^2 x a_{10} + e_{1,1} e_{4,1} t s^2 y a_{13} + 
   e_{2,1} e_{3,1} t y x^2 a_{18} + e_{3,1} e_{4,1} t^2 s x a_{21}\, , \\ d_5 =&& t s y x a_{15}\, , \\ 
 d_6 =& e_{1,1} e_{2,1} e_{3,1} e_{4,1} &\left(e_{1,1} e_{2,1} s y^2 x a_{26} + e_{1,1} e_{4,1} t s^2 y a_{29} + 
     e_{2,1} e_{3,1} t y x^2 a_{32} + e_{3,1} e_{4,1} t^2 s x a_{35}\right)\, , \\  
 d_7 =&& e_{1,1}^2 e_{2,1} e_{4,1} s^2 y^2 a_{24} + e_{1,1} e_{2,1}^2 e_{3,1} y^2 x^2 a_{27} + 
   e_{1,1} e_{2,1} e_{3,1} e_{4,1} t s y x a_{30}\\ && + e_{1,1} e_{3,1} e_{4,1}^2 t^2 s^2 a_{33} + 
   e_{2,1} e_{3,1}^2 e_{4,1} t^2 x^2 a_{36}\, , \\ 
 d_8 =&& e_{1,1} e_{2,1} s y^2 x a_{25} + e_{1,1} e_{4,1} t s^2 y a_{28} + 
   e_{2,1} e_{3,1} t y x^2 a_{31} + e_{3,1} e_{4,1} t^2 s x a_{34}\, , \\  
 d_9 =&& e_{1,1} e_{2,1}^2 e_{3,1} y^2 x^2 a_{1} + e_{2,1} e_{3,1}^2 e_{4,1} t^2 x^2 a_{2} + 
   e_{1,1}^2 e_{2,1} e_{4,1} s^2 y^2 a_{5} \\ && + e_{1,1} e_{3,1} e_{4,1}^2 t^2 s^2 a_{6} + 
   e_{1,1} e_{2,1} e_{3,1} e_{4,1} t s y x a_{37} \, .
  \end{array}
  \end{align}
\subsection{Base sections of Example~3}
\label{app:SectionsExample3}
Here we present the polynomial dependence of the base sections $b_i$ of  the biquadric fiber Equation~\eqref{eq:biquadric} of Example~3 presented in Section~\ref{sec:example3} in terms of the base coordinates $(X,Y,Z,e_1)$ of the $\mathbb{P}^{1,1,2}/\mathbb{Z}_2$ base:
\begin{align}
\label{eq:sectionsExample3}
\begin{array}{lrl}
b^{(+)}_1 =&& Z^2 a_{1} + e_1^2 X^4 a_{7} + e_1^2 Y^4 a_{8} + e_1^2 X^2 Y^2 a_{27} + 
  e_1 X Y Z a_{40}\, , \\ b^{(-)}_2 =& 
 e_1 &\left(e_1 X Y^3 a_{22} + e_1 X^3 Y a_{23} + Y^2 Z a_{35} + 
    X^2 Z a_{36}\right)\, , \\ b^{(+)}_3 =& &
 Z^2 a_{2} + e_1^2 X^4 a_{9} + e_1^2 Y^4 a_{10} + e_1^2 X^2 Y^2 a_{16} + 
  e_1 X Y Z a_{31}\, , \\ b^{(-)}_5 =& 
 e_1 &(e_1 X Y^3 a_{25} + e_1 X^3 Y a_{26} + Y^2 Z a_{38} + 
    X^2 Z a_{39})\, , \\ b^{(+)}_6 =& &
 e_1^2 Y^4 a_{19} + e_1^2 X^2 Y^2 a_{20} + e_1^2 X^4 a_{21} + 
  e_1 X Y Z a_{34} + Z^2 a_{41}\, , \\ b^{(-)}_7 =& 
 e_1 &(e_1 X Y^3 a_{14} + e_1 X^3 Y a_{15} + Y^2 Z a_{29} + 
    X^2 Z a_{30})\, , \\ b^{(+)}_8 =& &
 Z^2 a_{3} + e_1^2 X^4 a_{11} + e_1^2 Y^4 a_{12} + e_1^2 X^2 Y^2 a_{24} + 
  e_1 X Y Z a_{37}\, , \\ b^{(-)}_9 =& 
 e_1 &(e_1 X Y^3 a_{17} + e_1 X^3 Y a_{18} + Y^2 Z a_{32} + 
    X^2 Z a_{33})\, , \\ b^{(+)}_{10} =& &
 Z^2 a_{4} + e_1^2 Y^4 a_{5} + e_1^2 X^4 a_{6} + e_1^2 X^2 Y^2 a_{13} + 
  e_1 X Y Z a_{28} \, ,
  \end{array}
\end{align}
with complex coefficients $a_i$.
Sections with a $(-)$ superscript transform odd under $\Gamma_{2,b}$ and vanish over fixed points. Note that odd sections factor out a $e_1$ coordinate. Performing the hyperconifold transitions described in Section~\ref{sec:Example3Conifold} the base becomes smooth with four additional resolution divisors $A_x, A_y, B_x, B_y$. The nine base sections from above then obtain the form
\begin{align}
\label{eq:sectionsExample3b}
\begin{array}{lrl}
 b_1 =& A_x A_y B_x B_y &(A_x^2 A_y^2 B_x B_y e_1^2 X^2 Y^2 a_{1} + 
      B_x B_y Z^2 a_{3} + A_x A_y^3 B_x^2 e_1^2 Y^4 a_{6}\\ && + 
      A_x^3 A_y B_y^2 e_1^2 X^4 a_{7} + A_x A_y B_x B_y e_1 X Y Z a_{38})\, , \\ 
  b_2 =& A_x A_y B_x B_y e_1 &(A_x A_y^2 B_x e_1 X Y^3 a_{21} + 
      A_x^2 A_y B_y e_1 X^3 Y a_{22} + A_y B_x Y^2 Z a_{33} + 
      A_x B_y X^2 Z a_{34})\, , \\ 
  b_3 =&& B_x B_y Z^2 a_{4} + A_x A_y^3 B_x^2 e_1^2 Y^4 a_{8} + 
    A_x^3 A_y B_y^2 e_1^2 X^4 a_{9}\\ && + A_x^2 A_y^2 B_x B_y e_1^2 X^2 Y^2 a_{15} + 
    A_x A_y B_x B_y e_1 X Y Z a_{29}\, , \\ 
  b_5 =& A_x A_y B_x B_y e_1 &(A_x A_y^2 B_x e_1 X Y^3 a_{24} + 
      A_x^2 A_y B_y e_1 X^3 Y a_{25} + A_y B_x Y^2 Z a_{36} + 
      A_x B_y X^2 Z a_{37})\, , \\ 
  b_6 =&& A_x A_y^3 B_x^2 e_1^2 Y^4 a_{18} + 
    A_x^2 A_y^2 B_x B_y e_1^2 X^2 Y^2 a_{19} + A_x^3 A_y B_y^2 e_1^2 X^4 a_{20}\\ && +
     A_x A_y B_x B_y e_1 X Y Z a_{32} + B_x B_y Z^2 a_{39}\, , \\ 
  b_7 =& e_1 &(A_x A_y^2 B_x e_1 X Y^3 a_{13} + A_x^2 A_y B_y e_1 X^3 Y a_{14} + 
      A_y B_x Y^2 Z a_{27} + A_x B_y X^2 Z a_{28})\, , \\ 
  b_8 =&& B_x B_y Z^2 a_{5} + A_x A_y^3 B_x^2 e_1^2 Y^4 a_{10} + 
    A_x^3 A_y B_y^2 e_1^2 X^4 a_{11}\\ && + A_x^2 A_y^2 B_x B_y e_1^2 X^2 Y^2 a_{23} +
     A_x A_y B_x B_y e_1 X Y Z a_{35}\, , \\ 
  b_9 =& e_1 &(A_x A_y^2 B_x e_1 X Y^3 a_{16} + A_x^2 A_y B_y e_1 X^3 Y a_{17} + 
      A_y B_x Y^2 Z a_{30} + A_x B_y X^2 Z a_{31})\, , \\ 
  b_{10} =& e_1 X Y &(A_x A_y e_1 X Y a_{2} + A_x A_y e_1 X Y a_{12} + Z a_{26}) \, ,
 \end{array}
\end{align}
where we highlighted again the factorized form of the polynomials.

 \section{Conditions on $6$-Dimensional Anomaly Cancellation}
\label{app:Anomalies}
 
In this appendix we give a brief overview of the $6$-dimensional SUGRA relations
obeyed by any anomaly-free theory, following the notations of 
\cite{Taylor:2011wt,Park:2011wv}, which we refer to for more details. In Section~\ref{section4} we check that those conditions all apply, when descending to a quotient theory. Similarly they are checked in the explicit examples of sections \ref{sec:example1}-\ref{sec:example3}. 
For an effective SUGRA theory in $6$ dimensions, the anomaly cancellation 
conditions read: 
\begin{align}
\label{eq:6dAnomalies}
 \text{tr}R^4&:&H-V+ 29T=273\,,\quad  (\text{tr}R^2)^2:\,\,9-T=a\cdot a \,\,\,\,\,\,\,\,(\text{Pure gravitational})\nn\\
  \text{tr}F_\kappa^2\text{tr}R^2&:& \textstyle{ -\frac{1}{6}}\left( A_{adj_\kappa}-\sum_{\mathbf{R}} x_{\mathbf{R}} A_{\mathbf{ R}}\right)=a \cdotp \left( \frac{b_{\kappa}}{\lambda_\kappa}\right) \qquad\qquad\,\,\,\,\,\, (\text{Non-Abelian-gravitational})\nn \\
   F_mF_n\text{tr}R^2&:&-\textstyle{\frac{1}{6}}\sum_{\underline{q}} x_{q_m, q_n} q_m q_n=a\cdotp b_{mn} \qquad\qquad\qquad\qquad\,\,\,\,\,\,\,\,\, (\text{Abelian-gravitational})\nn\\
 \text{tr}F_\kappa^4&:& B_{adj_\kappa} - \sum_{\mathbf{R}} x_{\mathbf{R}} B_{\mathbf{R}} = 0\,,\qquad\qquad\qquad\qquad\qquad\,\,\,\,\,\,\,\,\,\,  (\text{Pure non-Abelian})\nn \\ 
\text{tr}F_\kappa^2\text{tr}F_\kappa^2&:&  \textstyle{\frac{1}{3}} \left( \sum_{\mathbf{R}} x_{\mathbf{R}} C_{\mathbf{R}}-  C_{adj_\kappa}  \right) = \left( \frac{b_\kappa}{\lambda_\kappa}\right)^2 \,,\qquad\qquad \qquad\qquad\,\,\,\,\,\,\,\,\,\,  \phantom{(\text{Pure non-Abelian})} \nn \\
  	F_m F_nF_kF_l&:&\sum_{\underline{q}} x_{q_m,q_n,q_k,q_l} q_m q_n q_k q_l=b_{(mn} \cdotp b_{kl)}\qquad\qquad\qquad\quad\,\,\,\,\,\,\,\,\, (\text{Pure Abelian})\nn \\
   F_{m}F_{n}\text{tr}F_\kappa^2&:&\sum_{\mathbf{R},q_m,q_n} x_{\mathbf{R},q_m,q_n} q_m q_n A_{\mathbf{R}} = \left( \frac{b_\kappa}{\lambda_{\kappa}} \right) \cdotp b_{mn}\,\,\qquad\qquad\quad(\text{Non-Abelian-Abelian} )\nn\\
F_{m}\text{tr}F_\kappa^3&:&\sum_{\mathbf{R},q_m} x_{\mathbf{R},q_m} q_i E_{\mathbf{R}} = 0\,. \qquad\qquad\qquad\qquad\qquad\,\,\,\,\,\,\,\,\,\,  \phantom{(\text{Pure non-Abelian})} 
\end{align}
We have given the terms of the $6$-dimensional anomaly 
polynomial, whose coefficients are the respective anomalies.
The Ricci tensor we denote by $R$ as well as the gauge field strengths $F_\kappa$ and $F_m$
of gauge group factor $G_\kappa$ and the $m$-th U(1). The numbers of hypers, vectors and tensors are denoted
as $H$, $V$ and $T$, respectively. The multiplicities of hypermultiplets 
in the representation $\mathbf{R}$ with m-th U(1) charge $q_m$ is given by $x_{\mathbf{R},q_m}$.

The right hand side of the equations represent their respective GS counter-terms $a$, $b_{\kappa}$ and $b_{mn}$. These transform as ${\rm SO}(1,T)$ vectors, and are 
determined by the underlying microscopic theory.
In context of F-theory compactifications these coefficients are interpreted in terms of geometrical objects that are
\beq \label{eq:bmnSU5}
a=[K_B]\,,\qquad b_\kappa=\mathcal{S}^b_{G_\kappa}\,,\qquad b_{mn}= 
-\pi({\sigma}({\hat s}_n)\cdot {\sigma}({\hat s}_m))\, ,
\eeq
where $K_B$ is the canonical divisor of $B$, 
$\mathcal{S}_{G_\kappa}^b$ is the divisor on $B$  supporting the non-Abelian group $G_\kappa$ 
and $\pi({\sigma}({\hat s}_n)\cdot {\sigma}({\hat s}_m))$ is 
the N\'eron-Tate height pairing.
Under these identifications, the inner product in \eqref{eq:6dAnomalies} with $\Omega_{\alpha,\beta}$ 
is replaced by the intersection pairing on the base $B$.

In addition, in the anomalies \eqref{eq:6dAnomalies}, we have made use 
of several group theory relations between different representations $\mathbf{R}$. Their explicit form can be found in \cite{Taylor:2011wt}. For this work, their explicit values are not relevant, apart from the SU(2) case that we summarize in the following table 
\beq
\text{
\begin{tabular}{|c|c|c|c|c|c|} \hline
 Representation & Dimension & $A_{\mathbf{R}}$ & $B_{\mathbf{R}}$ & $C_{\mathbf{R}}$ & $E_{\mathbf{R}}$  \\ \hline
 Fundamental & $2$ &1 & 0 & 0& 0 \rule{0pt}{1Em}\\ \hline 
  Adjoint & $3$ &$4$ & $0$  & 8 & 0 
 \\ \hline
\end{tabular}
} \, .
\eeq

\section{Jacobians of Genus One Fibers}
\label{app:cubicinWSF}
We summarize the Weierstrass coefficients $f$ and $g$ of the three different genus one fibrations considered in the main text that have, constructed in \cite{AnKimMarshall,artin_tate}.
The generic cubic, given in Equation~\eqref{eq:cubic} has ten sections $s_i$. The Weierstrass coefficients of the associated Jacobian are given as
\begin{align}
\begin{split}
\label{eq:fcubic}
f&=\frac{1}{48} (-(s_6^2 - 4 (s_5 s_7 + s_3 s_8 + s_2 s_9))^2 + 
    24 (-s_6 (s_{10} s_2 s_3 - 9 s_1 s_{10} s_4 + s_4 s_5 s_8 \\
&\phantom{=}+ s_2 s_7 s_8 + s_3 s_5 s_9 +
          s_1 s_7 s_9) + 
      2 (s_{10} s_3^2 s_5 + s_1 s_7^2 s_8 + s_2 s_3 s_8 s_9 + s_1 s_3 s_9^2 \\
&\phantom{=}+  s_7 (s_{10} s_2^2 - 3 s_1 s_{10} s_3 + s_3 s_5 s_8 + s_2 s_5 s_9) + 
         s_4 (-3 s_{10} s_2 s_5 + s_2 s_8^2 + (s_5^2 - 3 s_1 s_8) s_9))))\, , 
\end{split}
\end{align}

\begin{align}
\label{eq:gcubic}
\begin{split}
g&=\frac{1}{864} ((s_6^2 - 4 (s_5 s_7 + s_3 s_8 + s_2 s_9))^3 - 
   36 (s_6^2 - 4 (s_5 s_7 + s_3 s_8 + s_2 s_9)) \\
&\phantom{=}\times (-s_6 (s_{10} s_2 s_3 - 9 s_1 s_{10} s_4 + 
         s_4 s_5 s_8 + s_2 s_7 s_8 + s_3 s_5 s_9 + s_1 s_7 s_9) \\
&\phantom{=}+  2 (s_{10} s_3^2 s_5 + s_1 s_7^2 s_8 + s_2 s_3 s_8 s_9 + s_1 s_3 s_9^2 + 
         s_7 (s_{10} s_2^2 - 3 s_1 s_{10} s_3 + s_3 s_5 s_8 + s_2 s_5 s_9) \\
&\phantom{=}+  s_4 (-3 s_{10} s_2 s_5 + s_2 s_8^2 + (s_5^2 - 3 s_1 s_8) s_9))) + 
   216 ((s_{10} s_2 s_3 - 9 s_1 s_{10} s_4 + s_4 s_5 s_8 \\
&\phantom{=}+ s_2 s_7 s_8 + s_3 s_5 s_9 + 
        s_1 s_7 s_9)^2 + 4 (-s_1 s_{10}^2 s_3^3 - s_1^2 s_{10} s_7^3 - 
         s_4^2 (27 s_1^2 s_{10}^2 + s_{10} s_5^3 \\
&\phantom{=}+ s_1 (-9 s_{10} s_5 s_8 + s_8^3)) + 
         s_{10} s_3^2 (-s_2 s_5 + s_1 s_6) s_9 - s_1 s_3^2 s_8 s_9^2 \\
&\phantom{=}-  s_7^2 (s_{10} (s_2^2 s_5 - 2 s_1 s_3 s_5 - s_1 s_2 s_6) + 
            s_1 s_8 (s_3 s_8 + s_2 s_9)) \\
&\phantom{=}-  s_3 s_7 (s_{10} (-s_2 s_5 s_6 + s_1 s_6^2 + s_2^2 s_8 + 
               s_3 (s_5^2 - 2 s_1 s_8) + s_1 s_2 s_9) \\
&\phantom{=}+  s_9 (s_2 s_5 s_8 - s_1 s_6 s_8 + s_1 s_5 s_9)) + 
         s_4 (-s_{10}^2 (s_2^3 - 9 s_1 s_2 s_3) \\
&\phantom{=}+ s_{10} (s_6 (-s_2 s_5 s_6 + s_1 s_6^2 + s_2^2 s_8) + 
               s_3 (s_5^2 s_6 - s_2 s_5 s_8 - 3 s_1 s_6 s_8)) \\
&\phantom{=}+ (s_{10} (2 s_2^2 s_5 + 3 s_1 s_3 s_5 - 
                  3 s_1 s_2 s_6) + 
               s_8 (-s_3 s_5^2 + s_2 s_5 s_6 - s_1 s_6^2 - s_2^2 s_8 + 
                  2 s_1 s_3 s_8)) s_9 \\
&\phantom{=}+ (-s_2 s_5^2 + s_1 s_5 s_6 + 
               2 s_1 s_2 s_8) s_9^2 - s_1^2 s_9^3 + 
            s_7 (s_{10} (2 s_2 s_5^2 - 3 s_1 s_5 s_6 + 3 s_1 s_2 s_8 + 
                  9 s_1^2 s_9)\, . \\
&\phantom{=}- s_8 (s_2 s_5 s_8 - s_1 s_6 s_8 + s_1 s_5 s_9)))))) \, ,
\end{split}
\end{align}
For a biquadratic polynomial \eqref{eq:biquadric} i.e. a genus one curve in $\mathbb{F}_0$ we have:
\begin{equation} 
\begin{split}\label{eq:f-F2}
f&= \frac{1}{48} [ -(-4 b_1 b_{10} + b_6^2 - 4 (b_5 b_7 + b_3 b_8 + b_2 b_9))^2 
+24 (-b_6 (b_{10} b_2 b_5 + b_2 b_7 b_8 \\
&\phantom{=}+ b_3 b_5 b_9 + b_1 b_7 b_9) 
+ 2 (b_{10}  (b_1 b_5 b_7 + b_2^2 b_8 + b_3 (b_5^2 - 4 b_1 b_8) + b_1 b_2 b_9)  \\
 &\phantom{=}        + b_7 (b_1 b_7 b_8 + b_2 b_5 b_9) +  b_3 (b_5 b_7 b_8 + b_2 b_8 b_9 + b_1 b_9^2)))]\,,
\end{split}
\end{equation}
\begin{equation} 
\begin{split}\label{eq:g-F2}
g&= \frac{1}{864} [ (-4 b_1 b_{10} + b_6^2 - 4 (b_5 b_7 + b_3 b_8 + b_2 b_9))^3 - 36 (-4 b_1 b_{10} + b_6^2 - 
      4 (b_5 b_7\\
&\phantom{=} + b_3 b_8 + b_2 b_9))(-b_6 (b_{10} b_2 b_5 + b_2 b_7 b_8 + b_3 b_5 b_9 + b_1 b_7 b_9) + 
      2 (b_{10} (b_1 b_5 b_7 + b_2^2 b_8 \\
&\phantom{=}+ b_3 (b_5^2 - 4 b_1 b_8) + b_1 b_2 b_9) + b_7 (b_1 b_7 b_8 + b_2 b_5 b_9) + b_3 (b_5 b_7 b_8 + b_2 b_8 b_9 + b_1 b_9^2)))
\\ 
  &\phantom{=} + 216 ((b_{10} b_2 b_5 + b_2 b_7 b_8 + b_3 b_5 b_9 +
  b_1 b_7 b_9)^2 - 4 (b_2 b_3 b_5 b_7 b_8 b_9
\\
  &\phantom{=} + b_1^2 b_{10} (-4 b_{10} b_3 b_8 + b_7^2 b_8 + b_3 b_9^2) + 
         b_{10} (b_3^2 b_5^2 b_8 + b_2^2 b_5 b_7 b_8 + b_2 b_3 (-b_5 b_6 b_8 + b_2 b_8^2 
\\ 
  &\phantom{=} + b_5^2 b_9)) + b_1 (b_{10}^2 (b_3 b_5^2 + b_2^2 b_8) + b_2 b_7^2 b_8 b_9 + b_3^2 b_8 b_9^2 + b_3 b_7 (b_7 b_8^2 - b_6 b_8 b_9 + b_5 b_9^2) \\ 
  &\phantom{=}  + b_{10} (-4 b_3^2 b_8^2 + b_3 b_6 (b_6 b_8 - b_5 b_9) + b_2 b_7 (-b_6 b_8 + b_5 b_9))))) ]\, .
\end{split}
\end{equation}
For the quartic polynomial, the genus one curve in $\mathbb{P}^{1,1,2}$ given in Equation~\eqref{eq:quartichypersurface} there is:
\begin{equation}
\begin{split}
f&=\tfrac{1}{48} [-24 d_9 (-2 d_5 d_6^2 + d_4 d_6 d_7 - 2 d_3 d_6 d_8 + d_2 d_7 d_8 - 
     2 d_1 d_8^2 - 2 d_2 d_4 d_9 + 8 d_1 d_5 d_9) \\
&\phantom{=}
     - (d_7^2 - 4 (d_6 d_8 + d_3 d_9))^2]\,,
\end{split}
\end{equation}
\begin{equation}
\begin{split}
g&=\tfrac{1}{864} [36 d_9 (-2 d_5 d_6^2 + d_4 d_6 d_7 - 2 d_3 d_6 d_8 + d_2 d_7 d_8 - 
     2 d_1 d_8^2 - 2 d_2 d_4 d_9 + 8 d_1 d_5 d_9) \\
&\phantom{=}\times
     (d_7^2 - 4 (d_6 d_8 + d_3 d_9)) \\
&\phantom{=} + (d_7^2 - 4 (d_6 d_8 + d_3 d_9))^3 + 
  216 d_9^2 [4 d_2 d_5 d_6 d_7 - 4 d_1 d_5 d_7^2 
+ d_2^2 d_8^2 + d_4 (-2 d_2 d_6 d_8 + 4 d_1 d_7 d_8) \\
&\phantom{=} - 4 d_2^2 d_5 d_9 + 
     d_4^2 (d_6^2 - 4 d_1 d_9) - 4 d_3 (d_5 d_6^2 + d_1 d_8^2 - 4 d_1 d_5 d_9)]]\, .
\end{split}
\end{equation}
  
    \section{Another Description of the Bicubic}
 \label{app:GIODescription}
It is easy, using the conventional description of the quotiented bicubic, to check that the fibers of our fibration at generic points over the base are smooth as would be expected. However, we might wish to check whether or not the fibers over the orbifold fixed points in the base are multiple in the sense of being non-reduced (that is, everywhere singular). This is a somewhat subtle computation to carry out in the conventional description of the manifold as the group action is mapping fibers at such points to themselves. Given this, let us obtain another description of the quotient of the bicubic by the toric $\mathbb{Z}_3$ action in which the symmetry action is explicitly taken into account and is not imposed in addition to the defining relations. This can provide us with a different perspective on this aspect of the geometry.

We begin by constructing a generating set of monomials that are invariant under the symmetry action \cite{Luty:1995sd}.
\begin{eqnarray} \label{giodefs}
&& g_1 = x_0 \;,\; g_2 =y_0 \;,\; g_3= x_1 y_2 \;,\; g_4 = x_2 y_1 \;,\; g_5 = x_1 x_2 \;,\; \\ \nonumber
&& g_6 =y_1 y_2 \;,\; g_7 = x_1^3 \;,\; g_8 = x_2^3 \;,\; g_9 = y_1^3 \;,\; g_{10} = y_2^3 \;,\; \\ \nonumber
&& g_{11}= x_1^2 y_1 \;,\; g_{12} = x_1 y_1^2 \;,\; g_{13} = x_2^2 y_2 \;,\; g_{14} =x_2 y_2^2
\end{eqnarray}
If one takes the ideal generated as follows,
\begin{eqnarray} 
\hat{I} = \left< g_1 - x_0, g_2-y_0,g_3 - x_1y_2, \ldots \right> \;,
\end{eqnarray} 
and eliminates the original coordinates, one gets an algebraic description of the quotiented ambient space $(\mathbb{P}^2 \times \mathbb{P}^2 )/ \mathbb{Z}_3$.
\begin{eqnarray}  \label{ambmess}
\hat{J} &=& \hat{I} \cap \mathbb{C} [g_1,\ldots,g_{14}] \\ \nonumber
&=& \left< g_3 g_8-g_5 g_{13},g_4 g_7-g_5 g_{11},g_3 g_{13}-g_5 g_{14},g_3 g_{14}-g_5 g_{10},g_3 g_{11}-g_6 g_7, \right. \\ \nonumber && \left.
 g_3 g_4-g_5 g_6,g_{13}^2-g_8 g_{14},g_{13} g_{14}-g_8 g_{10},g_4 g_{13}-g_6 g_8,g_{14}^2-g_{10} g_{13}, \right. \\ \nonumber &&  \left. g_6 g_{14}-g_4 g_{10},g_4 g_{14}-g_6 g_{13},g_{12}^2-g_9 g_{11},g_{11} g_{12}-g_7 g_9,g_6 g_{12}-g_3 g_9,  \right. \\ \nonumber &&  \left. g_4 g_{12}-g_5 g_9,g_{11}^2-g_7 g_{12},g_6 g_{11}-g_3 g_{12},g_4 g_{11}-g_5 g_{12},g_5^3-g_7 g_8,g_3 g_5^2-g_7 g_{13}, \right. \\ \nonumber  && \left.  g_5^2 g_6-g_{11} g_{13},g_4 g_5^2-g_8 g_{11},g_3^2 g_5-g_7 g_{14},g_3 g_5 g_6-g_{11} g_{14},g_5 g_6^2-g_{12} g_{14},  \right. \\ \nonumber  &&  \left.  g_4 g_5 g_6-g_{12} g_{13},g_4^2 g_5-g_8 g_{12},g_6 g_7 g_8-g_5 g_{11} g_{13},g_6 g_7 g_{13}-g_5 g_{11} g_{14},  \right. \\ \nonumber &&  \left. g_3^3-g_7 g_{10},g_3^2 g_{12}-g_6^2 g_7,g_3^2 g_6-g_{10} g_{11},g_3 g_6^2-g_{10} g_{12},g_6^2 g_{13}-g_4^2 g_{10},  \right. \\ \nonumber  && \left. g_6^3-g_9 g_{10},g_4 g_6^2-g_9 g_{14},g_4^2 g_6-g_9 g_{13},g_4^3-g_8 g_9 \right>
\end{eqnarray} 

Note that this method describes the ambient space as a non-complete intersection in a space that inherits non-trivial scalings of coordinates from its parent product of projective spaces. In particular
\begin{eqnarray}  \label{mrscalings}
(g_1 : g_2: g_3:g_4: g_5:g_6: g_7: g_8: g_9 : g_{10} : g_{11}: g_{12} :g_{13}: g_{14} ) \sim \\ \nonumber
(\lambda_1 g_1 : \lambda_2 g_2:  \lambda_1 \lambda_2 g_3: \lambda_1 \lambda_2 g_4: \lambda_1^2 g_5: \lambda_2^2 g_6: \lambda_1^3 g_7:  \lambda_1^3 g_8:  \lambda_2^3 g_9 : \lambda_2^3 g_{10}  \\ \nonumber : \lambda_1^2 \lambda_2 g_{11}:  \lambda_1 \lambda_2^2 g_{12} : \lambda_1^2 \lambda_2 g_{13}: \lambda_1 \lambda_2^2 g_{14} )
\end{eqnarray}
where $\lambda_1$ and $\lambda_2$ are two scalings inherited from the scalings of the original ambient space. Note that the orbifold singularities in the quotiented ambient space are now encoded, not in an explicit $\mathbb{Z}_3$ action, but rather in the usual singularities that occur in such weighted projective spaces. There is also an associated Stanley Reisner Ideal which we neglect to write out here.

\vspace{0.1cm}

We are, of course, not interested in the ambient space but rather the Calabi-Yau hypersurface inside it. This can be computed in a very similar manner.  Let us choose a random complex structure, consistent with the $\mathbb{Z}_3$ action for the initial upstairs description of the threefold defining relation.
\begin{eqnarray}  \label{exdefnrel}
p=58 x_0^3 y_0^3 + 49 x_1^3 y_0^3 + 86 x_0 x_1 x_2 y_0^3 + 51 x_2^3 y_0^3 + 
 20 x_0 x_1^2 y_0^2 y_1\\ \nonumber + 44 x_0^2 x_2 y_0^2 y_1 
 + 34 x_1 x_2^2 y_0^2 y_1 + 15 x_0^2 x_1 y_0 y_1^2 + 55 x_1^2 x_2 y_0 y_1^2 
\\ \nonumber   + 24 x_0 x_2^2 y_0 y_1^2 + 22 x_0^3 y_1^3 
 + 86 x_1^3 y_1^3 + 94 x_0 x_1 x_2 y_1^3 + 68 x_2^3 y_1^3  \\ \nonumber + 73 x_0^2 x_1 y_0^2 y_2 
 + 29 x_1^2 x_2 y_0^2 y_2
 + 95 x_0 x_2^2 y_0^2 y_2 + 63 x_0^3 y_0 y_1 y_2  \\ \nonumber + 11 x_1^3 y_0 y_1 y_2 
 + 30 x_0 x_1 x_2 y_0 y_1 y_2
 + 90 x_2^3 y_0 y_1 y_2 + 55 x_0 x_1^2 y_1^2 y_2  \\ \nonumber + 20 x_0^2 x_2 y_1^2 y_2 + 
 66 x_1 x_2^2 y_1^2 y_2
 + 69 x_0 x_1^2 y_0 y_2^2 + 3 x_0^2 x_2 y_0 y_2^2  \\ \nonumber + 49 x_1 x_2^2 y_0 y_2^2 
 + 78 x_0^2 x_1 y_1 y_2^2
 + 51 x_1^2 x_2 y_1 y_2^2 + 11 x_0 x_2^2 y_1 y_2^2 + 38 x_0^3 y_2^3 
\\ \nonumber  + 20 x_1^3 y_2^3 + 100 x_0 x_1 x_2 y_2^3 + 37 x_2^3 y_2^3
\end{eqnarray} 
We then simply perform the following elimination to obtain an algebraic description of the quotiented Calabi-Yau threefold.
\begin{eqnarray}
I &=& \left< g_1 - x_0, g_2-y_0,g_3 - x_1y_2, \ldots , p \right>  \\
J &=& I \cap \mathbb{C} [g_1,\ldots,g_{14}] 
\end{eqnarray}
Performing this computation we arrive at the following.
\begin{eqnarray} \label{mrtheideal}
J= \left< g_3 g_8-g_5 g_{13},g_4 g_7-g_5 g_{11},g_3 g_{13}-g_5 g_{14},g_3 g_{14}-g_5 g_{10},g_3 g_{11}-g_6 g_7, \right. \\ \nonumber \left.  g_3 g_4-g_5 g_6,g_{13}^2-g_8 g_{14},g_{13} g_{14}-g_8 g_{10},g_4 g_{13}-g_6 g_8,g_{14}^2-g_{10} g_{13},\right. \\ \nonumber \left.  g_6 g_{14}-g_4 g_{10},g_4 g_{14}-g_6 g_{13},g_{12}^2-g_9 g_{11},g_{11} g_{12}-g_7 g_9,g_6 g_{12}-g_3 g_9,\right. \\ \nonumber \left.  g_4 g_{12}-g_5 g_9,g_{11}^2-g_7 g_{12},g_6 g_{11}-g_3 g_{12},g_4 g_{11}-g_5 g_{12},g_5^3-g_7 g_8,g_3 g_5^2-g_7 g_{13},\right. \\ \nonumber \left.  g_5^2 g_6-g_{11} g_{13},g_4 g_5^2-g_8 g_{11},g_3^2 g_5-g_7 g_{14},g_3 g_5 g_6-g_{11} g_{14},g_5 g_6^2-g_{12} g_{14},\right. \\ \nonumber \left.  g_4 g_5 g_6-g_{12} g_{13},g_4^2 g_5-g_8 g_{12},g_6 g_7 g_8-g_5 g_{11} g_{13},g_6 g_7 g_{13}-g_5 g_{11} g_{14},\right. \\ \nonumber \left.  g_3^3-g_7 g_{10},g_3^2 g_{12}-g_6^2 g_7,g_3^2 g_6-g_{10} g_{11},g_3 g_6^2-g_{10} g_{12},g_6^2 g_{13}-g_4^2 g_{10},\right. \\ \nonumber \left.  g_6^3-g_9 g_{10},g_4 g_6^2-g_9 g_{14},g_4^2 g_6-g_9 g_{13},g_4^3-g_8 g_9,\right. \\ \nonumber \left. 58 g_2^3 g_1^3+63 g_2 g_6 g_1^3+22 g_9 g_1^3+38 g_{10} g_1^3+73 g_2^2 g_3 g_1^2+44 g_2^2 g_4 g_1^2 \right. \\ \nonumber \left.+78 g_3 g_6 g_1^2+20 g_4 g_6 g_1^2+15 g_2 g_{12} g_1^2+3 g_2 g_{14} g_1^2+69 g_2 g_3^2 g_1+24 g_2 g_4^2 g_1 \right. \\ \nonumber \left.+86 g_2^3 g_5 g_1+30 g_2 g_5 g_6 g_1+94 g_5 g_9 g_1+100 g_5 g_{10} g_1+20 g_2^2 g_{11} g_1+55 g_3 g_{12} g_1\right. \\ \nonumber \left.+95 g_2^2 g_{13} g_1+11 g_6 g_{13} g_1+29 g_2^2 g_3 g_5+34 g_2^2 g_4 g_5+49 g_2^3 g_7+11 g_2 g_6 g_7\right. \\ \nonumber \left.+51 g_2^3 g_8+90 g_2 g_6 g_8+86 g_7 g_9+68 g_8 g_9+20 g_7 g_{10}+37 g_8 g_{10}+55 g_2 g_5 g_{12}\right. \\ \nonumber \left.+66 g_{12} g_{13}+49 g_2 g_5 g_{14}+51 g_{11} g_{14} \right>
\end{eqnarray}
Note that the first 8 lines here reproduce the description of the quotiented ambient space that we obtained in (\ref{ambmess}). The remaining generator describes the Calabi-Yau as a hypersurface within this ambient space.

\vspace{0.1cm}

Let us now examine this description of the downstairs manifold and observe the fibration structure and the nature of the fibers over the fixed points in the base. We begin by describing the fibration itself in this language.

\vspace{0.1cm}

The projection map for the fibration is given by 
\begin{eqnarray}
g_4\to 0\;,\;g_3\to 0\;,\;g_5\to 0\;,\;g_{11}\to 0\;,\;g_{12}\to 0\;,\; \\\nonumber g_{14}\to 0\;,\;g_{13}\to 0\;,\;g_1\to 0\;,\;g_7\to 0\;,\;g_8\to 0
\end{eqnarray}
Taking the image of the entire manifold under this map we obtain the following defining relation for the base
\begin{eqnarray} \label{basedef}
g_6^3-g_9 g_{10} =0
\end{eqnarray}
in the weighted coordinates $(g_2 : g_6 : g_9 : g_{10})\sim(\lambda_2 g_2 : \lambda_2^2 g_6 : \lambda_2^3 g_9 : \lambda_2^3 g_{10})$. By studying the nature of the gauge invariant operators (GIOs) $g_2 =y_0, g_6 = y_1 y_2, g_9=y_1^3$ and $g_{10}=y_2^3$ we can see that this description of the base is simply the description of $\mathbb{P}^2/\mathbb{Z}_3$ that would be obtained by using the same formalism that we have employed above to describe the quotient of the total space. This $2$-dimensional base clearly has orbifold singularities thanks to the non-homogeneous scalings.

Perhaps the easiest orbifold fixed point to see explicitly corresponds to $(y_0:y_1:y_2)=(0,0,1)$, or $(g_2:g_6:g_9:g_{10})=(0:0:0:1)$ in 
the current description (note if you set these $g$'s to zero, the ideal (\ref{mrtheideal}) then implies that $g_4=g_{11}=g_{12}=0$ also). If one perturbs slightly away from this point and makes $g_2,g_6$ and $g_9$ slightly non-zero then we see that we identify three sets of homogeneous coordinates with the scaling while leaving $g_{10}$ unchanged (by taking $\lambda_2$ to be a third root of unity). As we take $g_2,g_6$ and $g_9$ back to zero these three identified points coalesce - giving us a triple fixed point. Note that, in performing this analysis, we have simply swapped the use of the symmetry and two scalings to see fixed points and multiple fibers (as employed in the upstairs picture) for just two scalings (at the price of those scalings becoming inhomogeneous).

What does the fibre look like over such a singular point in the base? We can see from the above analysis that we have three identified fibers coalescing at this one point - and thus in that sense we have a triple fiber. Algebraically we can find an expression for the fiber by simply substituting the values $(g_2:g_6:g_9:g_{10})=(0:0:0:1)$ into the ideal (\ref{mrtheideal}). Upon doing this and performing some trivial algebra, we arrive at the following description of the fibre in terms of the coordinates $g_1,g_3$ and $g_{14}$.
\begin{eqnarray} \label{fibdef1}
38 g_1^3 + 37 g_{14}^3 + 100 g_1 g_{14} g_3 + 20 g_3^3
\end{eqnarray}
Notice from (\ref{mrscalings}) that all three of these variables scale linearly with $\lambda_1$. Thus the fiber is described by a cubic in $\mathbb{P}^2$ and is manifestly an elliptic curve as expected (a more careful analysis of (\ref{giodefs}) shows that this $\mathbb{P}^2$ is identical to the first $\mathbb{P}^2$ in the original description and in particular therefore has the correct Stanley Reisner ideal). Notice also that the ideal describing the fiber over the singular point is primary and thus the fiber is irreducible. The ``triple'' nature of the fiber can only be seen by the argument of the proceeding paragraphs. A straightforward and standard analysis shows that this elliptic curve is smooth everywhere. This is as expected, as in these examples the orbifold singularities in the ambient space miss the Calabi-Yau. 

A similar analysis to that presented above can also be carried out when we tune the complex structure of the manifold to a singular point where we regain a section to the fibration. This could be done, for example by setting all of the coefficients of $x_2^3$ in (\ref{exdefnrel}) to zero. We then find the following description for the fiber over a fixed point in the base, replacing (\ref{fibdef1}).
\begin{eqnarray} \label{fibdef2}
38 g_1^3+100 g_3 g_{14} g_1+20 g_3^3=0
\end{eqnarray}
This fiber is singular at the point $g_1=g_3=0$, which is not unexpected as the Calabi-Yau is singular after such a tuning of complex structure. Note, however, that the fiber is still not singular everywhere and thus the multiple nature of the fiber over the orbifold fixed point can only be seen by considering the action of the scalings, even in this limit.

To add a final insight into this somewhat convoluted description of the multiple fiber, it should be noted that we could perform the analysis above for the standard Enriques quotient of $K3$. In a close analogy to the bi-cubic, consider a $K3$ surface defined as a $\{2,2,2\}$ hypersurface in a $\mathbb{P}^1 \times \mathbb{P}^1 \times \mathbb{P}^1$ ambient space. Then the toric $\mathbb{Z}_2$ action, $x_i \to (-1)^i x_i$, in each ambient $\mathbb{P}^1$ yields a smooth surface with a non-trivial, finite first fundamental group. In addition, the Atiyah-Singer index of the resulting quotiented surface is $\text{Ind}(K3/\mathbb{Z}_2)=1$. Hence the quotient produces an Enriques surface. Repeating the analysis above provides an identical description of the well-known two multiple fibers of the Enriques surface \cite{beauville} realized by $\mathbb{Z}_2$ scalings as above.
 

\end{document}